\documentclass[usenatbib,a4paper,fleqn,twocolumn]{mnras}

\usepackage{txfonts}
\usepackage{amssymb}	% Extra maths symbols
\usepackage{graphicx}
\usepackage{txfonts}
\usepackage{adjustbox}
\usepackage{caption}
\usepackage{float}
\usepackage[flushleft]{threeparttable}
\usepackage{longtable}
\usepackage{textcomp}
\usepackage{multirow}
\usepackage{hhline}
\usepackage{lscape}
\usepackage{pdflscape}
\usepackage{colortbl}
\usepackage{multicol}
\usepackage{natbib}

\title[SUDARE-VOICE Variability And Transient Search]{Variability and transient 
        search in the SUDARE-VOICE field: a new method to extract the light curves}
\author[Liu et al.]
           {Dezi Liu$^{1,2}$\thanks{E-mail: adzliu@ynu.edu.cn},
            Wenqiang Deng$^{1,3}$, 
            Zuhui Fan$^{1}$\thanks{E-mail: zuhuifan@ynu.edu.cn},
            Liping Fu$^{2}$\thanks{E-mail: fuliping@shnu.edu.cn},
            Giovanni Covone$^{4,5,6}$, 
            \newauthor
            Mattia Vaccari$^{7,8}$,
            Mario Radovich$^{9}$,
            Massimo Capaccioli$^{5}$,
            Demetra De Cicco$^{10,11}$,
            \newauthor
            Aniello Grado$^{4}$,
            Lucia Marchetti$^{12,7,8}$, 
            Nicola Napolitano$^{4}$,
            Maurizio Paolillo$^{5,4,6}$,
            \newauthor
            Giuliano Pignata$^{13,10}$,
            Fabio Ragosta$^{5,4,6}$
\\
$^{1}$South-Western Institute for Astronomy Research, Yunnan University, Kunming 650500, China \\
$^{2}$The Shanghai Key Lab for Astrophysics, Shanghai Normal University, 100 Guilin Road, Shanghai 200234, China \\
$^{3}$School of Physics and Astronomy, Yunnan University, Kunming 650500, China \\
$^{4}$INAF--Osservatorio Astronomico di Capodimonte, Salita Moiariello 16, Napoli 80131, Italy \\
$^{5}$Department of Physics, University of Napoli “Federico II”, via Cinthia 9, 80126 Napoli, Italy \\
$^{6}$INFN, Sezione di Napoli, Napoli 80126, Italy \\
$^{7}$Department of Physics \& Astronomy, University of the Western Cape, Robert Sobukwe Road,
7535 Bellville, Cape Town, South Africa \\
$^{8}$INAF - Istituto di Radioastronomia, via Gobetti 101, 40129 Bologna, Italy \\
$^{9}$INAF--Osservatorio Astronomico di Padova, vicolo dell'Osservatorio 5,  Padova 35122, Italy \\
$^{10}$Millennium Institute of Astrophysics (MAS), Nuncio Monse\~nor S\'otero Sanz 100, Providencia, Santiago, Chile \\
$^{11}$Instituto de Astrof\'{i}sica, Pontificia Universidad Cat\'{o}lica de Chile, Av. Vicu\~{n}a Mackenna 4860, 7820436 Macul, Santiago, Chile\\
$^{12}$ Department of Astronomy, University of Cape Town, Private Bag X3, Rondebosch 7701, Cape Town, South Africa\\
$^{13}$Departemento de Ciencias Fisicas, Universidad Andres Bello, Avda. Republica 252, Santiago, Chile\\
}

\date{Accepted XXX. Received YYY; in original form ZZZ}

\pubyear{2019}

\begin{document}
\label{firstpage}
\pagerange{\pageref{firstpage}--\pageref{lastpage}}
\maketitle

\begin{abstract}
The VST Optical Imaging of the CDFS and ES1 Fields (VOICE) Survey, in synergy with the SUDARE 
survey, is  a deep optical $ugri$ imaging of the CDFS and ES1 fields using the VLT Survey Telescope 
(VST). The observations for the CDFS field comprise about 4.38 deg$^2$ down to $r\sim26$\,mag. 
The total on-sky time spans over four years in this field, distributed over four adjacent sub-fields. In 
this paper, we use the multi-epoch $r$-band imaging data to measure the variability of the detected 
objects and search for transients. We perform careful astrometric and photometric calibrations and point 
spread function (PSF) modeling.  A new method, referring to as differential running-average photometry, 
is proposed to measure the light curves of the detected objects. With the method, the difference of PSFs 
between different epochs can be reduced, and the background fluctuations are also suppressed. 
Detailed uncertainty analysis and detrending corrections on the light curves are performed. We visually 
inspect the light curves to select variable objects, and present some objects with interesting light curves. 
Further investigation of these objects in combination with multi-band data will be presented in our 
forthcoming paper.
\end{abstract}


\section{Introduction}
Time-domain astronomy opens a new window to study the properties of astronomical objects.
Properly timed observations allow to obtain their light curves which represent the flux variations as a function 
of time. The shapes of light curves contain physical information of different types of objects, such as the pulsating 
stars, active galactic nuclei (AGNs),  
supernovae (SNe), tidal disruption events (TDEs), and so forth \citep{2009MNRAS.400.2070S,2000A&A...357..816C,
2012NewAR..56..122W,2015pust.book.....C,2016ASPC..505..107L}. In addition to providing clues on the nature and 
origin of these variable objects, they can also be used as tools for astrophysical applications. For example, the 
period-luminosity relations of Cepheids and RR Lyrae stars are crucial anchors in distance measurements 
\citep{2007AJ....133.1810B,2018ApJ...869...82R}. The characteristic light-curve behaviors of Type Ia supernovae 
make them standardizable candles to measure distance on cosmological scales, and thus to probe the expansion 
history of the Universe \citep{1998AJ....116.1009R,1999ApJ...517..565P,2019ApJ...872L..30A}. The brightness drop 
of stars caused by their transiting planets leads to abundant discoveries of exoplanet candidates 
\citep{2018ApJS..235...38T,2018AJ....156..102S}. Given the importance, many surveys have been dedicated to 
time-domain observations, e.g. the Panoramic Survey Telescope and Rapid Response System (Pan-STARRS; 
\citet{2004SPIE.5489...11K}), the Catalina Real-Time Surveys 
(CRTS; \citet{2009ApJ...696..870D}), the Palomar Transient Factory (PTF/iPTF; \citet{2009PASP..121.1395L}), the 
High Cadence Transient Survey (HiTS; \citet{2016ApJ...832..155F,2018AJ....156..186M}), the SkyMapper Transient 
Survey \citep{2017PASA...34...30S}, and the Zwicky Transient Facility (ZTF; \citet{2019PASP..131g8001G}). The 
upcoming facilities, such as the Large Synoptic Survey Telescope (LSST; \citet{2019ApJ...873..111I}), will also 
conduct time-domain observations with about a half sky coverage to faint magnitudes. 

To identify the variable objects and measure their light curvers, different methods have been applied, 
including the difference imaging analysis \citep{1998ApJ...503..325A,2015AJ....149...50O,2016ApJ...830...27Z}, 
and point spread function (PSF) homogenization \citep{2017ApJ...849..110S} \textit{etc}. 
The forward modeling of the entire image (galaxy+transient) in a non-parametric manner 
\citep{Fabbro2001PhD,2006A&A...447...31A} has also been applied to obtain high precision light curves for known 
transients without involving explicit image subtractions. Since its first application to microlensing 
surveys \citep{1998ApJ...503..325A}, the difference imaging (or image subtraction) method has been widely used 
in many surveys, such as PTF/iPTF, HiTS and ZTF. To perform image subtraction, a reference image should be first 
built which can either be the image with the best seeing or the coadded image from multiple exposures.  A newly 
observed image is then subtracted from the reference one after PSF homogenization so that the flux variations of 
the objects therein can be detected from the residual image. 
The implementation of the image subtraction method is relatively easy and fast. This makes it particularly well suited to 
search for transients in wide sky surveys. In practice, however, this method suffers from several 
limitations. Firstly, the PSF varies  spatially  over the entire image. For telescopes with a large field of view, the 
PSF variations are generally significant from the center to the edge in the focal plane. Therefore, accurately 
modeling the PSF and performing homogenization between the reference and new image are challenging. 
Secondly, the variable background noise between different exposures makes the observed depths different. 
The subtraction between the reference and new image will further magnify the background noise. The two 
facts can lead to large uncertainties of the measured variability or even spurious detections \citep{2016ApJ...830...27Z}.

In this paper, we propose a new method, referring to as differential running-average photometry (\texttt{drap}), 
to measure the variability of objects and apply it to the SUDARE-VOICE $r$-band imaging data by taking the 
advantage of the long time baseline and high image quality of the survey. This method can moderately
mitigate  the difference of PSFs between different exposures, and suppress the background fluctuations, making 
it applicable to data with relatively long time accumulations and a large enough number of exposures.  
The paper is organized as follows. In section~\ref{sec:voice}, we describe the SUDARE-VOICE observations and 
detailed data reduction methods.  The methodology for light curve extraction 
is presented in section~\ref{sec:drap}. We shows some typical results in section~\ref{sec:res}. Finally, summaries are 
given in section~\ref{sec:sum}. In Appendix~\ref{appen}, we explain the PSF variations in \texttt{drap}. All magnitudes 
quoted in this paper are in the AB system.

\section{The SUDARE-VOICE Survey}\label{sec:voice}
The VOICE survey\footnote{http://www.mattiavaccari.net/voice/} (PIs: Giovanni Covone \& Mattia Vaccari; 
\citet{2016heas.confE..26V}),  in synergy with the SUDARE survey \citep{2015A&A...584A..62C,2017A&A...598A..50B}, 
was proposed to cover about eight square degrees evenly split between the CDFS \citep{2001ApJ...551..624G,
2001ApJ...562...42T} and the ES1 \citep{2000MNRAS.316..749O,2004MNRAS.351.1290R,
2005MNRAS.358..397V} fields in four optical $ugri$ bands using VST/OmegaCam camera. 
The project also includes additional coverage of the COSMOS field (PI: Giuliano Pignata) with a 
smaller sky coverage but extended to longer baseline of 3 years \citep{2015A&A...574A.112D,
2019A&A...627A..33D}.
The VST, located at  Cerro Paranal, Chile, is a 2.6-m 
modified Ritchey-Chretien alt-az  telescope designed for wide-field optical imaging. The equipped 
OmegaCAM \citep{2011Msngr.146....8K} is a mosaic of $8 \times 4$ CCD chips, each with 4k\,$\times$\,2k pixels. 
It covers 1$^{\circ}$\,$\times$\,1$^{\circ}$ field-of-view with a pixel scale of 0.214$\arcsec$.

The  SUDARE-VOICE survey aims at providing deep optical images in the targeted 
fields to enable various astrophysical studies in conjunction with other existing data covering different 
wavelengths \citep{2015fers.confE..27V,2016ASSP...42...71V,2010A&A...518L..20V}. The imaging observations 
of the CDFS field have been completed. The entire field was divided into four tiles (CDFS1--4), with each about 
one square degree. Over one hundred exposures, spanning almost two years, with a single exposure time of 
360 seconds, were obtained for each tile \citep{2015A&A...579A.115F}. 
Observations were performed in dithering 
mode, made of at least five consecutive exposures in one night ( or one epoch), to cover the detector gaps.
The images were preprocessed (including instrumental effect removal, flat fielding, 
CCD gain harmonization, and illumination correction etc.) with the VST-Tube pipeline \citep{2012MSAIS..19..362G}. 
With the multi-epoch imaging data, many astrophysical topics have been investigated, such as the studies of the SN 
explosion rate \citep{2015A&A...584A..62C,2017A&A...598A..50B}, the variability-based selections of AGNs  
\citep{2015A&A...579A.115F,2020arXiv200102560P}, as well as the weak gravitational lensing shear measurements 
and cosmological analyses \citep{2018MNRAS.tmp.1498F,2018MNRAS.478.2388L}.

The $r$-band observations were taken with a cadence (i.e. the time interval between two consecutive epochs) of about 
3-4 days, avoiding the ten days around the full moon. The $g$- and $i$-band observations were taken every seven days, 
and the $u$-band observations did not have a specific cadence.  Because of their best cadence and 
image quality, we focus on the analyses of $r$-band data in the present study. Totally, there are 35, 25, 34, and 
30 epochs for CDFS1-4 fields, respectively. The average 5$\sigma$ limiting magnitude of individual epochs is about 24.3\,mag for 
point source within 1.0 arcsec aperture radius. In the rest of this section, we will give detailed description on the 
image processing procedures.

\subsection{Image Reduction}
As mentioned above, the $r$-band single exposure images have been preprocessed by the VST-Tube 
pipeline \citep{2012MSAIS..19..362G}. For accurate variability analyses, we start with the images after 
removing the instrumental effects by the VST-Tube pipeline, and continue to perform additional calibrations 
by using our customized routines, including cosmic-ray removal, background subtraction,
astrometric and photometric calibrations. 

\subsubsection{Cosmic Ray Removal and Background Subtraction}
Careful removal of the cosmic rays is crucial because any residuals on the detected astronomical 
objects may lead to spurious variabilities. We use a modified \texttt{Python} 
code\footnote{http://www.astro.yale.edu/dokkum/lacosmic/} that implements the L.A. 
Cosmic algorithm to detect and remove cosmic rays. The algorithm is based 
on a variant of Laplacian edge detection \citep{2001PASP..113.1420V}. It is capable of detecting and rejecting 
cosmic rays with arbitrary shape by convolving a 2D Laplacian kernel which is sensitive to variations on small 
scales. By applying this algorithm, however, we find that the peak values of some unsaturated bright point objects 
(typically, about $\sim$10-20 such point objects on each CCD chip) can be misclassified as cosmic rays, especially 
those observed under good seeing conditions. Our analysis shows that their peak values are systematically larger than 
half of the saturation level of the CCD chips. To overcome the problem, we slightly modify the code to include 
additional information from a flag map. To create the flag map, we first run \texttt{SExtractor} 
(version 2.19.5; \citet{1996A&AS..117..393B}) on each exposure for object detection, and then assign the isophotal 
pixels of the point objects with peak values larger than half of the saturation level to zero. Other pixels in the  flag map are 
set to be one. With this map, the bright point objects will not be considered for cosmic ray detections. We determine the best 
parameters by visual inspection of the mask images, and remove the cosmic rays through iterating the algorithm 
three times. Residual cosmic rays, including those potentially superposing on the bright point objects, 
will be further rejected in our following reduction procedures.

We run \texttt{SExtractor} to subtract the background for each 
CCD chip separately. To construct the background map, \texttt{SExtractor} estimates the local 
background in each mesh of a grid (64$\times$64 pixels) that covers the entire CCD chip. In the presence 
of bright or saturated stars, however, the local background will be overestimated, hence leading to an underestimate 
of the fluxes of real objects. Therefore, we perform the background subtraction in two steps. 
We first create a preliminary background-subtracted image with \texttt{SExtractor} and detect the objects 
using a low detection threshold (i.e. \texttt{DETECT\_THRESH}=1.5). Then these objects are masked from the original image. To reduce the effect 
of the residual light which is below the detection threshold, the mask region of each object is slightly enlarged. 
Using the same method as described in \citet{2017AJ....153...53L}, for a specific object-masked region, the 
median value $a$ and variance $\sigma^2$ are calculated through its adjacent pixels (at least 900 unmasked pixels). 
The masked pixels  are then filled with random numbers sampled by the Gaussian distribution $N(a, \sigma^2)$. 
Compared to the conventional interpolation method, this procedure preserves the local statistical properties and 
eliminates many artificial effects. Secondly, we re-run \texttt{SExtractor} on the object-masked image to construct 
the background map  and subtract it from the original image. Our analyses show that this method can produce better 
local background estimate.

\subsubsection{Astrometric and Photometric Calibrations}
We use \texttt{SCAMP} (version 2.2.6; \citet{2006ASPC..351..112B}) for astrometric and photometric calibrations. 
The calibrations are performed on every epoch individually. For astrometric calibration, the \textit{Gaia} DR1 
catalog \citep{2016A&A...595A...2G} is used as reference. 
%The \texttt{SCAMP} is run twice in order to derive the 
%accurate astrometric solution. In the first run we consider the detector positions to be independent between 
%exposures (i.e. \texttt{MOSAIC\_TYPE = LOOSE}). In that case, the astrometric calibration is performed 
%separately for each exposure. Then we run \texttt{SCAMP} again to derive a common and median relative positioning 
%of the 32 chips within the focal plane (\texttt{MOSAIC\_TYPE = FIX\_FOCALPLANE}). 
The final rms offsets of the astrometry are less than 0.06 arcsec along both right ascension and declination axes.

Homogeneous photometric calibration between different epochs is essential for accurate variability measurements. 
Taking into account potential zeropoint variations between different CCD chips and different exposures for a given 
epoch, we first run \texttt{SCAMP} to perform relative (internal) photometric calibration between different exposures 
so that the mean of the relative flux scaling parameter (\texttt{FLXSCALE}) is close to 1.0. Then we run \texttt{SWarp} 
(version 2.38.0; \citet{2010ascl.soft10068B}) with median mode  to stack the individual exposures, and create a single-epoch  
image as well as the corresponding inverse variance weight map. The median stacking can further reject the 
residual cosmic rays. We run \texttt{SExtractor} to detect the bright objects 
in these single-epoch images and match the corresponding  catalogs individually with the \textit{Gaia} DR1 star 
catalog to generate the star samples for all epochs. The magnitudes of the stars are then restricted to be in the range of 
17.5\,mag to 21.5\,mag. Quantitative comparison of the instrumental magnitudes of the common stars between 
any two different epochs shows that the median of the magnitude difference varies. For a few epochs, the difference can be 
even larger than 0.1\,mag. This can be attributed to either the impact of different airmass or the non-photometric conditions. 
In this work, we do not distinguish these different effects and simply regard  them as zeropoint variations.
To eliminate such difference, we set the epoch with the best seeing in CDFS1 sub-field as reference, and scale the 
fluxes of other single-epoch images to the reference.  The flux scaling factors are derived by comparing the instrumental fluxes of stars 
between the reference and other images. The partial overlap between the four sub-fields enables us to homogenize the 
zeropoints of all the single-epoch images to the reference. We find that the minimum overlap region between two adjacent sub-fields is 
about 110 arcmin$^{2}$, resulting in $\sim$150 common stars with good quality. According to the procedures, the final dispersion 
of the photometric calibration between different epochs is smaller than 0.02\,mag.

\subsubsection{Image Coaddition and Photometry}\label{subsec:imgcp}
To assess the quality of each single-epoch image, we calculate the full width at half maximum (FWHM) and the  elongation 
of stars, and the background fluctuation $\sigma_\mathrm{bkg}$. We firstly exclude 
the epochs with median elongation larger than 0.1. The large elongation most probably results from the tracking instability of the 
telescope during the observation. The epochs with $\mathrm{FWHM}>1.2\arcsec$ and $\sigma_\mathrm{bkg}>15.0$\,ADUs 
are also rejected from the following analyses in order to reduce the object blending effect and positional uncertainty, as well as to 
optimize the signal-to-noise ratio (SNR) of objects. The excluded epochs have either large seeing or shallow limiting magnitude. Finally, 
we have 27, 21, 24 and 26 epochs for the CDFS1-4 sub-fields, respectively. Since the four sub-fields 
partially overlap  with each other, the overlapping sky regions can have a larger number of observed epochs than 
the other regions. The very central sky region, covering about 1.7$\times$6.2\,arcmin$^2$ by the four sub-fields, has almost 
98 observed epochs.

We then stack all the remaining single-epoch images using median combination method to create the final mosaic image 
(hereafter \textit{det} image). Again, the median coaddition  enables us a further removal of  residual cosmic rays, 
satellite tracks and other image defects remaining in the single-epoch image. In total, The \textit{det} image covers 
4.38\,deg$^2$ and it is used for objects detections and selections (see Section \ref{sec:drap}). The initial absolute 
photometric calibration was calculated by comparing the observed magnitudes of standard stars with photometric 
reference magnitudes. This was performed by the SUDARE-VOICE team based on the observation on  July 30, 2012 . 
In the present work, we directly compare the instrumental magnitudes of stars in the \textit{det} image with the calibrated 
magnitudes to determine the final zeropoint and apply it to the individual epochs. The derived 5$\sigma$ limiting magnitude 
of the \textit{det} image is about 26.3\,mag for point source within 1.0 arcsec aperture radius.

Saturated stars and their surrounding halos can systematically affect the photometry on the nearby objects. 
We therefore visually identify all these regions and mask them from the \textit{det} image. The area of such mask 
regions accounts for about 7 per cent of the original image. 

\begin{figure}
\centering
\includegraphics[width=0.48\textwidth]{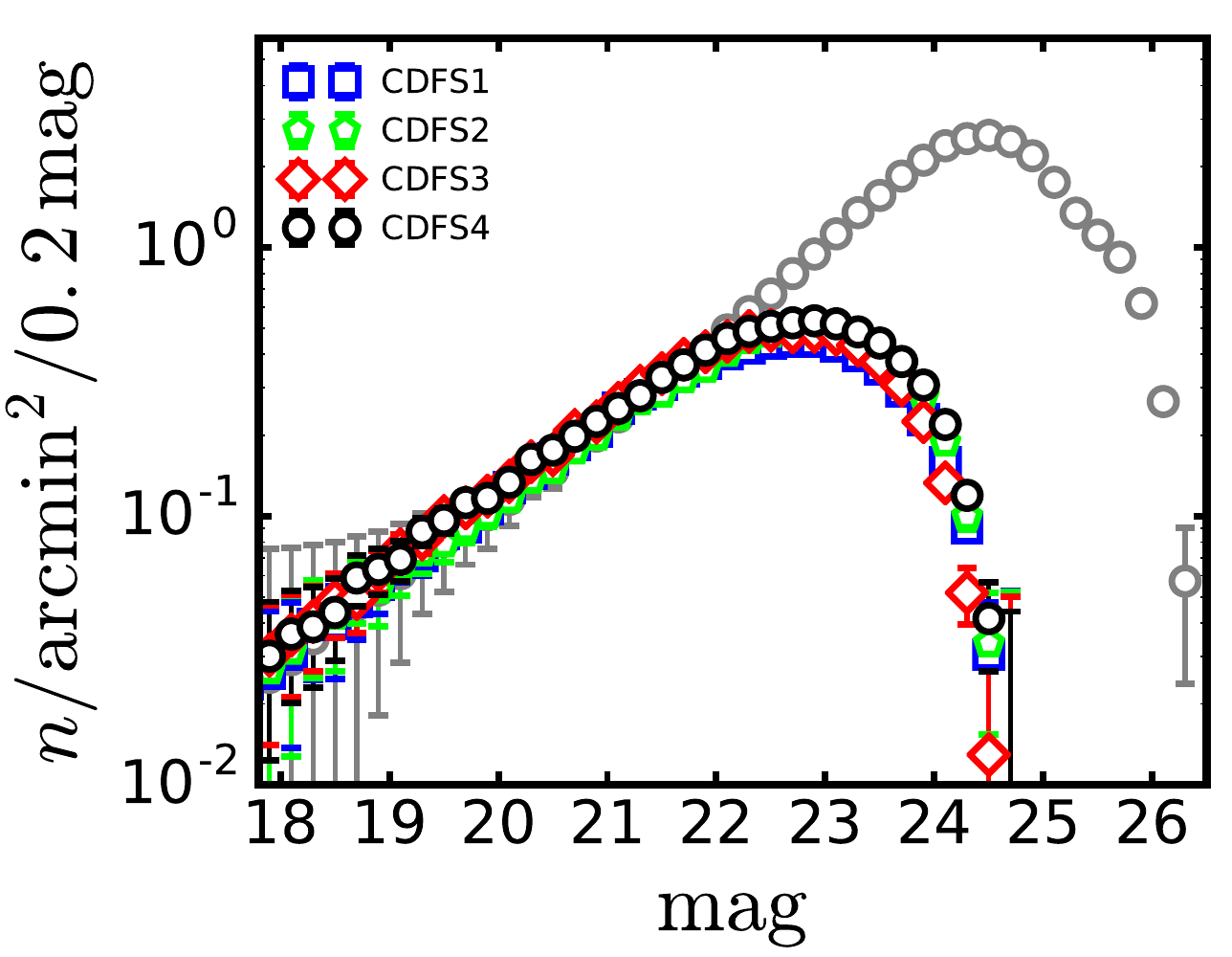}
\caption{The $r$-band magnitude distributions with errorbars estimated from the Poisson statistics. 
The grey points represent the distribution of the \textit{det} image, while the other four color-encoded distributions 
are the results by averaging the individual catalogs of all epochs for the sub-fields, respectively.}
\label{fig:magdist}
\end{figure}

We run \texttt{SExtractor} on the \textit{det} image for object detection and photometry. The detection 
threshold is set to be 2.0$\sigma$ above the background, and at least three connected pixels are required 
for a detection. For photometry of the blended objects, we set the number of deblending threshold 
to be \texttt{DEBLEND\_NTHRESH}=32 and the low contrast parameter to be \texttt{DEBLEND\_MINCONT}=0.002. 
In total, 381,937 objects are detected. The same configuration is also applied for the photometry on individual 
epochs. Figure \ref{fig:magdist} shows the $r$-band magnitude (\texttt{MAG\_AUTO}) distributions of objects 
detected in the \textit{det} image and the average of individual epochs for the four sub-fields. We can see that 
the peak of the magnitude distribution from the \textit{det} image is about 24.5\,mag which it close to the limiting 
magnitude of single epoch images.

\subsection{PSF Modeling}\label{subsec:psf}
For every epoch, we construct the spatially varied PSF model using \texttt{PSFEx} software 
(version 3.17.4; \citet{2013ascl.soft01001B}). Because each single-epoch image covers  about 
1.0$\times$1.0 deg$^2$ and results from stacking single exposures, the spatial 
variations of the PSF over the entire image are significant \citep{2018MNRAS.tmp.1498F}. 
To accurately model the PSF variations with polynomial interpolation and reduce the impact 
of discontinuities at the CCD edges due to the stack of individual exposures, we firstly split the 
image into 4$\times$2 sub-images of uniform size. Basically, each sub-image covers the 
area of about 2$\times$2 CCD chips. The PSF model is then constructed for each 
sub-image individually. 

\begin{figure}
\centering
\includegraphics[width=0.48\textwidth]{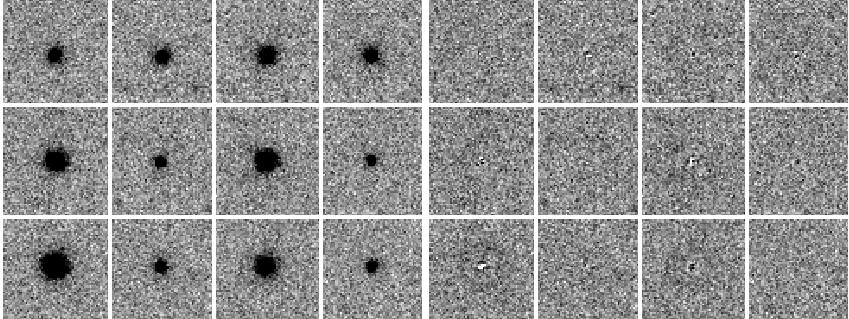}
\caption{A typical example of PSF model constructed by \texttt{PSFEx}. The first four columns are star 
stamps selected for PSF modeling. While the last four columns show the corresponding residuals by 
subtracting the PSF models. The size of each stamp is 12.2\,$\times$\,12.2 arcsec$^2$ with pixel 
scale of 0.2 arcsec.}
\label{fig:psfmodel}
\end{figure}

To obtain a clean star sample for PSF modeling, we generate the object catalog for each sub-image 
and then match it with the \textit{Gaia} DR1 catalog. Only unsaturated stars with SNR larger than 50 
and  \texttt{SExtractor} parameter \texttt{FLAGS=0} are selected. To reduce the non-linearity effect, 
stars with peak counts larger than half of the saturated values are also rejected. These criteria result 
in over 100 isolated and unsaturated stars for each sub-image as \texttt{PSFEx} input. We fix the image 
size of the PSF model (\texttt{PSF\_SIZE}) to be 31$\times$31 pixels. To extract the principal components 
of the PSF model from Principal Component Analyses (PCAs), the basic vector parameter \texttt{BASIS\_TYPE} 
is set to be \texttt{PIXEL}.  A third-order polynomial function is applied to model the spatial variations. Finally, 
the PSF model at a given image position can be calculated by a linear combination of ten pixel basis 
vector images. Figure~{\ref{fig:psfmodel}} displays a typical example to illustrate the accuracy of the PSF 
construction. Statistically comparing the stars with the corresponding PSF models shows that our 
implementation can yield near-zero model residuals. 
However, there still exist systematic biases in 
the very central region of the bright stars, as shown in Figure~{\ref{fig:psfmodel}}, which probably 
bias the photometry of the measured light curves. Therefore, we further perform the detrending correction 
on the light curves in the following section.

\begin{figure}
\centering
\includegraphics[width=0.48\textwidth]{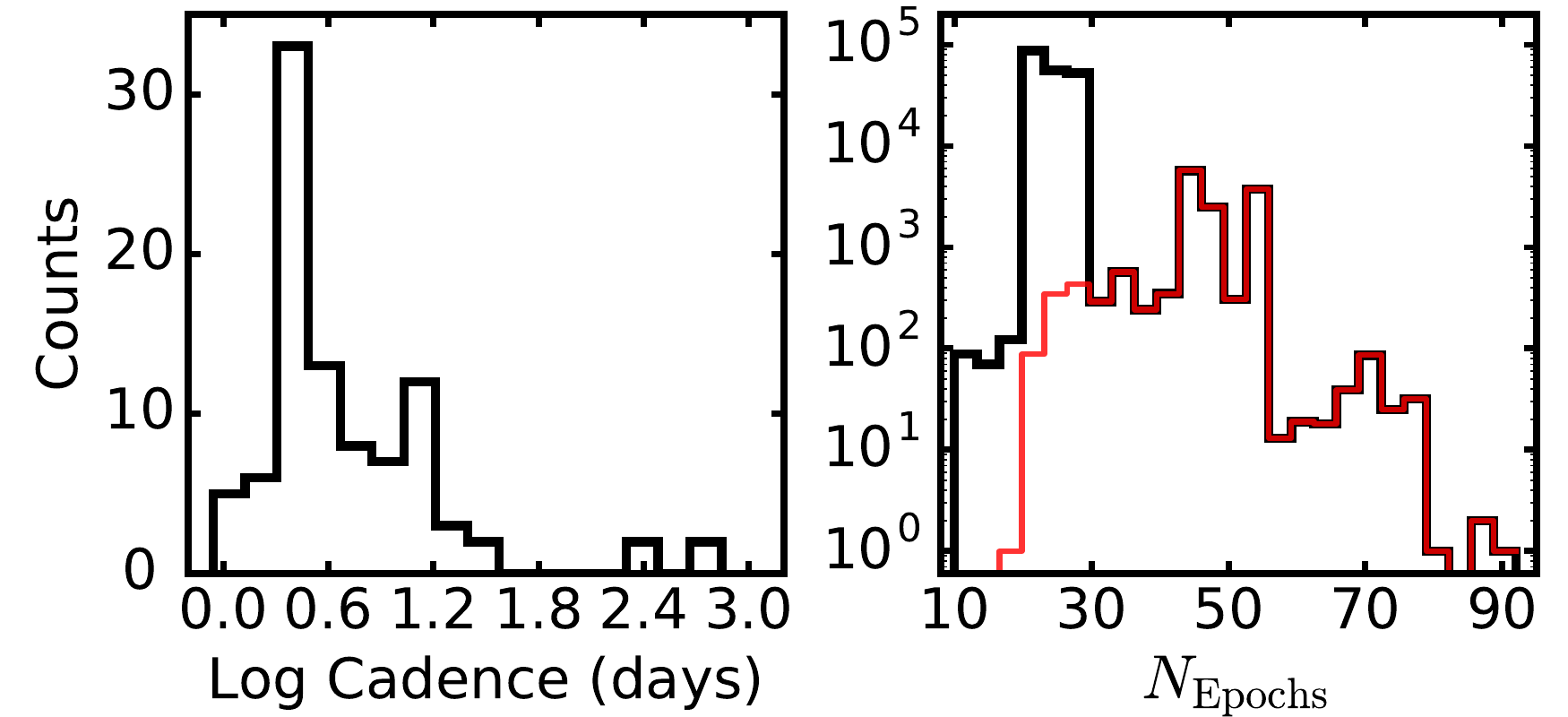}
\caption{Histograms of the observed cadence and number of epochs for objects in the \textit{clean} sample. 
The red line in the right figure represents the distribution of the epoch number for objects observed in at 
least two sub-fields.}
\label{fig:histcn}
\end{figure}

\section{Variability Measurements}\label{sec:drap}
In this section, we firstly describe the criteria to select objects for variability measurements. 
Then we  introduce the \texttt{drap} method to extract the light curves for the selected objects, 
and present detailed error analyses on the results. It is noted that current time-domain 
surveys mainly focus on point-like objects (e.g. stars and quasars) or special extended objects 
(e.g. AGNs and SNe with observable hosts). In our studies here using \texttt{drap}, we do not make 
priori type selections. Thus the objects we analyze consist of both point-like and extended objects. 

\subsection{Source Selection}
The catalog extracted from the \textit{det} image is used for initial object selection.  As shown 
in Figure \ref{fig:magdist}, most objects in the catalog are very faint and below the limiting depth of the 
individual epochs. 
%As described below, however, the \texttt{drap} method allows us to measure 
%the variability of objects with very faint detections in a single epoch as long as the SNR of these 
%objects in the \textit{det} image is large enough. 
Thus in our analyses, we conservatively select objects 
with SNR larger than 20.0 in the catalog, roughly corresponding to SNR$\sim$3.8 
in a single epoch. This criterion rejects about 44.3\% of the objects. We further 
exclude objects with $r$-band magnitude brighter than 16.0\,mag which are basically saturated stars.
Objects with bad photometry flagged by \texttt{SExtractor} are also rejected, but the blended pairs  
are allowed since they could be transient events (e.g supernova and its host galaxy).  
In addition, objects with the total number of observed epochs less than 
10 are removed. The selection criteria are summarized as follows: 
\begin{itemize}
\item  $r \geq 16.0$\,mag \& $\mathrm{SNR} \geq 20.0$
\item \texttt{FLAGS}\,$<=$\,3 (blending allowed)
\item $n_\mathrm{epoch}\geq 10$
\end{itemize}
Finally, 210,530 objects are selected for variability measurements, of which the faintest object 
reaches to $r\sim24.9$\,mag. We refer to these objects as the \textit{clean} 
sample. Figure~\ref{fig:histcn} shows the histograms of the observational  
cadence and the number of epochs for objects in the sample. The peak cadence is about 3 days, and the 
fraction of objects which are observed in at least two sub-fields is about 7.0\%.

For every object in the \textit{clean} sample, we cut the stamps from all single-epoch images with a uniform size 
of 65$\times$65 pixels, corresponding to 13$\times$13 arcsec$^2$. This size of stamp is about 11 times 
larger than the FWHM of the worst PSF (see Section~\ref{subsec:imgcp}). If the object is located at the edge 
(i.e. the distance between the center of the object and the edge of the image is less than 32 pixels) or inside a 
mask region in a certain epoch, that stamp is rejected. The PSF of an object is determined using the model 
constructed in Section~\ref{subsec:psf} and is normalized so that the sum of its pixel values is equal to one.

\subsection{Differential Running-Average Photometry (\texttt{DRAP})}
In this subsection, we introduce the detailed mathematics on the \texttt{drap} method.  For each object, 
the extracted stamps span many different photometric epochs. The background noise, the 
seeing conditions and the size and shape of the PSFs can vary between different epochs.
For a given stamp $i$, the two dimensional surface brightness distribution of an object and 
the corresponding PSF profile are denoted as $S_{i}$ and $\mathrm{P}_{i}$, respectively. Then 
stacking all the stamps by weighted average method, we obtain a master stamp 
\begin{eqnarray}
S_{0} = \sum_{i}^{n}w_{i}S_{i}/\sum_{i}w_{i},
\end{eqnarray}
where $w_{i}$ is the weight map of the stamp $i$ and $n$ is the total number of stamps. 
During the stack, the $\sigma$-clipping method is applied to reject pixels with values (e.g. 
residual cosmic rays) larger than 5 times of the standard deviation. The outlier pixel values 
are replaced by the median of the surrounding 5$\times$5 unmasked pixels.  
The same method is also used to stack the individual PSFs to yield the master PSF image, 
denoted as $\mathrm{P}_{0}$. Similarly, excluding the $j$th stamp, then we can generate 
the stacked stamp $\tilde{S}_{j}$ and corresponding PSF $\tilde{\mathrm{P}}_{j}$. We can 
expect that the flux difference between the two stacked stamps $S_{0}$ and $\tilde{S}_{j}$ 
results from the variability of the object in the $j$th stamp. In addition, 
in Appendix~\ref{appen}, we show that the two PSFs $\mathrm{P}_{0}$ and $\tilde{\mathrm{P}}_{j}$ 
are almost identical if $n$ is much larger than the difference of the pixel values between 
$\mathrm{P}_{j}$ and $\mathrm{P}_{0}$. The same conclusion also holds 
for the background fluctuations between the two stamps. 

\begin{figure}
\centering
\includegraphics[width=0.48\textwidth]{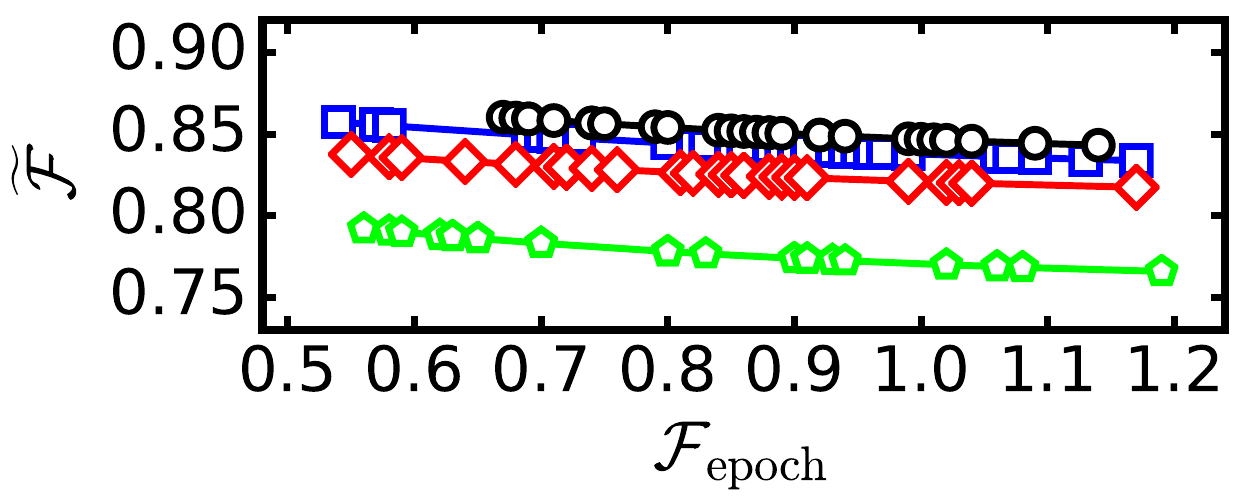}
\includegraphics[width=0.48\textwidth]{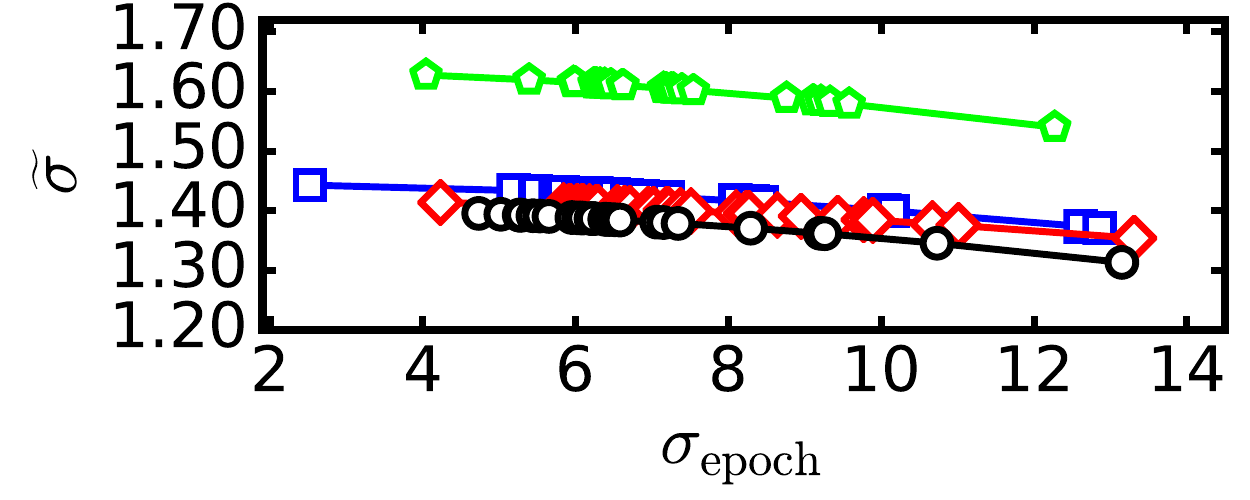}
\caption{\textit{Top panel}: comparison of the FWHM of PSF of each single epoch and the running-average result. 
Different colors, with the same notation as in Figure~\ref{fig:magdist}, corresponds to different sub-fields in 
SUDARE-VOICE survey. \textit{Bottom panel}: comparison of the background fluctuation of each epoch and the 
running-average result.}
\label{fig:fsigRel}
\end{figure}

To illustrate the advantage of \texttt{drap}, for simplicity, assuming the PSF of  each of our observational 
images is constant and follows the Gaussian 
profile with FWHM $\mathcal{F}_\mathrm{epoch}$ fixed to the observational value of the corresponding epoch,  
we calculate the FWHM $\widetilde{\mathcal{F}}$ of the running-average PSF for each epoch following the same 
procedure to obtain $\tilde{\mathrm{P}}_{j}$. As shown in the top panel of Figure~\ref{fig:fsigRel}, 
while the PSF varies significantly for the individual epochs, the running-average PSF  
keeps very stable with maximum change of only 3.4\% for CDFS2 sub-field. Similarly, the bottom panel of 
Figure~\ref{fig:fsigRel} compares the background fluctuation $\sigma_\mathrm{epoch}$ of each 
epoch and the corresponding  running-average value $\widetilde{\sigma}$ which is also very close  
to constant with maximum change of 6.1\% for CDFS4 sub-field. The stable PSF and the background from 
\texttt{drap} make it very suitable for variability studies. 

To measure the fluxes in the stacked stamps and precisely calculate the 
variability, we further perform PSF homogenization between the two stacked stamps although 
they are already rather stable. Different algorithms have been developed to construct the homogenization 
kernel, such as the deconvolution solution in Fourier space \citep{1995ASPC...77..297P}, regularization 
representation with a set of basis functions \citep{1998ApJ...503..325A,2008MNRAS.386L..77B} and so forth. 
As discussed in \citet{2016ApJ...830...27Z}, because of the effects of noise and other implementation issues,  
some homogenization operations can potentially lead to artifacts in the difference images. Here we homogenize the 
PSFs between the two stamps $S_{0}$ and $\tilde{S}_{j}$ using the cross-convolution method proposed by  
\citet{2008ApJ...680..550G}. As noted there,  this method can degrade both 
PSFs so that it may limit the detection of faint variable sources. However, since no deconvolution or 
regularization process is applied, it can be more numerically stable and leave less artifacts. In this case, 
the difference between the two stacked stamps is  derived as 
\begin{eqnarray}\label{eq:dimg}
D_{j} = S_{0} \otimes \tilde{\mathrm{P}}_{j} - \tilde{S}_{j} \otimes \mathrm{P}_{0},
\end{eqnarray}
where $\otimes$ represents the convolution operation.

We use aperture photometry to measure the residual flux, denoting as $F_{D_j}$,  in the 
difference image $D_{j}$. For the variability measurements, we fix the aperture 
radius for all objects to be 1.5 arcsec which is about 1.26 times of the worst FWHM. 
With the convolution operation in the above equation, it is non-trivial 
to obtain the analytical expression between $F_{D_j}$ and the true flux variation $\delta{F_{j}}$ 
in the $j$th stamp. However, as mentioned above, in case of large $n$, we  have 
$\mathrm{P}_{0} \approx \tilde{\mathrm{P}}_{j}$. This approximation holds for all the current 
and future time-domain surveys which usually have hundreds of exposures on the same sky region 
(e.g. LSST \citep{2019ApJ...873..111I}). In that case, the flux variation $\delta{F_{j}}$ in the $j$th stamp relative 
to the flux in the master stamp 
can be simply derived as 
\begin{eqnarray}
\delta{F_{j}} = s \times F_{D_j},
\end{eqnarray}
where $s$ represents the flux scaling factor which satisfies  
\begin{eqnarray}
Ns = \sum_{p,q}\sum_{i\neq{j}}w_{i}/w_{j},
\end{eqnarray}
where $N$ is  the total number of pixels within the photometric aperture, and $p$ and $q$ are the pixel 
indices referring to the column and row of the weight map. The 
summation is performed within the photometric region. When the weights are identical, $s$ reduces 
to $s=n-1$. The corresponding flux error $\sigma_{\delta{F_{j}}}$ is expressed as 
\begin{eqnarray}\label{eq:ferr}
\sigma_{\delta{F_{j}}} = s \times \sigma_{F_{D_j}} = s \times \sqrt{F_{D_j}/g + N\sigma^{2}_\mathrm{bkg}},
\end{eqnarray}
where $g$ is the gain in the difference image, and 
$\sigma_\mathrm{bkg}$ is the rms of the background, which can be derived by $n^{2}\sigma^2_\mathrm{bkg} = \sigma^2_{j} 
+ \widetilde{\sigma}^2_{j}$ where $\sigma_{j}$ and $\widetilde{\sigma}_{j}$ are the background fluctuations for $S_{j}$ 
and $\widetilde{S}_{j}$, respectively.

\begin{figure}
\centering
\includegraphics[width=0.48\textwidth]{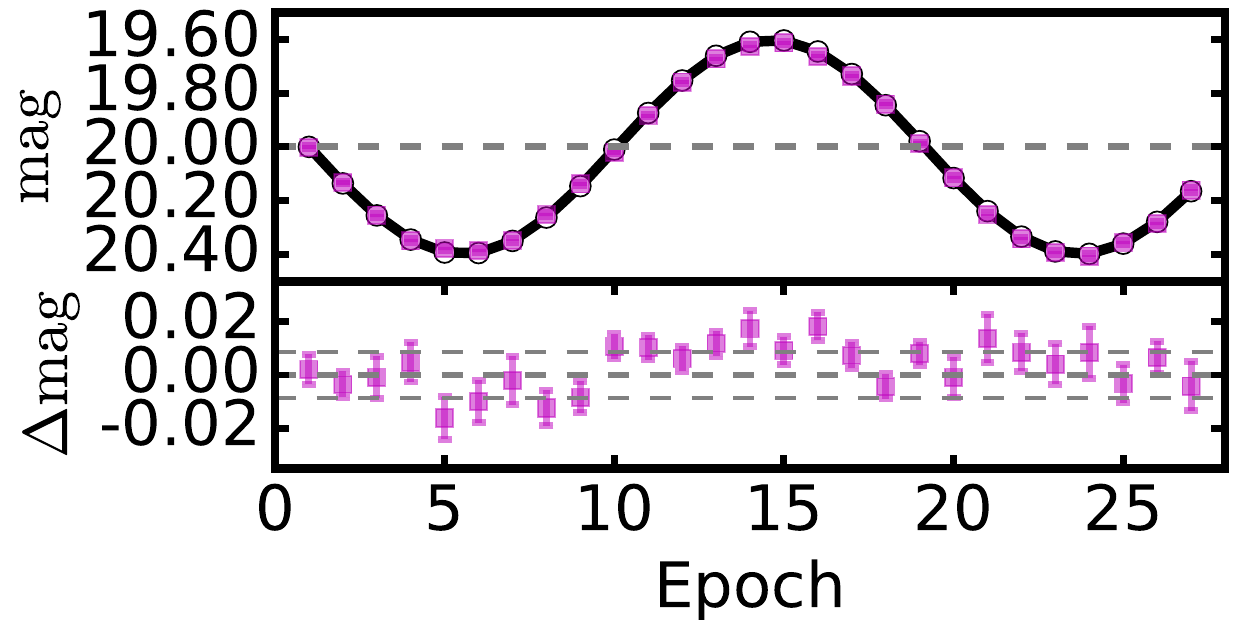}
\caption{Comparison of the simulated light curve of a star and the measured result by the \texttt{drap} method. 
The light curve of the star is assumed to be  sinusoidal, following $m(t) = A\sin(t) + m_{0}$ where $m(t)$ is the 
magnitude at time $t$, $A$ is the amplitude and $m_{0}$ is a constant magnitude. In the simulation, we fix 
$m_{0}$ = 20.0\,mag and  $A$ = 0.4\,mag. \textit{Top panel}: the black line is the light curve of the star. The black 
circles and the magenta squares with errorbars represent the input magnitudes in the simulation and the measured 
values by the \texttt{drap} method. \textit{Bottom panel}: the magenta squares with errorbars represent the difference 
between the input and measured magnitudes. The thin dashed lines indicate the 1.0$\sigma$ dispersion of the magnitude 
difference.}
\label{fig:mdrap}
\end{figure}

We generate a series of simulated image stamps of a star to validate the method. The light curve of the star is assumed 
to be  sinusoidal, following $m(t) = A\sin(t) + m_{0}$ where $m(t)$ is the magnitude at time $t$, $A$ is the amplitude 
and $m_{0}$ is a constant magnitude. In the simulation, we fix $m_{0}$ = 20.0\,mag and  $A$ = 0.4\,mag. We generate 
in total 27 stamps which is the same as the number of epochs in CDFS1 sub-field. The simulated PSFs follow Gaussian 
profile with FWHMs fixed to the observational values of individual epochs in the CDFS1 sub-field. Meanwhile, the 
background fluctuation values are also from the CDFS1 sub-field. The black curve in the top panel of Figure~\ref{fig:mdrap} 
shows the light curve of the star. The black circles represent the input magnitudes in the simulation, while the magenta 
squares with errorbars are the measured values by the \texttt{drap} method described above. The difference between 
the input and measured magnitudes is shown in the bottom panel of Figure~\ref{fig:mdrap}. We see that the difference is 
consistent with zero with $\sigma_{\Delta\mathrm{mag}} \sim 0.01$, meaning that the \texttt{drap} method can accurately 
recover the true light curve of the simulated star.

\subsection{Uncertainty Analyses}

\begin{figure}
\centering
\includegraphics[width=0.48\textwidth]{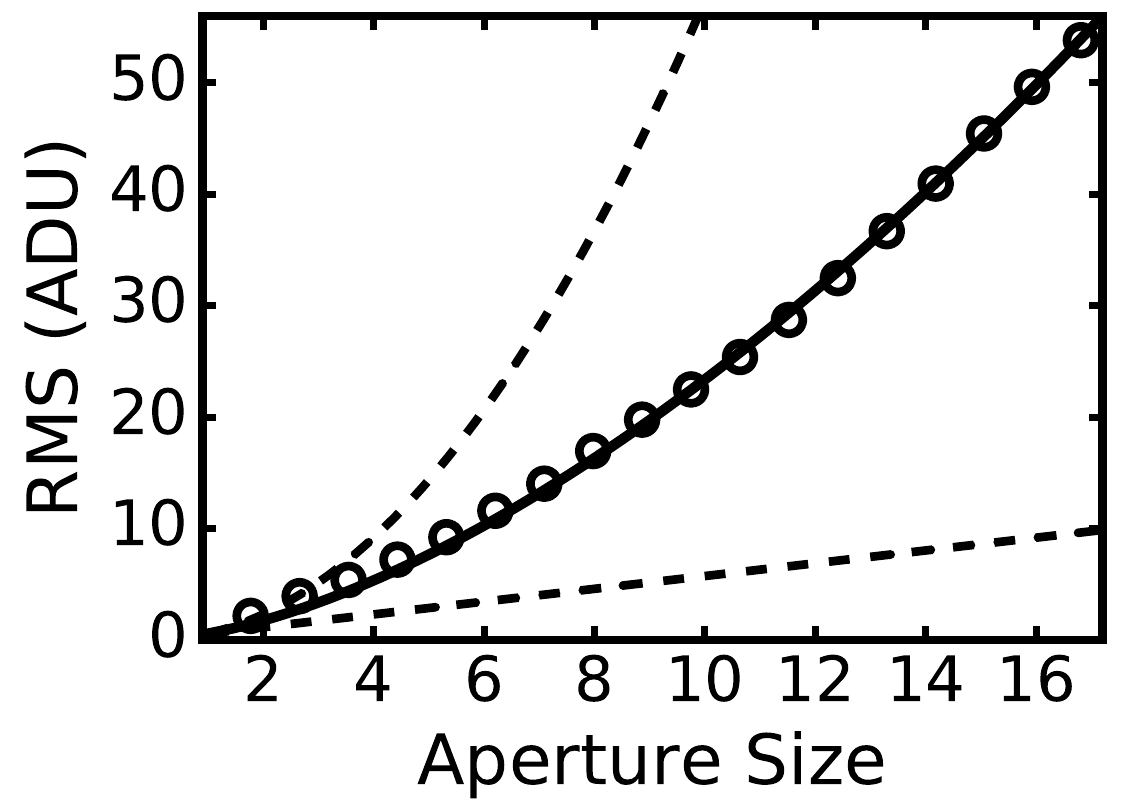}
\caption{Gaussian dispersion as a function of aperture size defined as $\sqrt{N}$. 
The black circles are measured from the difference image $D_{I}$. A power-law 
curve with free parameter $\beta$ (solid line) is used to fit the data points. 
For illustration, the bottom and top dashed curves represent two limiting cases: 
no pixel correlation and complete correlation 
in adjacent pixels.}
\label{fig:cnoise}
\end{figure}

Since the image stacking and PSF convolution procedures can introduce correlated noise in the difference 
image $D_{j}$, Equation~(\ref{eq:ferr}) may underestimate the uncertainty of the measured $F_{D_j}$. 
Without noise correlation, the background noise of a given image is determined by $\sigma^2 = \sigma^2_{0} N$, 
where $\sigma_{0}$ is the standard deviation of background noise and $N$ is the pixel number in the 
photometric aperture. In the presence of noise correlation, however, the background noise should be 
estimated by  $\sigma^2 = \sigma^2_{0} N^{2\beta}$, where $\beta$ is a free parameter within [0.5,\,1.0]. 
In the case of pure background noise dominated, $\beta=0.5$, while if the adjacent pixels are completely 
correlated, $\beta=1.0$ \citep{2017AJ....153...53L,2018AJ....156..186M}. Therefore, to take the noise correlation 
into account, the uncertainty of the measured $F_{D_j}$ can be generalized as 
\begin{eqnarray}\label{eq:mderr}
\sigma_{F_{D_{j}}} = \sqrt{F_{D_j}/g + N^{2\beta}\sigma^{2}_\mathrm{bkg}}.
\end{eqnarray}

We estimate the noise correlation as follows. For a given epoch, we obtain the corresponding running-average image 
$\widetilde{I}$. For simplicity, the PSF of the image  $\widetilde{I}$ is assumed to be Gaussian and spatially invariant. 
The size of the PSF is fixed to the value shown in Figure~\ref{fig:fsigRel}. Meanwhile, we can obtain the master image 
$I$ and Gaussian PSF by average-stacking all the single epoch images.  Then following the Equation~(\ref{eq:dimg}) 
we can generate the difference image $D_{I}$ for this epoch. To eliminate the potential impact of objects on estimating 
the noise, the positions in the difference image where objects are distinctly detected in the image $I$ are 
masked. We then select a set of about 2000 random positions on the object-masked difference image. These positions 
are selected to be non-overlapping with the mask regions within radius of 20 pixels. The fluxes are measured for each 
position using different apertures. For a given aperture, a Gaussian function is used to fit the histogram of the measured 
fluxes to derive the dispersion. Generally, larger apertures give larger Gaussian dispersion. Then we use the power-law 
equation described above to fit the relation between Gaussian dispersion and aperture size as displayed in 
Figure~{\ref{fig:cnoise}} for an example. The derived $\beta$ parameter, with value basically ranging from 0.6 to 0.8, 
is then applied to Equation~(\ref{eq:mderr}) to revise the uncertainty.

\subsection{Detrending}
Although the PSFs between different stamps in the \texttt{drap} approach is rather stable, and we 
further perform PSF homogenization in the variability measurements, we still need a detrending procedure for 
calibration. Systematic biases on the measured light curves can be introduced for instance by a not accurate 
modeling of the positional variation of the PSF by a polynomial function, or by errors in the photometric calibrations 
of different epochs.
%The spatial variations of PSF in observations are complicated \citep{2018MNRAS.tmp.1498F}. The 
%inaccuracy in modeling the spatial variation by a simple polynomial function and photometric calibration between 
%different epochs can lead to systematic biases, which are expected to be position-dependent, on the measured 
%light curves.} 
Therefore, it is necessary to correct for these biases. Non-variable objects with constant fluxes as a 
function of time are ideal for such correction because any deviation of the measured flux at a given epoch from 
the expected value can be attributed to the systematic effects. 

\begin{figure}
\centering
\includegraphics[width=0.48\textwidth]{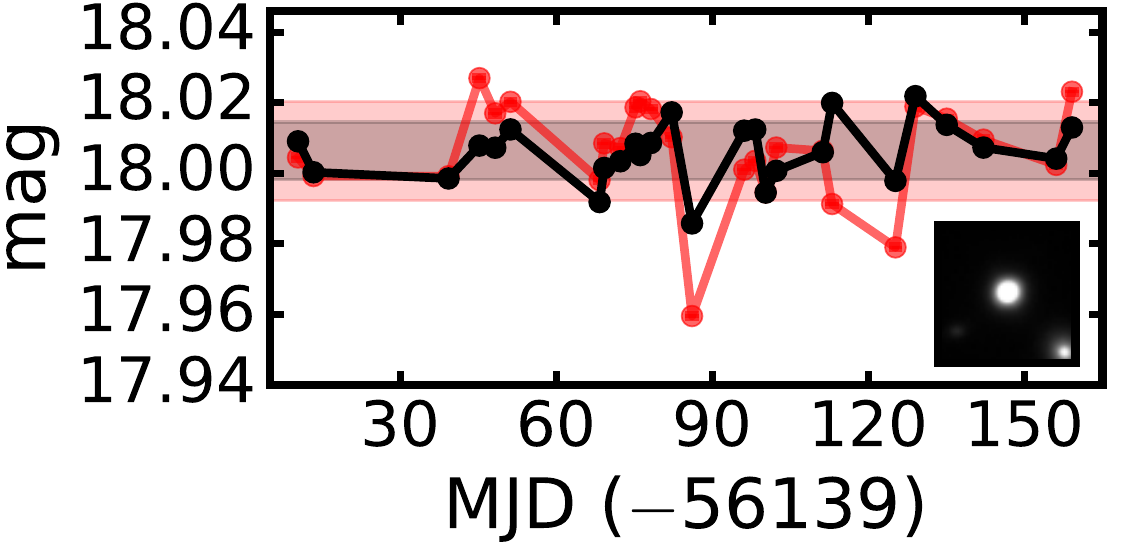}
\includegraphics[width=0.48\textwidth]{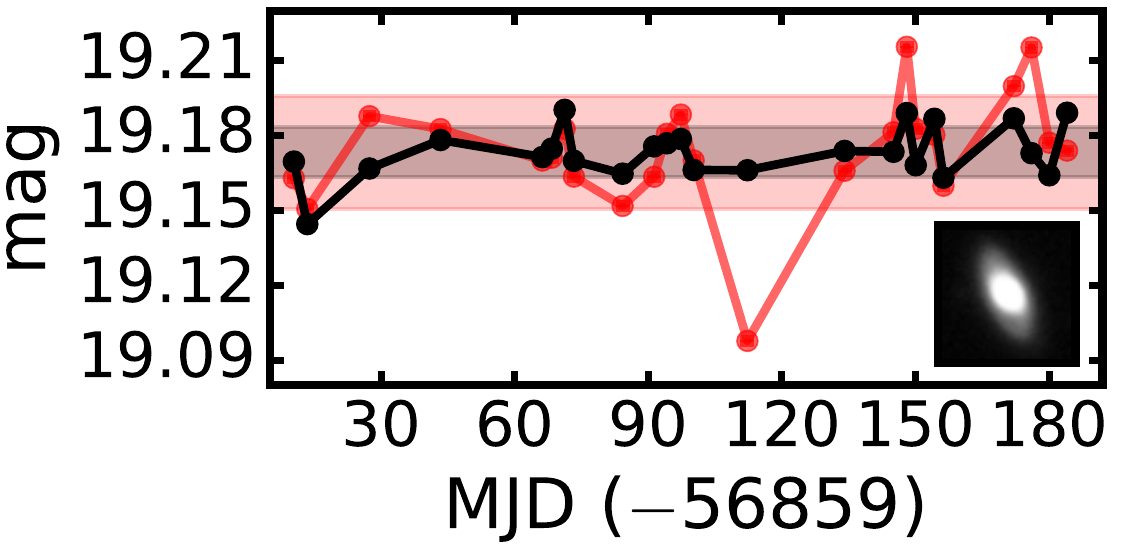}
\caption{\textit{Top panel}: Comparison between the light curves of a star before (red line) and after (black line) detrending correction. 
The shadow regions correspond to the 1.0$\sigma$ intervals around the mean magnitudes of the two light curves. The inset shows the 
stacked image stamp of the star. \textit{Bottom panel}: Comparison between the light curves of a bright galaxy before (red line) and after 
(black line) detrending correction.}
\label{fig:detrend}
\end{figure}

\begin{figure}
\centering
\includegraphics[width=0.48\textwidth]{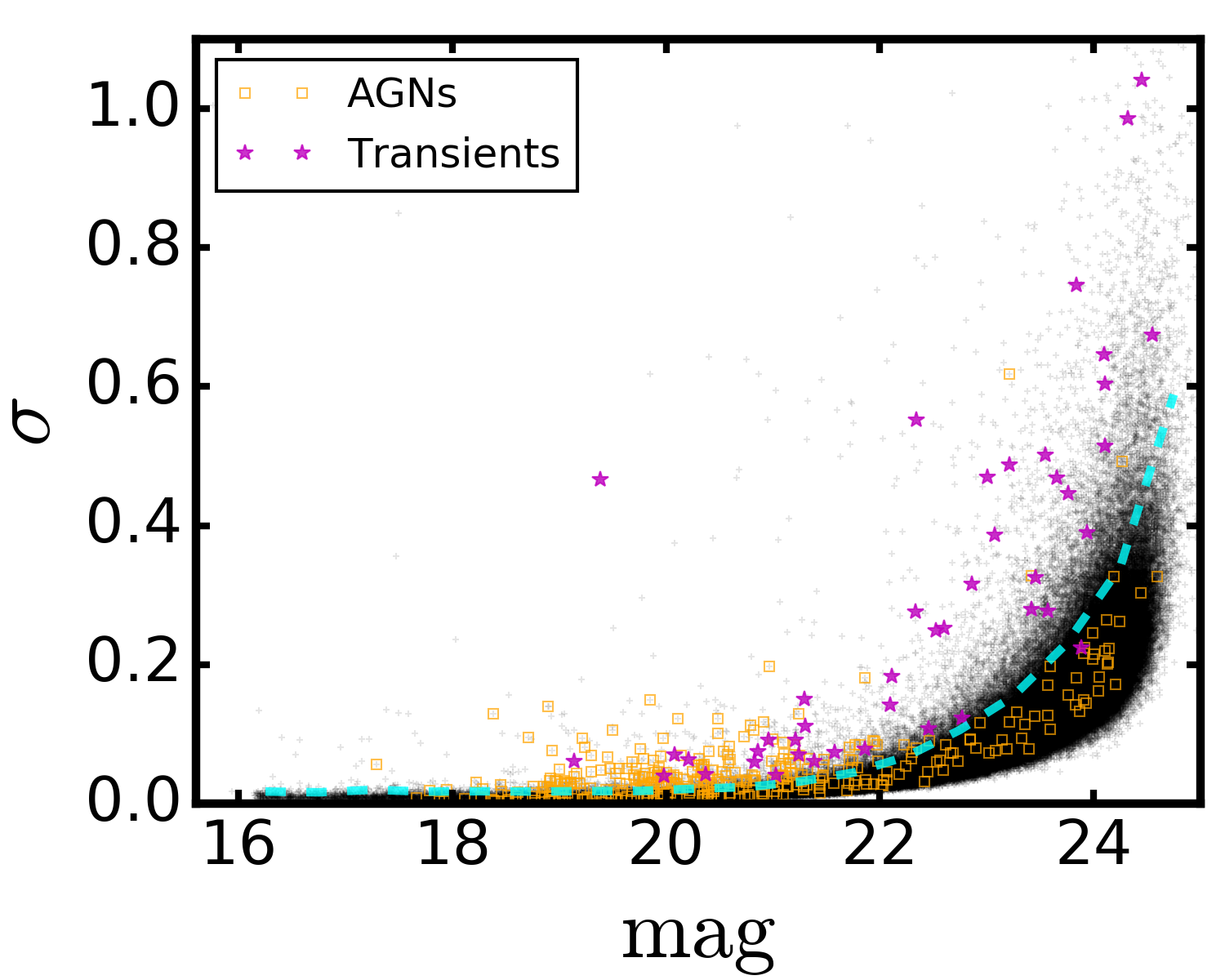}
\caption{The standard deviation as a function of average magnitude for the entire sample (black crosses).
The orange squares show the AGNs from the Million Quasar Catalog, and the magenta stars represent the 
transients identified by visual inspection. The dashed cyan line is the 3.0$\sigma_{[\sigma_{i}]}$ threshold of the standard deviation.}
\label{fig:magrms}
\end{figure}

As with the PSF modeling, the detrending correction is performed, using the PSF stars as 
described in Section~\ref{subsec:psf}, on individual sub-images for a given epoch. The 
\texttt{drap} method is applied to measure the magnitudes of these stars. To eliminate the impact of variables, 
a star is rejected if the standard deviation of the magnitudes at all epochs larger than the 3$\sigma$ 
limit of the standard deviation of the whole sample as illustrated in Figure~\ref{fig:magrms} 
(see Section~\ref{sec:res} for more details). For each remaining star, we calculate the flux ratio $s_{j}$ for 
the $j$th epoch relative to the flux measured on the master stamp $S_{0}$. Evidently, without the existence of 
systematic biases, the relative flux ratios of stars in a given sub-image should be equal to one. However, 
it is found that the relative flux ratios can be systematically as large as 5 per cents for some epochs (corresponding 
to the magnitude bias of about 0.06\,mag). We apply a second order polynomial on each sub-image to model the 
systematic biases, and the detrending equation is written as 
\begin{equation}
s_{j}(x,\,y) = p_{0} + p_{1}x + p_{2}y + p_{3}x^{2} + p_{4}xy + p_{5}y^2, 
\end{equation}
where $p_{i}$ ($i=1,2,...,5$) are free parameters, and $x$ and $y$ are the pixel coordinates of 
stars in the sub-image. The least square fitting method is used to derive the best-fit parameters.  
The top panel of Figure~{\ref{fig:detrend}} compares the light curves of a star before and after 
detrending correction. The shadow regions correspond to the standard deviations of the two light curves. 
After applying the detrending correction, the scatter of the light curve is decreased and well within the 
photometric accuracy. The similar result is also displayed for a bright non-AGN galaxy in the bottom panel.
We perform tests using higher order polynomial detrending. The results show no significant improvements 
comparing to the second order modeling. 

Besides detrending, we also use median smoothing method to further reject the outliers due to bad photometry for a 
given light curve.  The window size is set to be 5 consecutive data points. Then we 
subtract the smoothed light curve  from the raw light curve. The residuals are expected to follow a 
Gaussian distribution with zero mean. In the presence of outliers, however, the standard deviation 
of the residuals derived by the usual method can be overestimated. Thus, instead, we calculate the median absolute 
deviation $\sigma_\mathrm{mad}$\footnote{The median absolute deviation is defined in this work as 
$\sigma_\mathrm{mad}=1.483\times\mathrm{median}|x-\mathrm{median}(x)|$, where $x$ 
is the data series.} which is less sensitive to outliers \citep{2008MNRAS.386L..77B} and is equal 
to the standard deviation for a Gaussian distribution. If the residual magnitude of a certain epoch  
is larger than 5$\sigma_\mathrm{mad}$, it is marked as an outlier. We note that this procedure may also 
potentially reject real variability, especially those explosive events with the time scale less than the 
cadence of our observations. However, they are beyond the reach using our data. The outlier removal 
method adopted here is therefore suitable for our analyses focusing on relatively long-time scale 
variables, such as AGN and SNe.

\subsection{Comparison With Difference Imaging Method}
\citet{2015A&A...584A..62C} and \citet{2017A&A...598A..50B} studied the SN explosion rates using the same dataset 
but only in the CDFS1-2 sub-fields. The SN candidates were detected by the difference imaging method 
\texttt{hotpants}\footnote{https://github.com/acbecker/hotpants} which is an implementation of the algorithm described in 
\citet{1998ApJ...503..325A}. The PSF-fit photometry was then performed in the difference images to extract the light curves 
of these candidates. In this subsection, we compare the light curves measured by the difference imaging method and \texttt{drap}.

\begin{figure}
\centering
\includegraphics[width=0.48\textwidth]{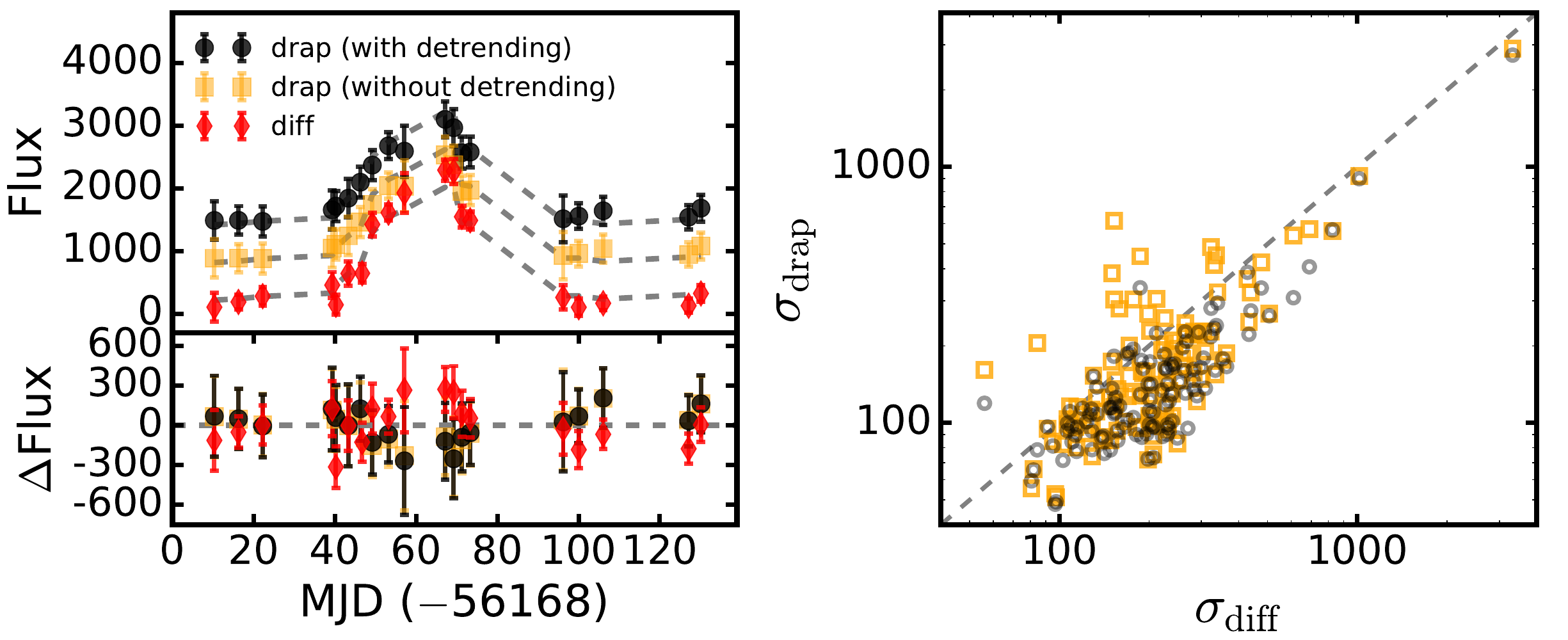}
\caption{\textit{Top left panel}: Comparison between the light curves measured by the difference imaging method 
and \texttt{drap}. The flux is in unit of analog-to-digital units (ADUs). The black dots (orange squares) represent the light curve 
measured by \texttt{drap} with (without) detrending correction. The red diamonds are the light curve measured by the difference 
imaging method \texttt{hotpants}. Arbitrary offsets are applied to the two light curves of  \texttt{drap} for clarity. The dashed grey 
lines represent the intermediate light curve (see text for detail). \textit{Bottom left panel}: Residuals after subtracting the intermediate 
light curve. \textit{Right panel}: Comparison between the standard deviations of the light curve residuals of the two methods. 
The black dots represent the comparison to \texttt{drap} with detrending correction, while the orange squares without detrending 
correction.}
\label{fig:comlc}
\end{figure}

Through matching the SN sample with our \textit{clean} sample, totally 116 common objects are acquired. The top left panel of 
Figure~\ref{fig:comlc} compares the light curves of one SN candidate measured by the 
two methods. As expected, visual inspection shows that the light curves of all these common objects exhibit the same peak 
structures. To further check the photometric accuracy, for each object we average the detrended light curve of \texttt{drap} 
and the corresponding one measured by the difference imaging method, and then derive an  
intermediate light curve by applying the median smoothing method to the average. The dashed gray curves in the top left panel of 
Figure~\ref{fig:comlc} shows the obtained intermediate light curve for this SN candidate, with the offsets between different lines the same as 
those for the data points. We subtract the intermediate light curve from the original light curves, as shown in 
the bottom left panel of Figure~\ref{fig:comlc}. The standard deviations ($\sigma_\mathrm{diff}$ and $\sigma_\mathrm{drap}$) of 
the residuals are calculated to quantify the photometric accuracy. The right panel of Figure~\ref{fig:comlc} compare the results 
for all the 116 common objects. The dashed grey line corresponds to the one-to-one relation. It can be seen that  the standard deviations 
measured by the difference imaging method are systematically larger than those of \texttt{drap}. Several reasons can be responsible for 
the results. Firstly, to perform image difference, the reference and new images used by \citet{2015A&A...584A..62C} are both from single 
epoch observations which suffer from larger background and Poisson noises compared to \texttt{drap}. The differencing operation 
enlarges the noise level in the difference image, and hence leads to significant photometric uncertainty in the measured 
fluxes. Secondly, inaccurate modeling of the spatially varied PSF kernels in the difference imaging algorithm can not only lead to 
false positives in the difference image \citep{2019A&C....2800284S}, but also affect the flux measurements of those real transients. 
Instead, the \texttt{drap} method can reduce  the difference of PSFs between different epochs. In short, the reduced sensitivity to PSF variations 
and the reduced noise in \texttt{drap} comparing to those of \texttt{hotpants} leads to better light curve determinations with less dispersions 
of the data points around the resulting intermediate light curve.

\begin{figure*}
\centering
\includegraphics[width=0.32\textwidth]{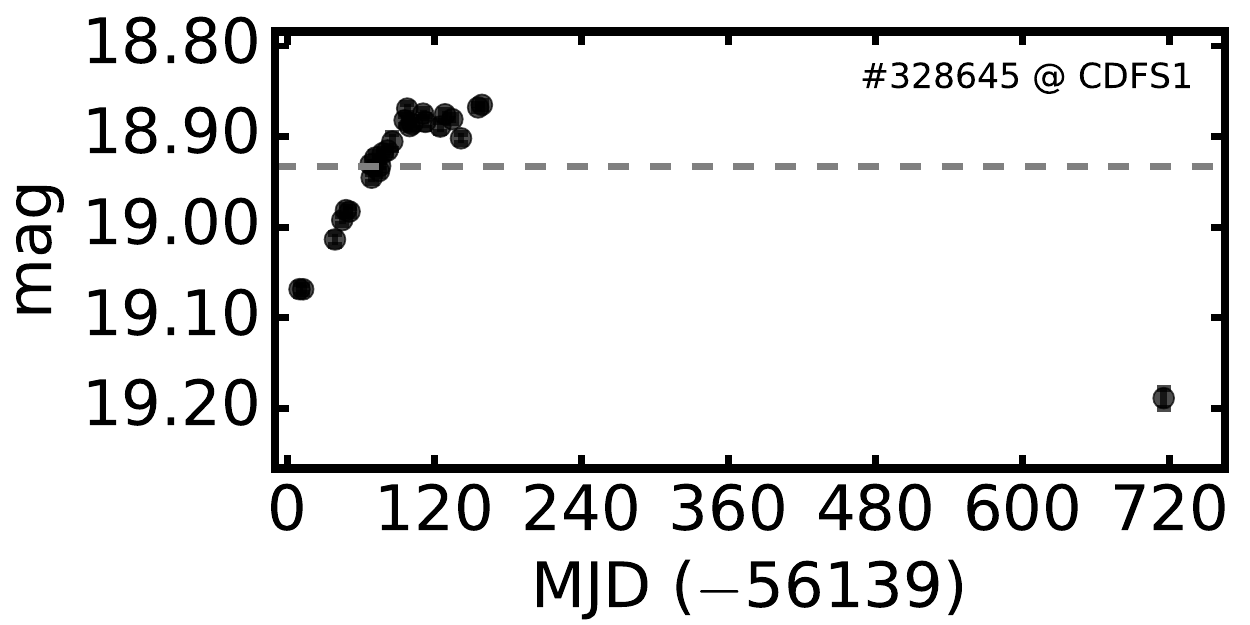}
\includegraphics[width=0.32\textwidth]{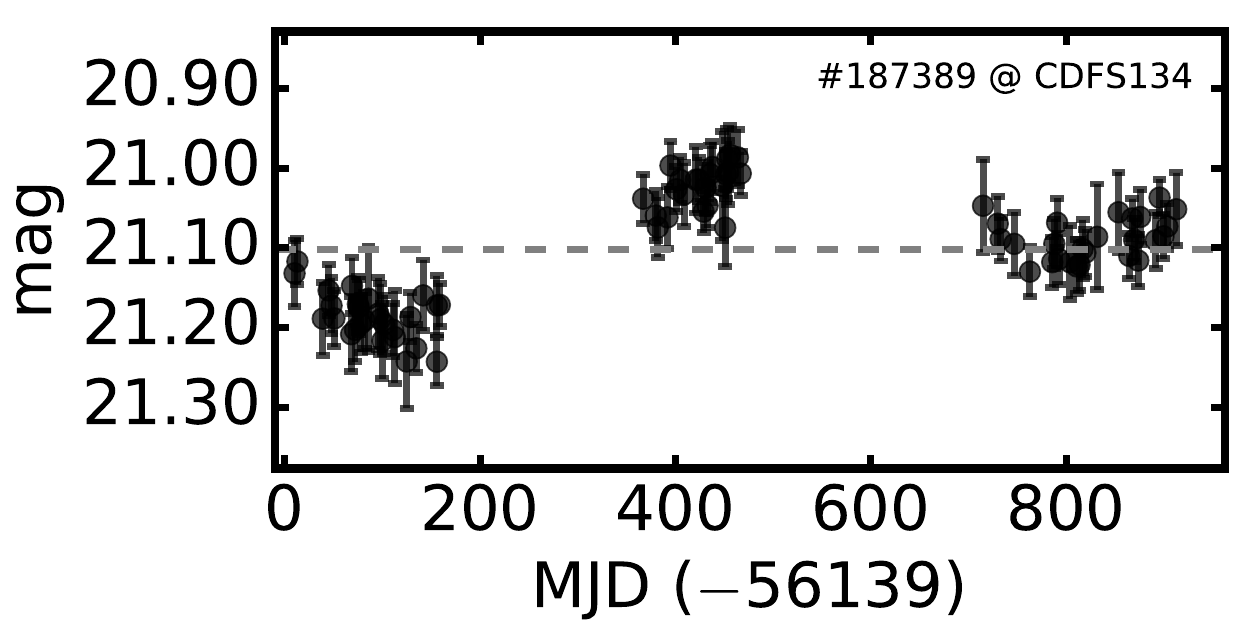}
\includegraphics[width=0.32\textwidth]{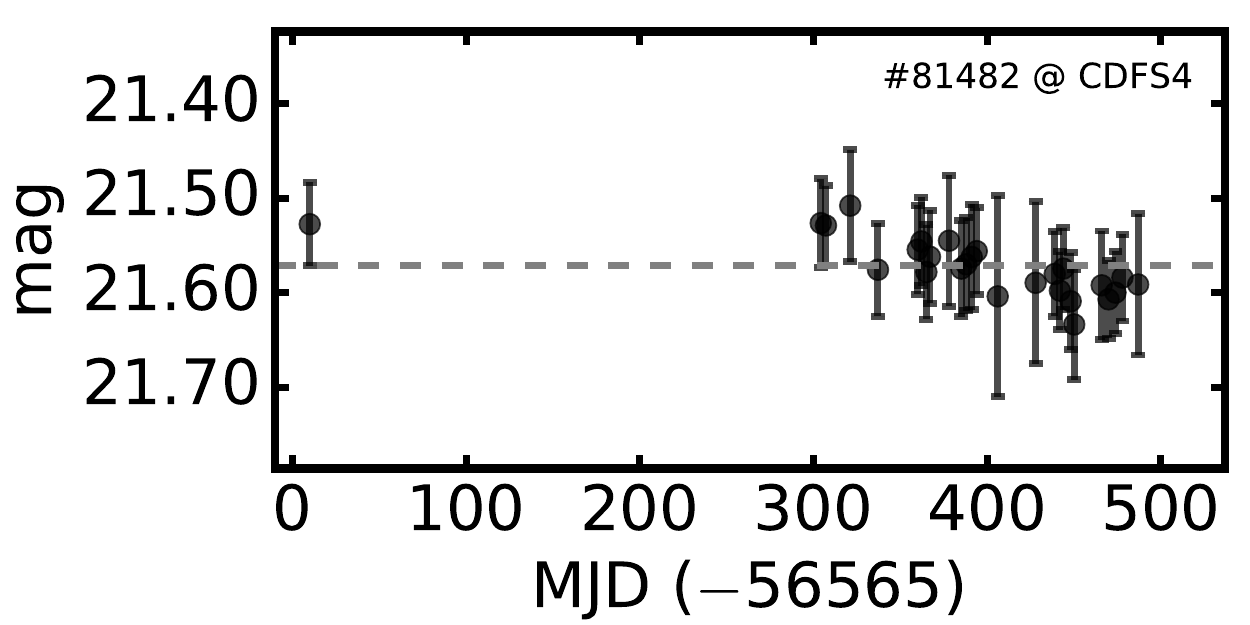}
\includegraphics[width=0.32\textwidth]{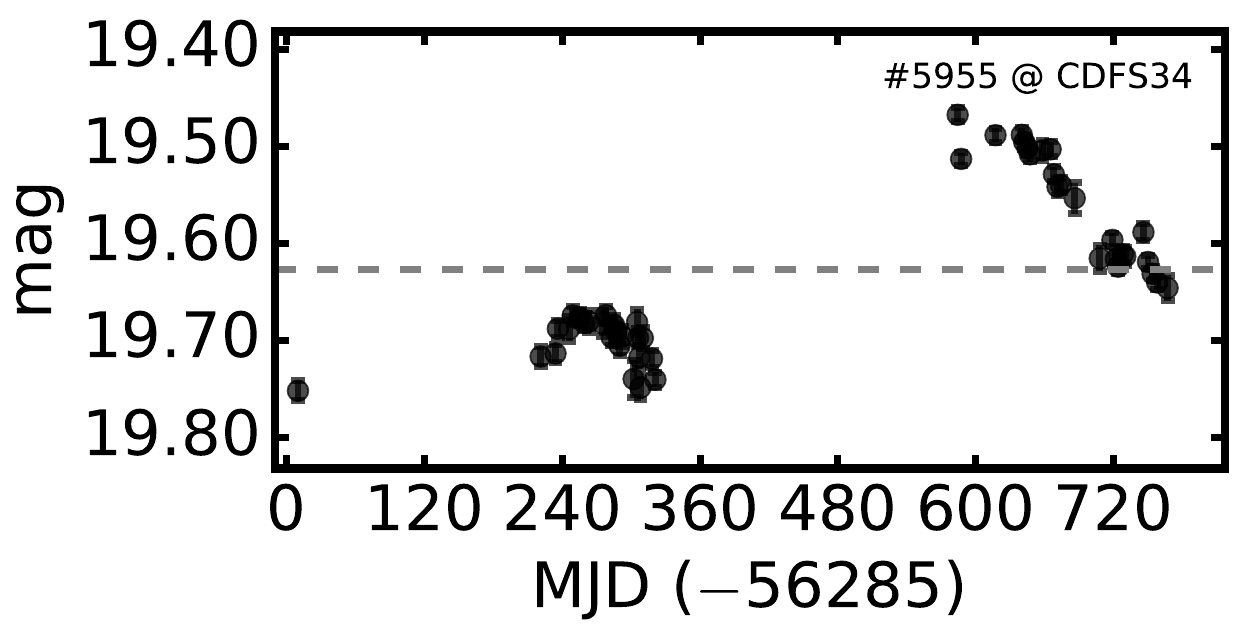}
\includegraphics[width=0.32\textwidth]{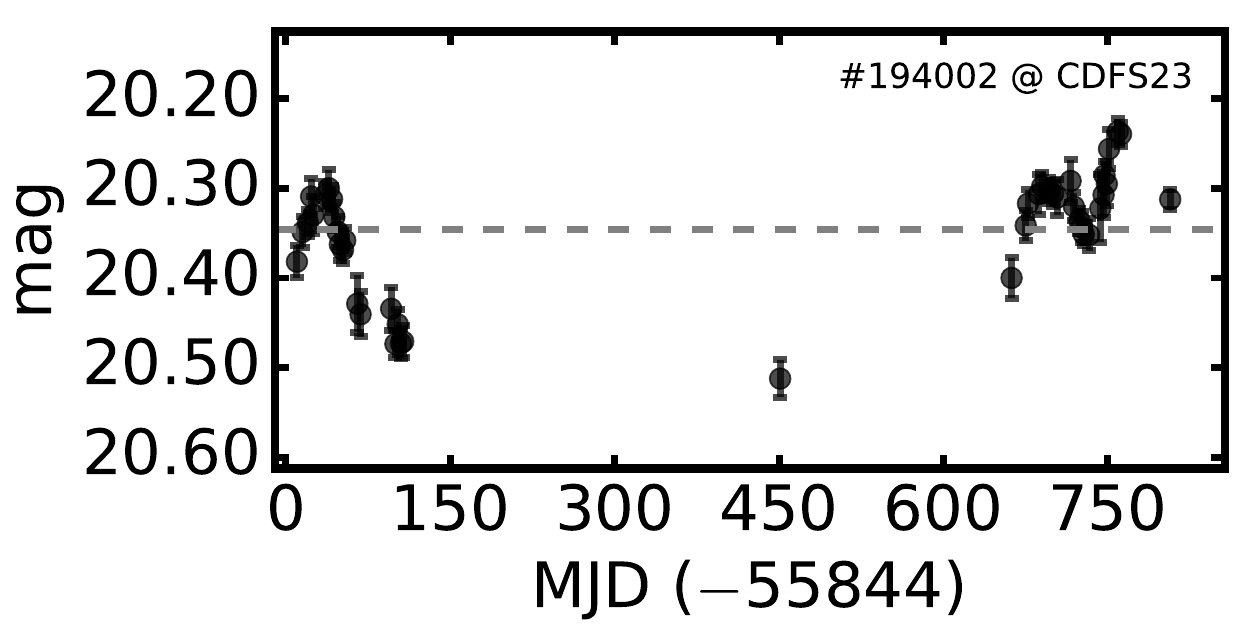}
\includegraphics[width=0.32\textwidth]{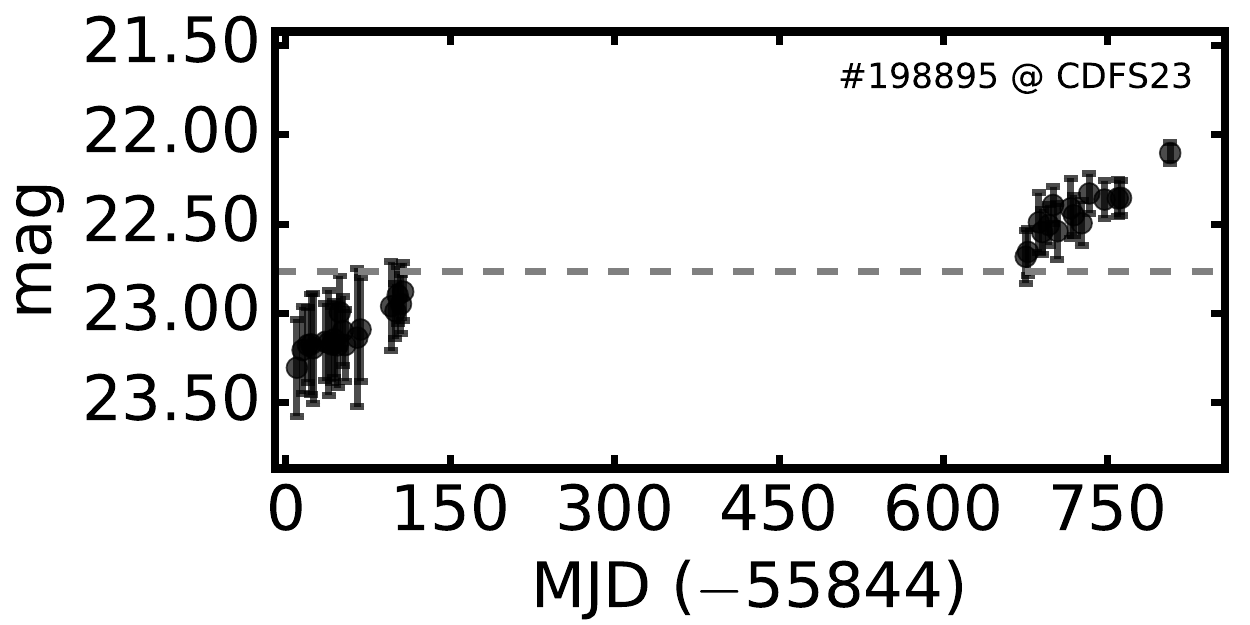}
\caption{Light curves of AGNs and AGN candidates. Top panel shows three confirmed AGNs from the 
Million Quasar Catalog at different redshifts (from left to right the redshift is $z$\,=\,1.025, 2.164, 3.872, 
respectively). The gray dashed line in each figure represents the magnitude measured in $S_{0}$, while 
the number is the ID in the \textit{clean} sample. The bottom panel gives three AGN candidates which are 
not spectroscopically confirmed by any current survey.}
\label{fig:agns}
\end{figure*}

\begin{figure*}
\centering
\includegraphics[width=0.23\textwidth]{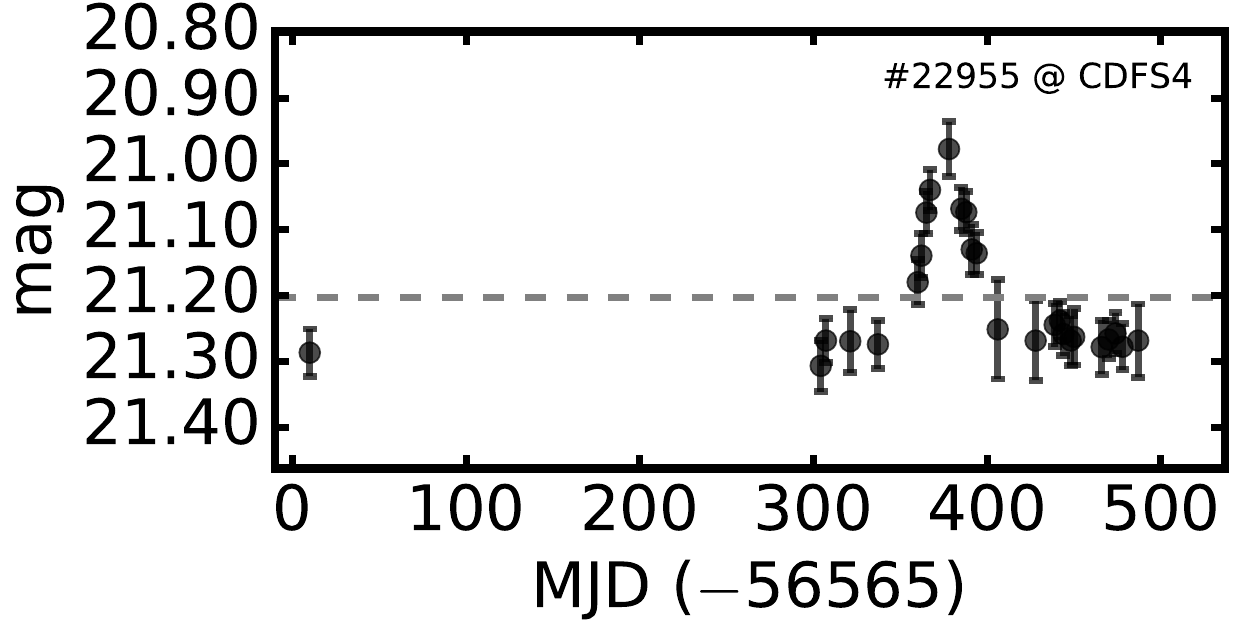}
\includegraphics[width=0.23\textwidth]{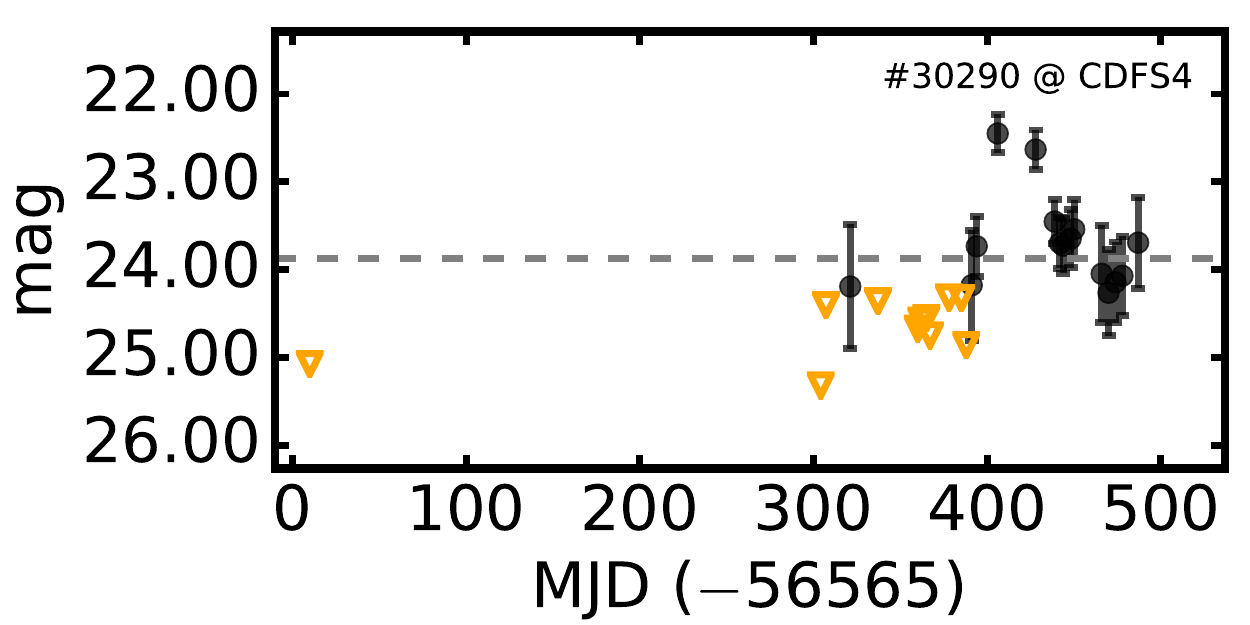}
\includegraphics[width=0.23\textwidth]{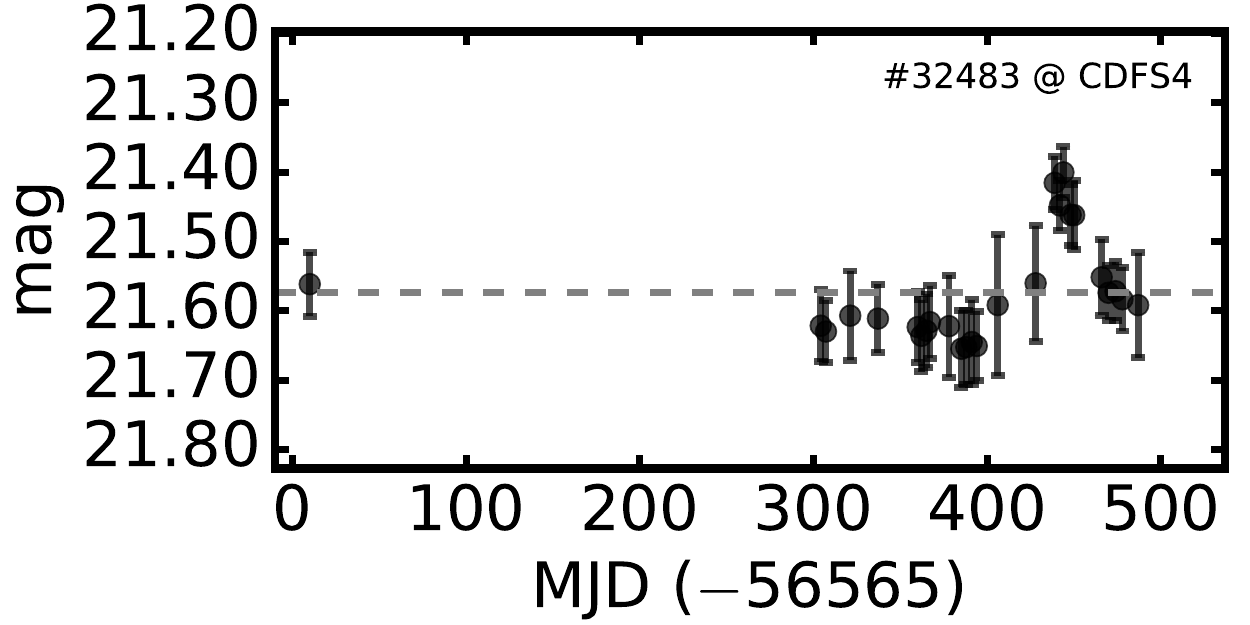}
\includegraphics[width=0.23\textwidth]{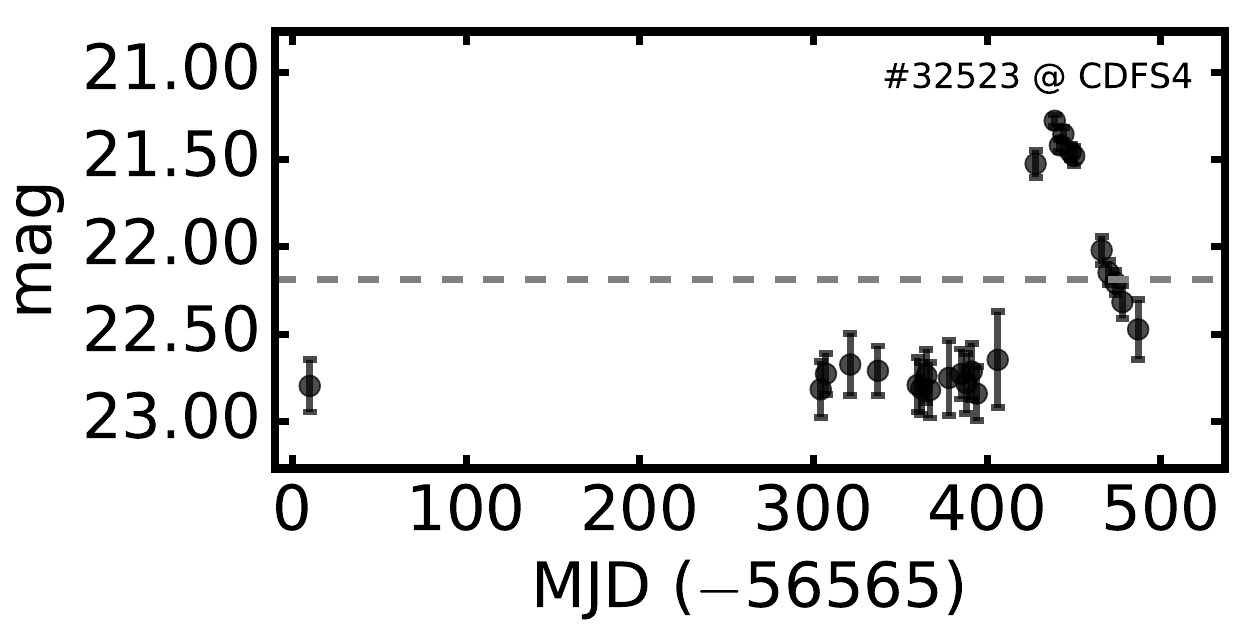}
\includegraphics[width=0.23\textwidth]{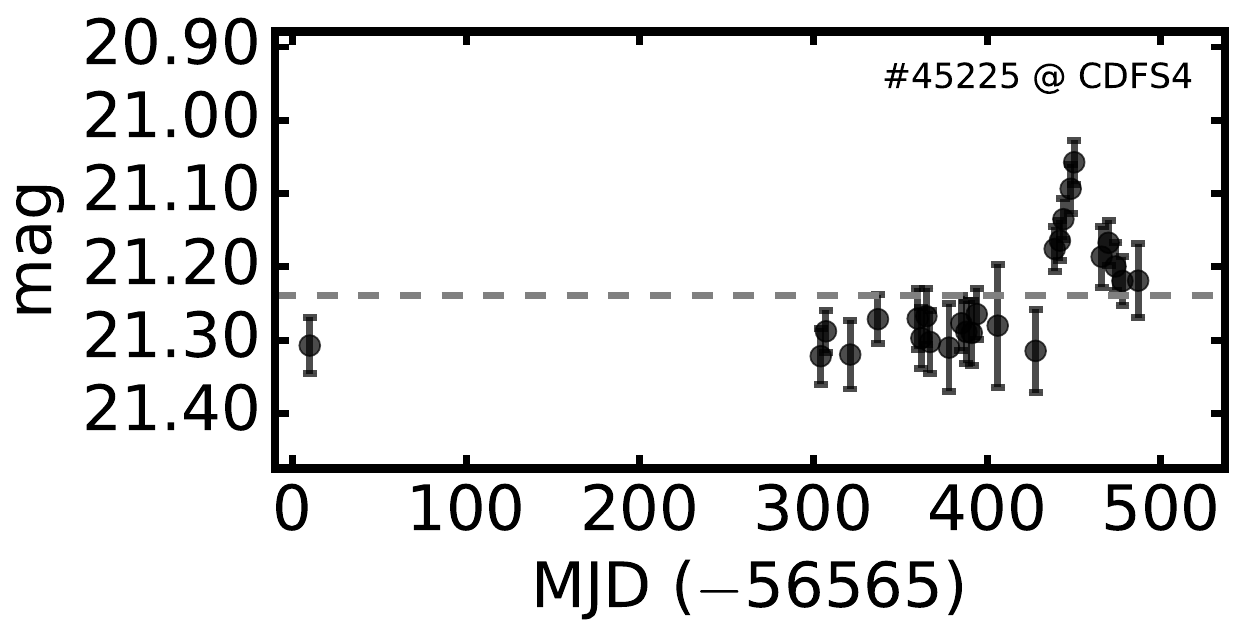}
\includegraphics[width=0.23\textwidth]{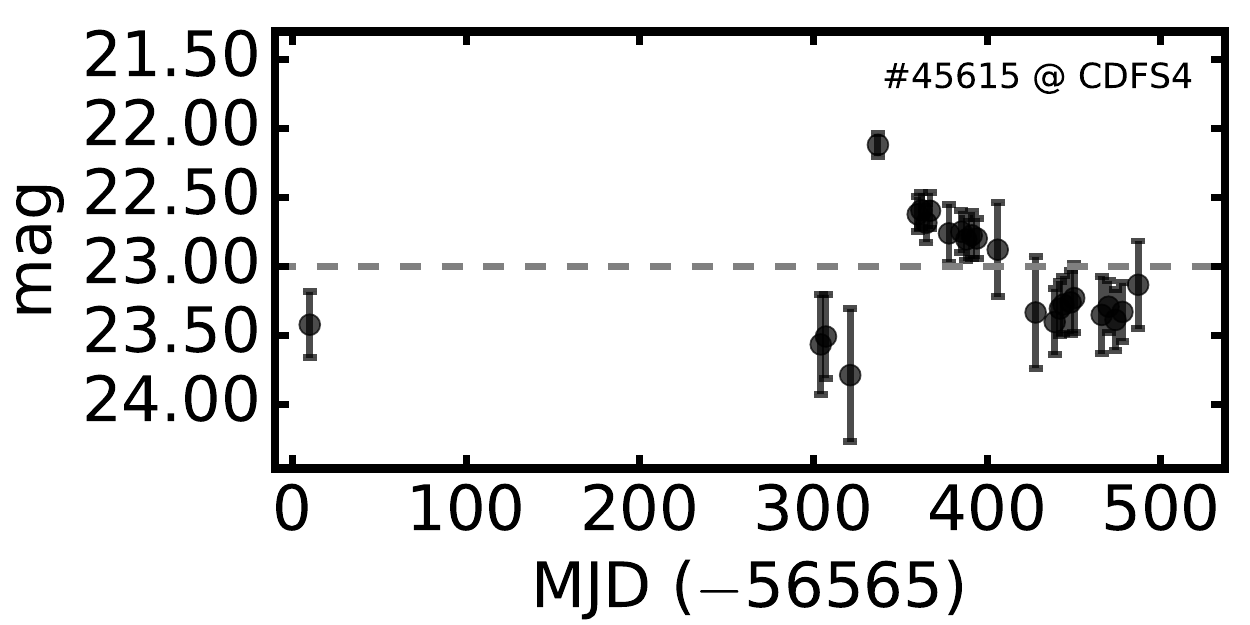}
\includegraphics[width=0.23\textwidth]{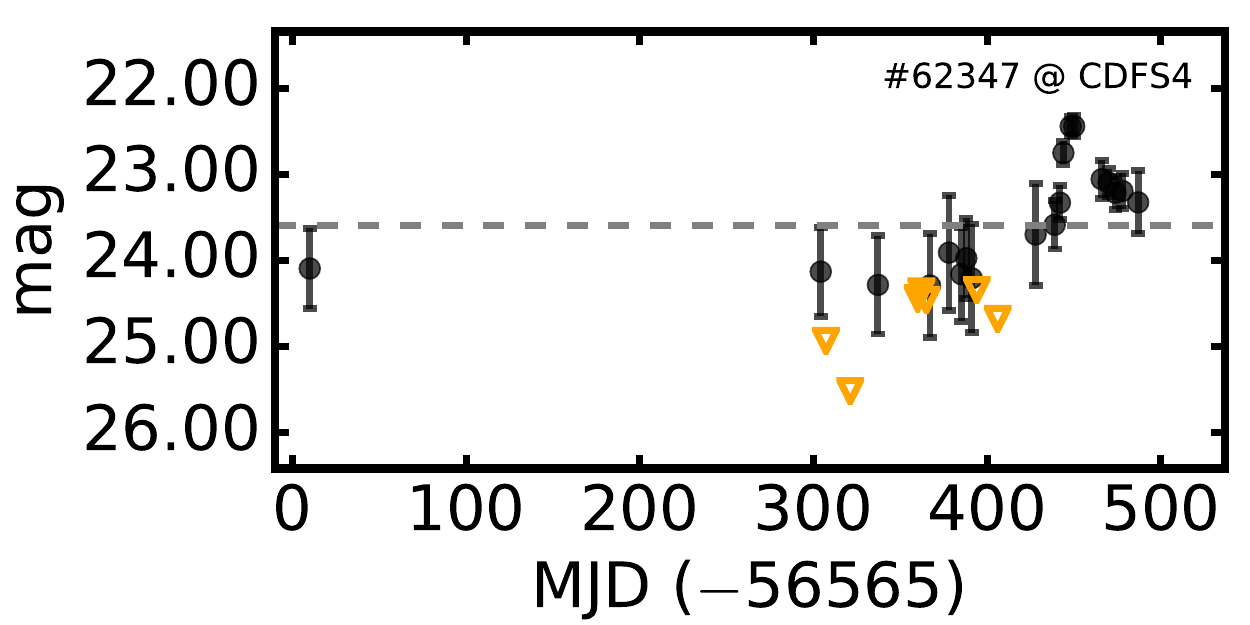}
\includegraphics[width=0.23\textwidth]{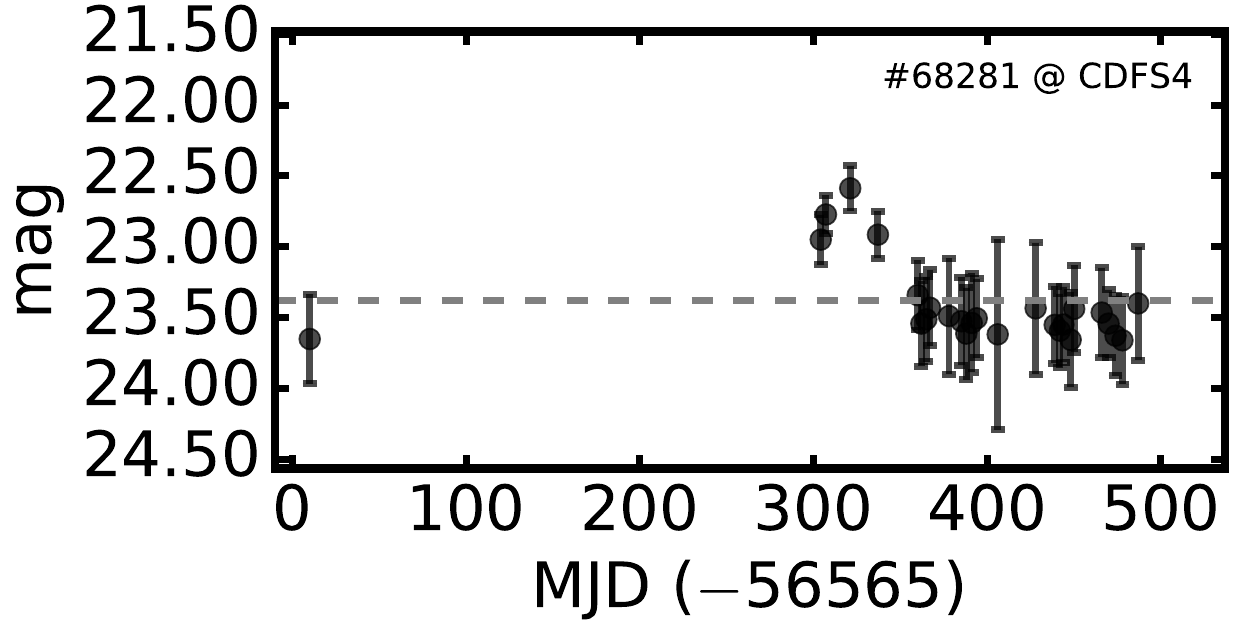}
\includegraphics[width=0.23\textwidth]{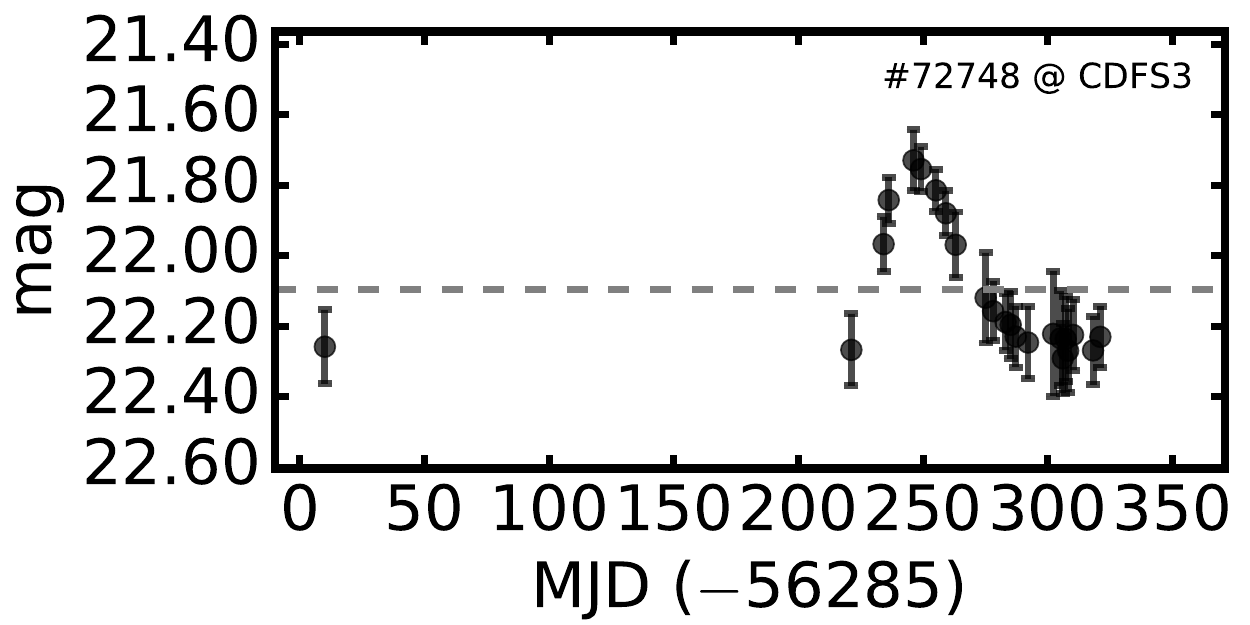}
\includegraphics[width=0.23\textwidth]{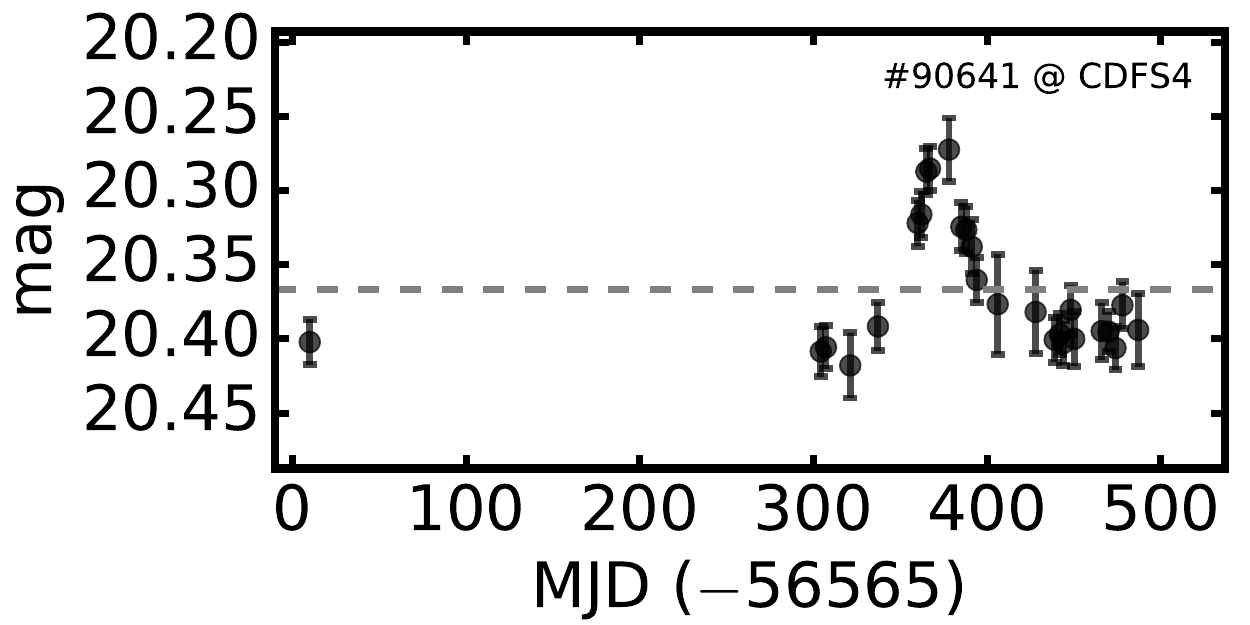}
\includegraphics[width=0.23\textwidth]{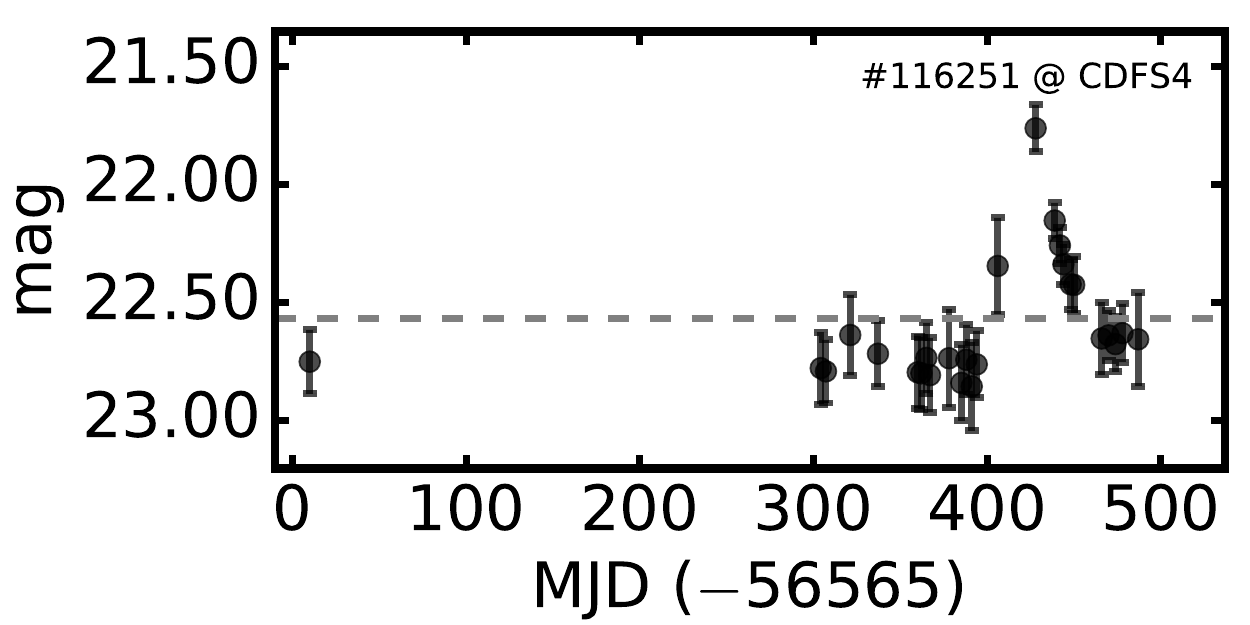}
\includegraphics[width=0.23\textwidth]{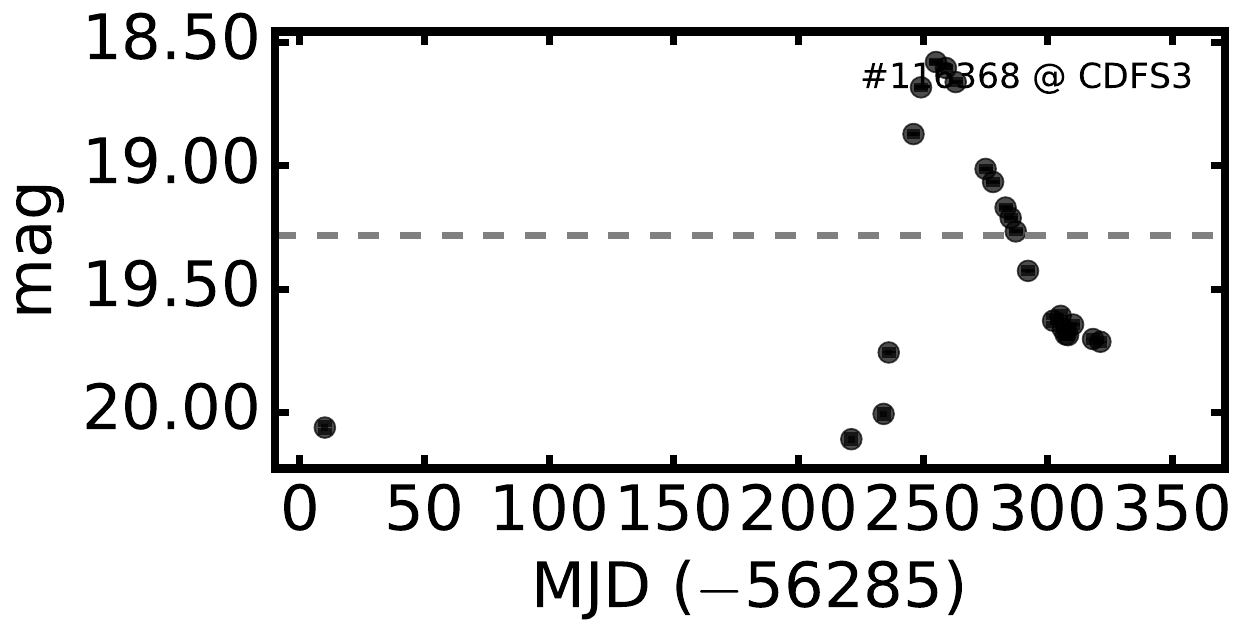}
\includegraphics[width=0.23\textwidth]{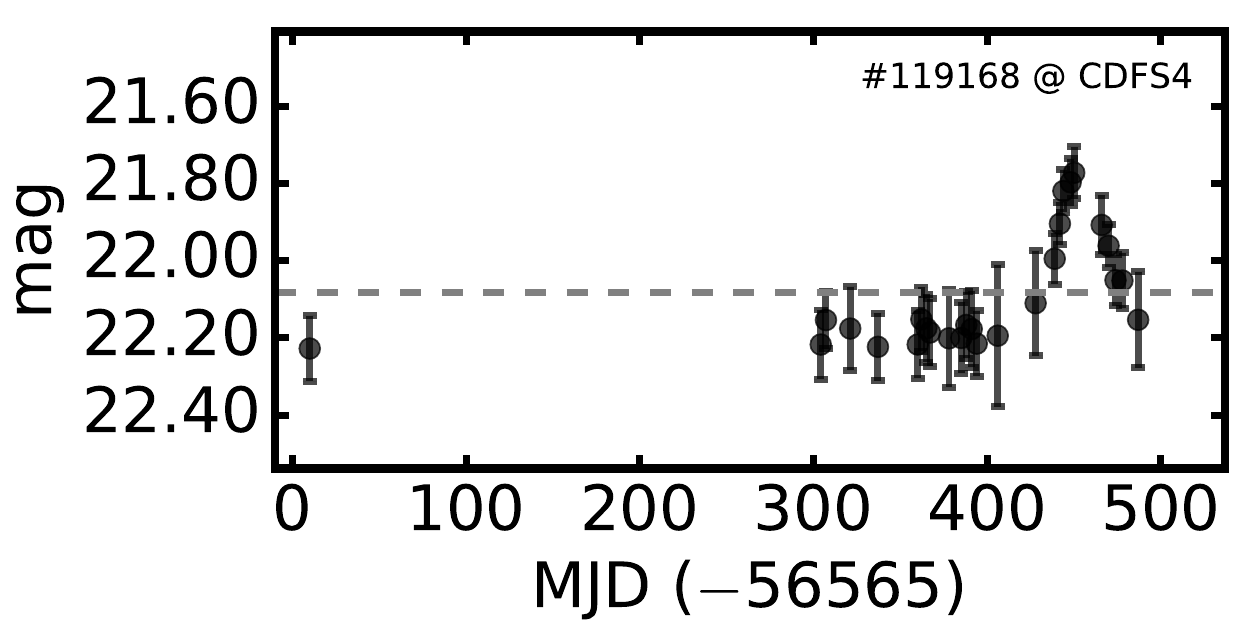}
\includegraphics[width=0.23\textwidth]{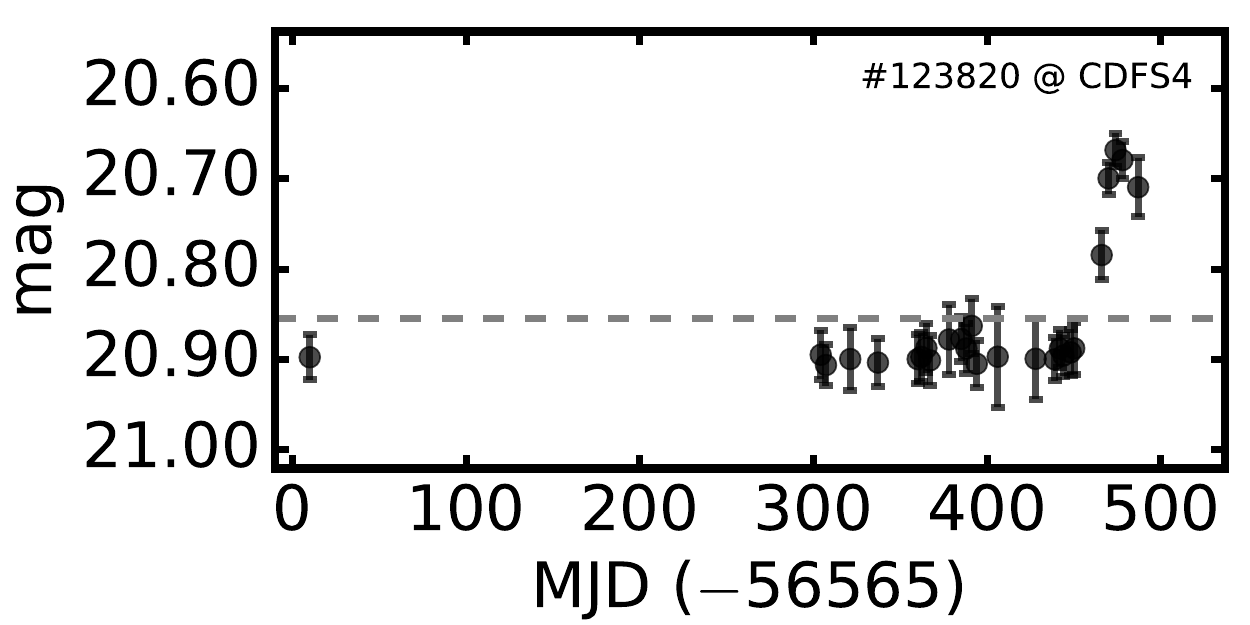}
\includegraphics[width=0.23\textwidth]{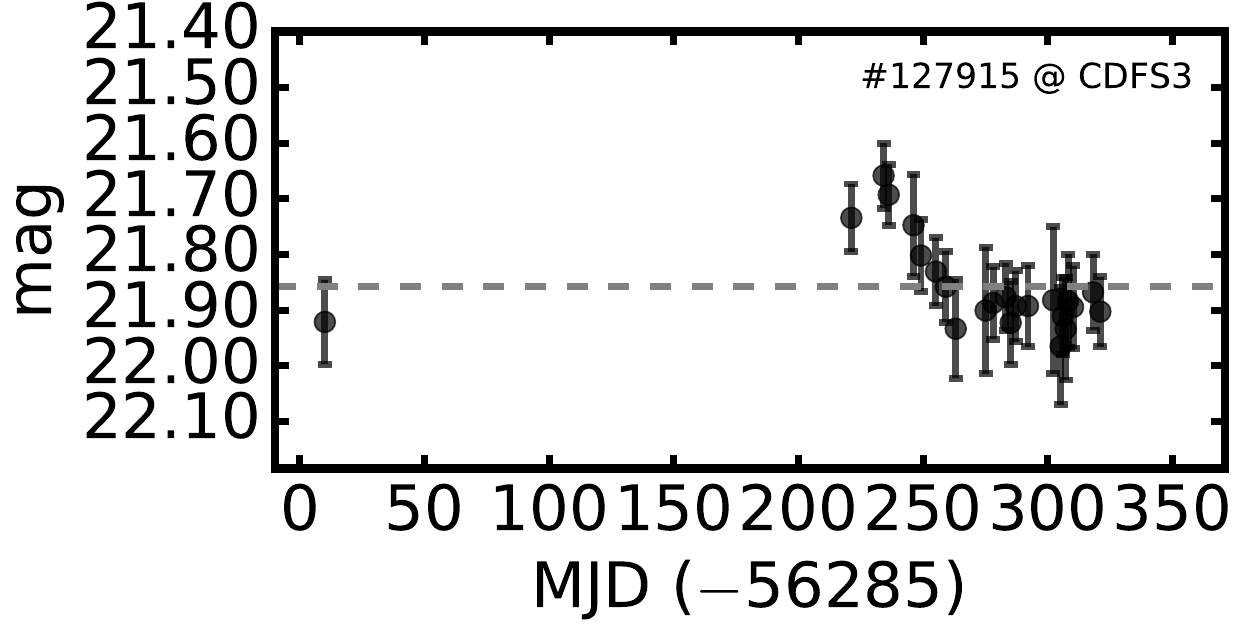}
\includegraphics[width=0.23\textwidth]{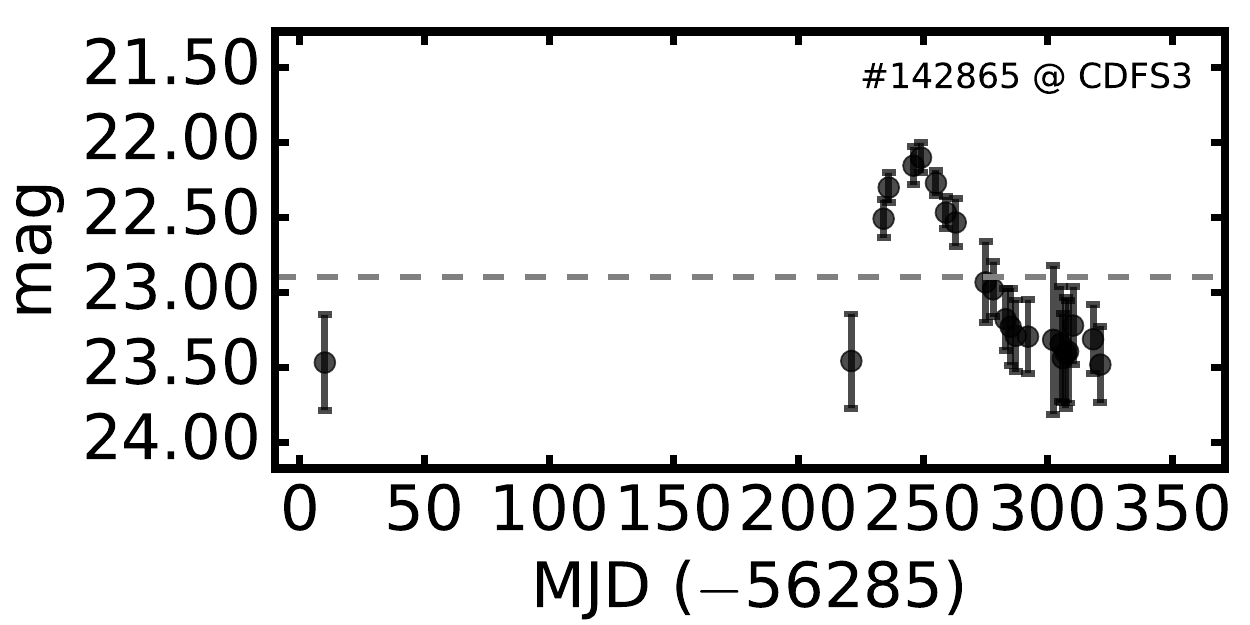}
\includegraphics[width=0.23\textwidth]{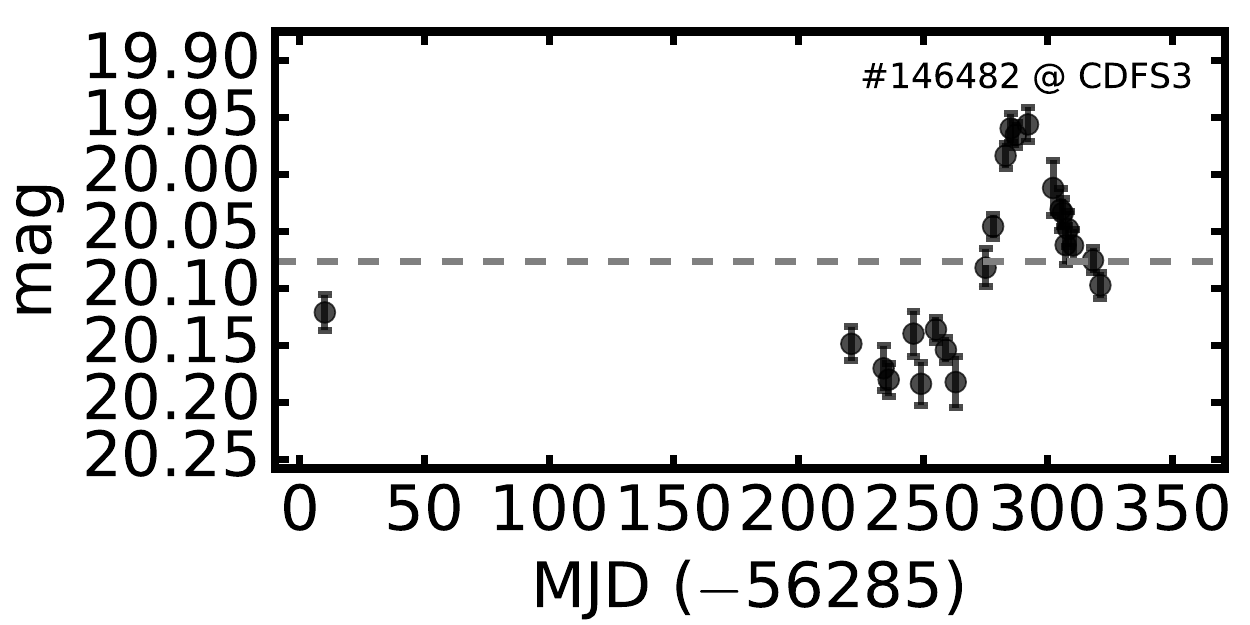}
\includegraphics[width=0.23\textwidth]{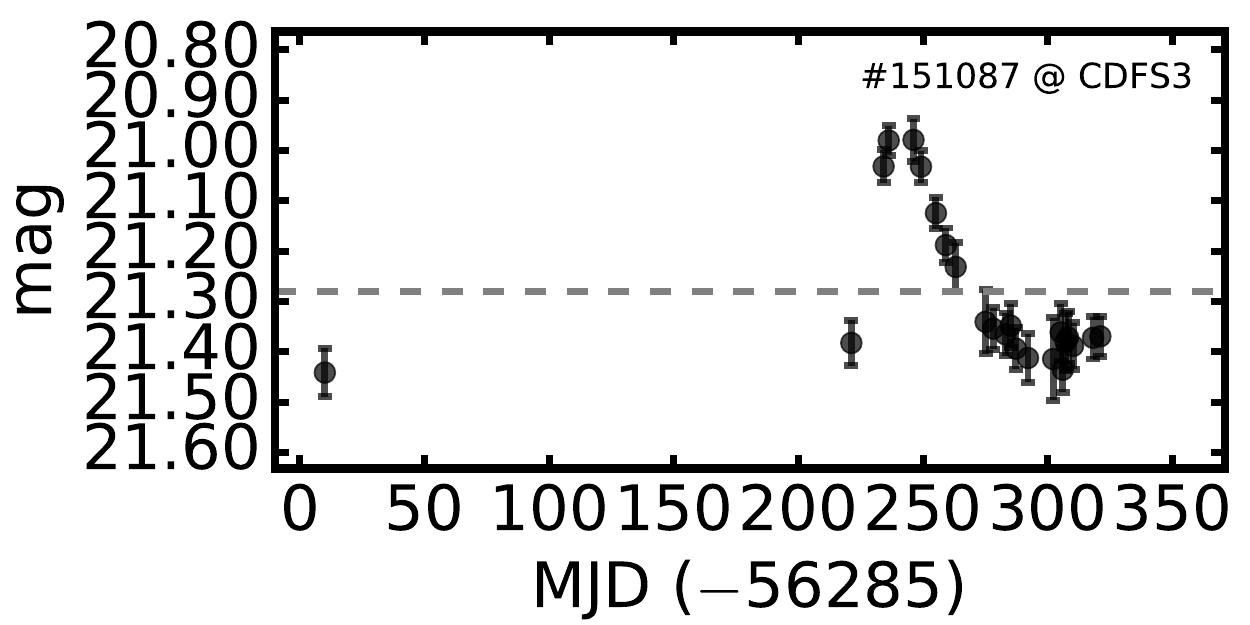}
\includegraphics[width=0.23\textwidth]{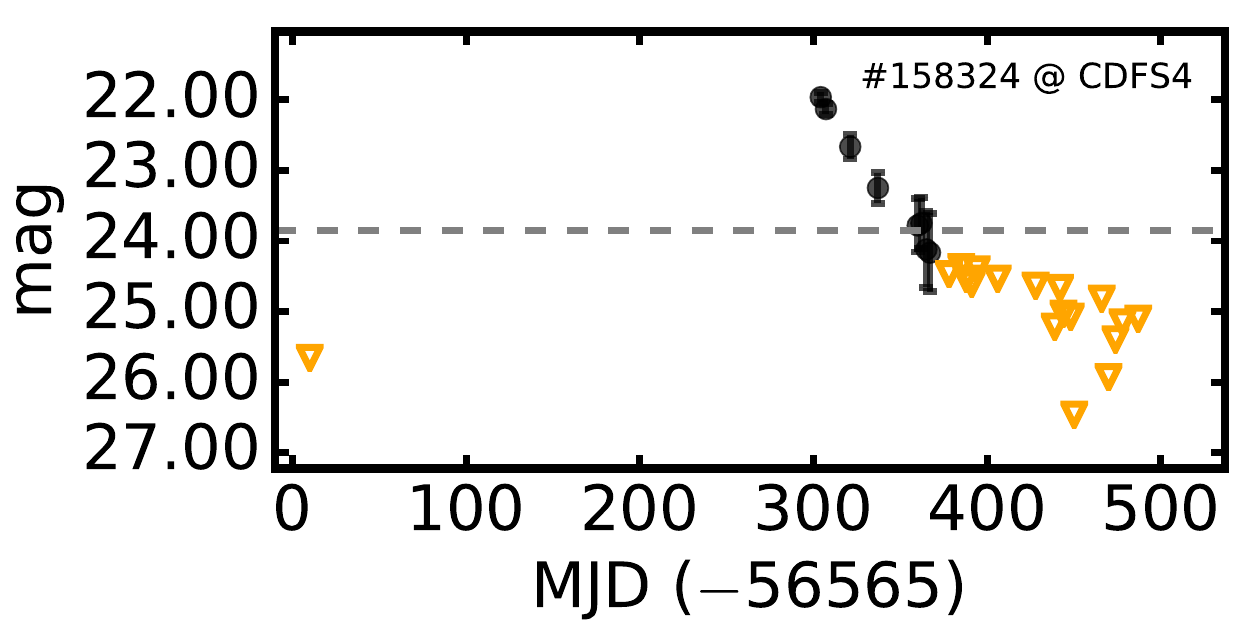}
\includegraphics[width=0.23\textwidth]{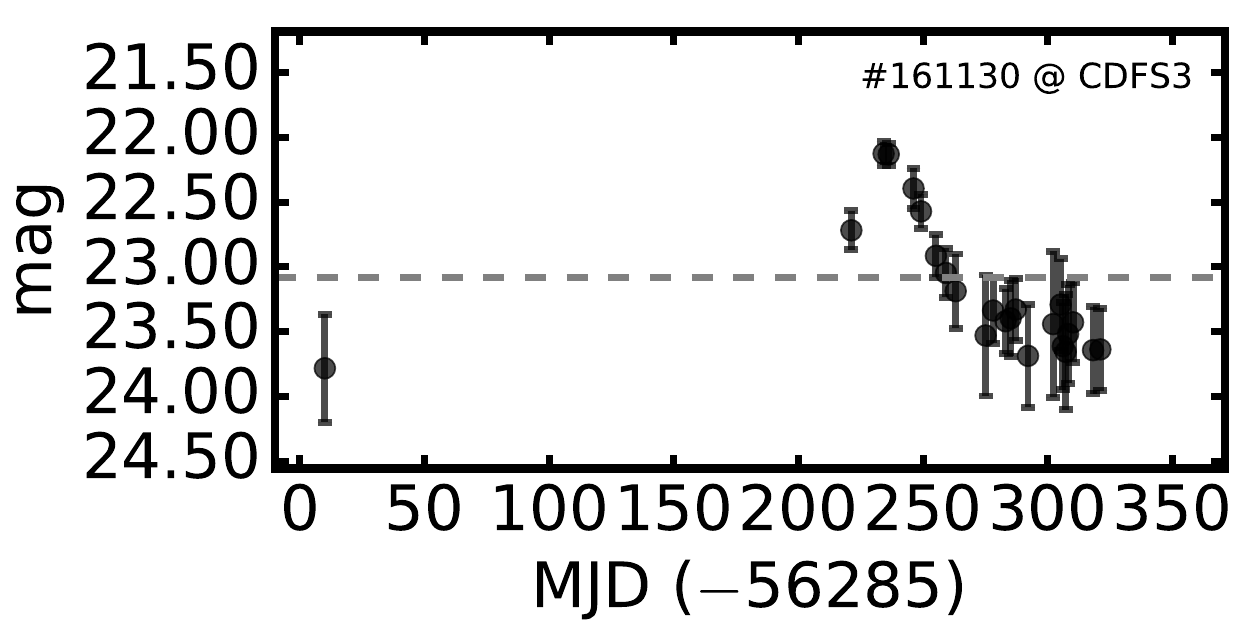}
\includegraphics[width=0.23\textwidth]{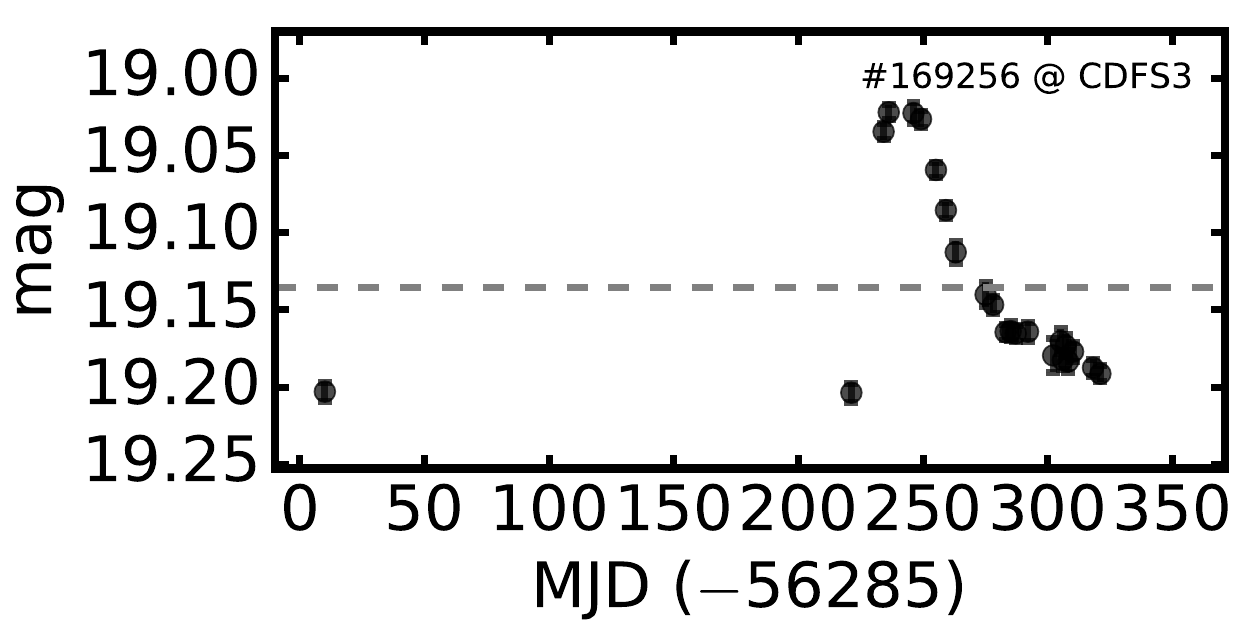}
\includegraphics[width=0.23\textwidth]{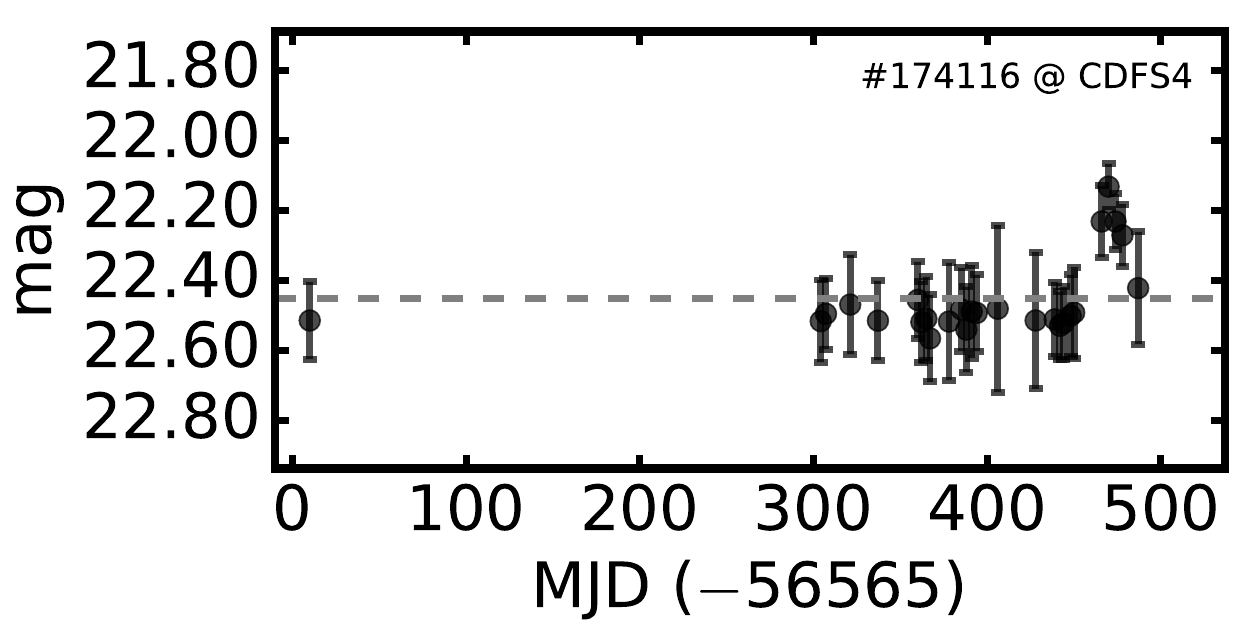}
\includegraphics[width=0.23\textwidth]{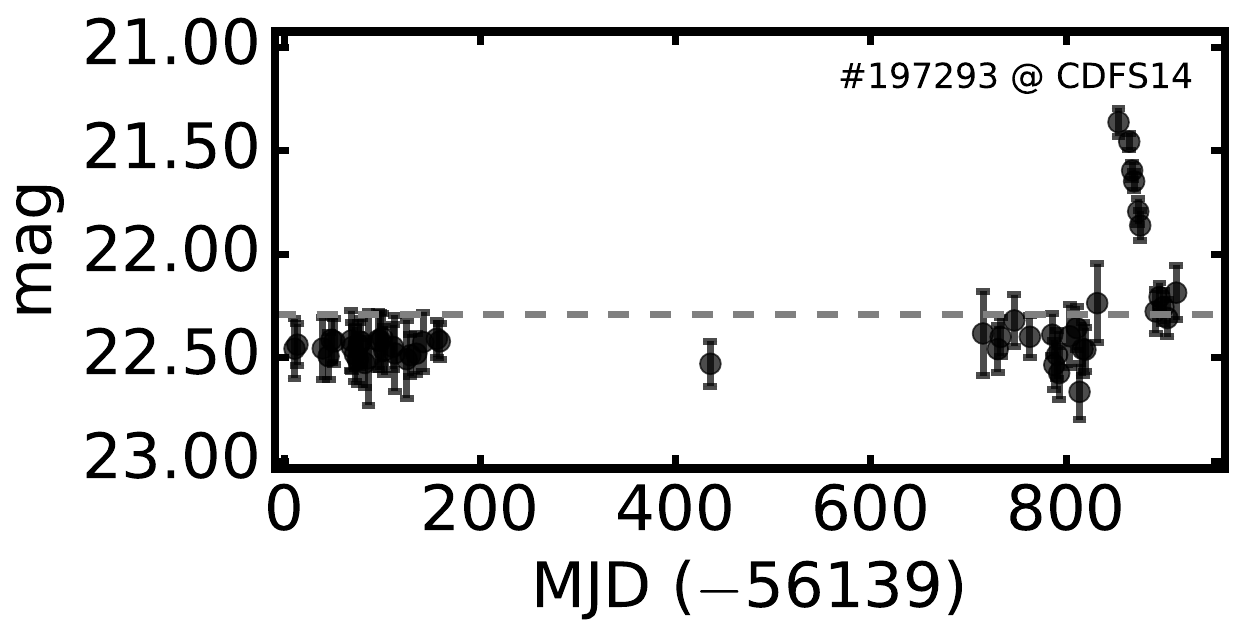}
\includegraphics[width=0.23\textwidth]{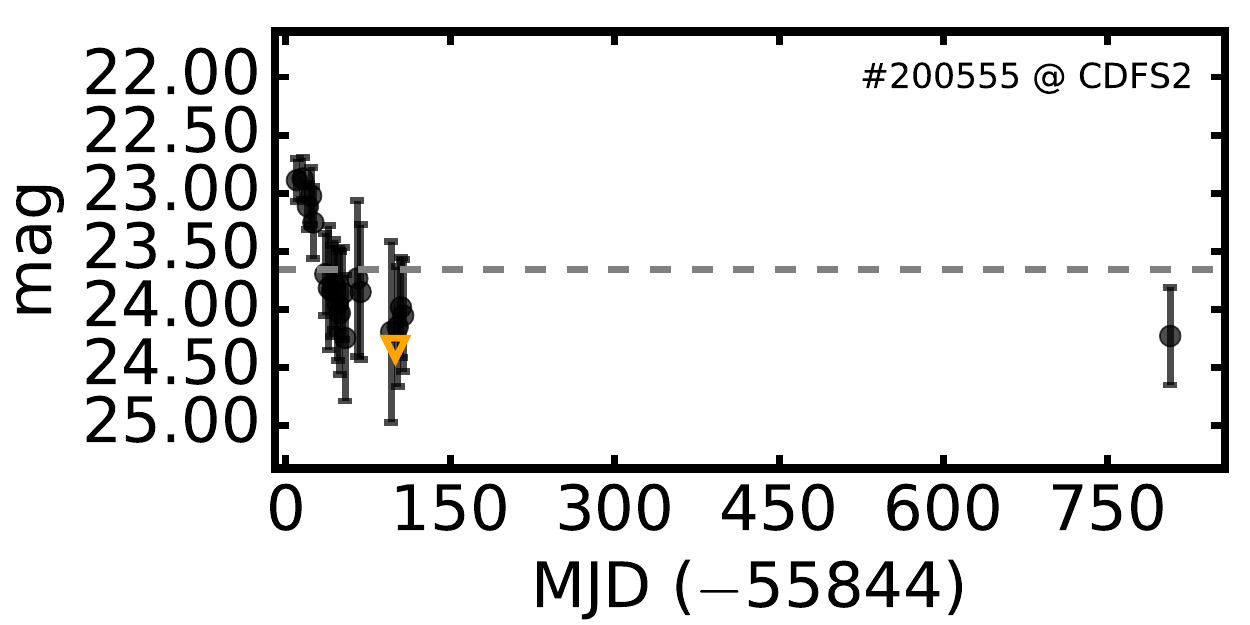}
\includegraphics[width=0.23\textwidth]{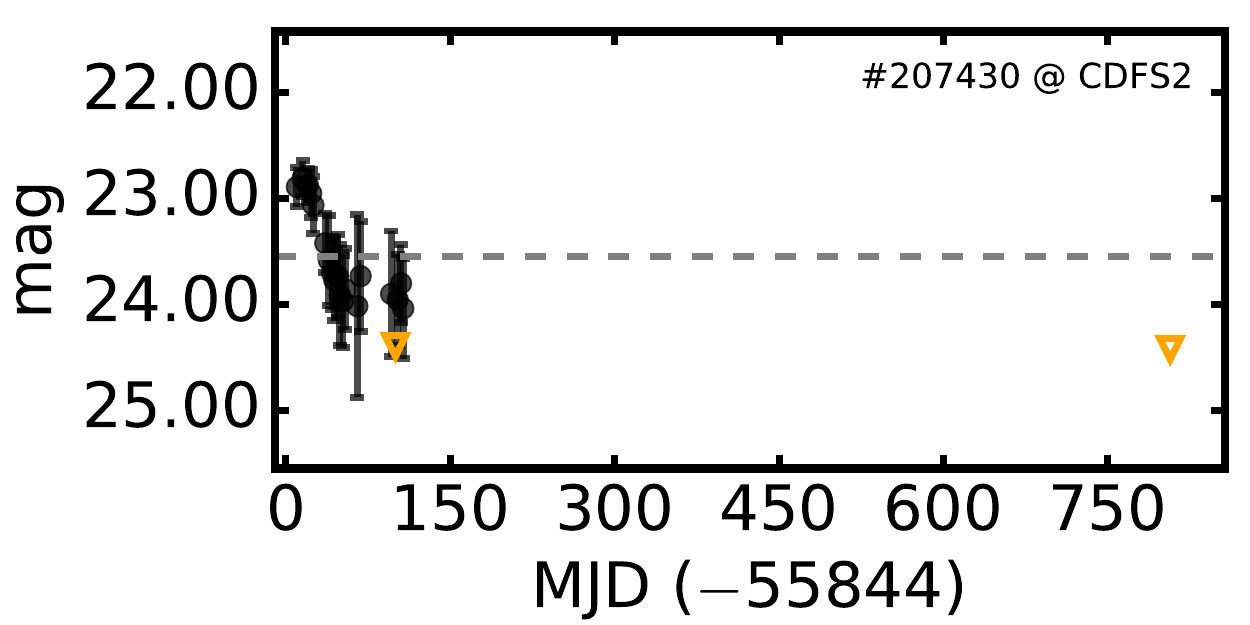}
\includegraphics[width=0.23\textwidth]{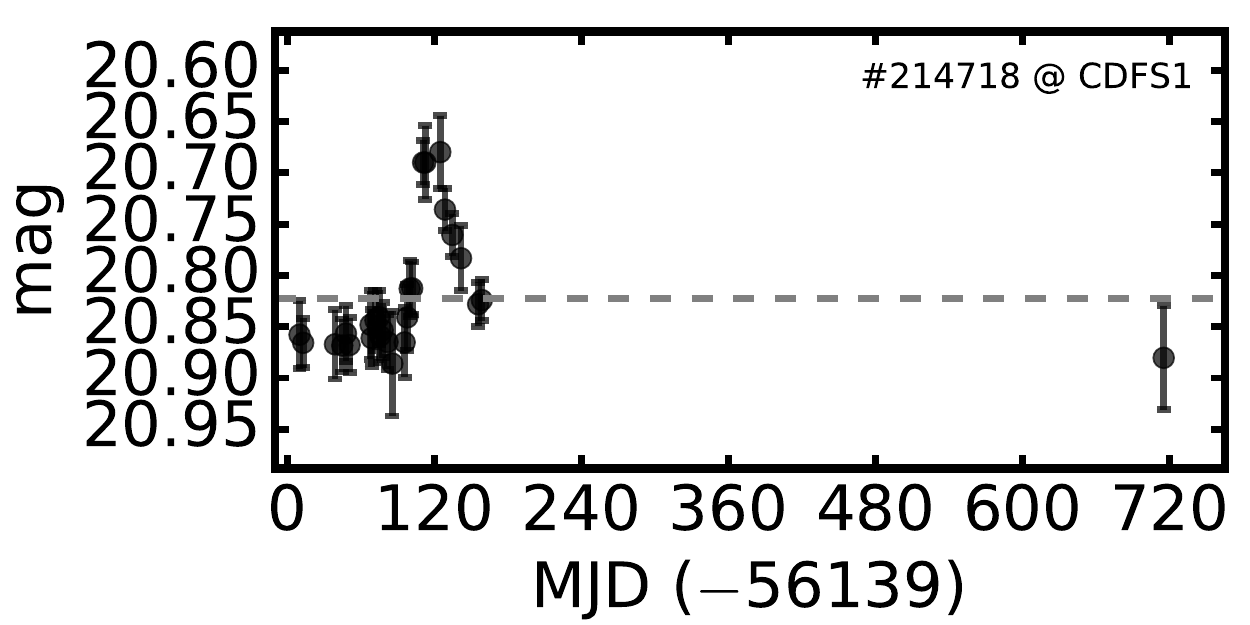}
\includegraphics[width=0.23\textwidth]{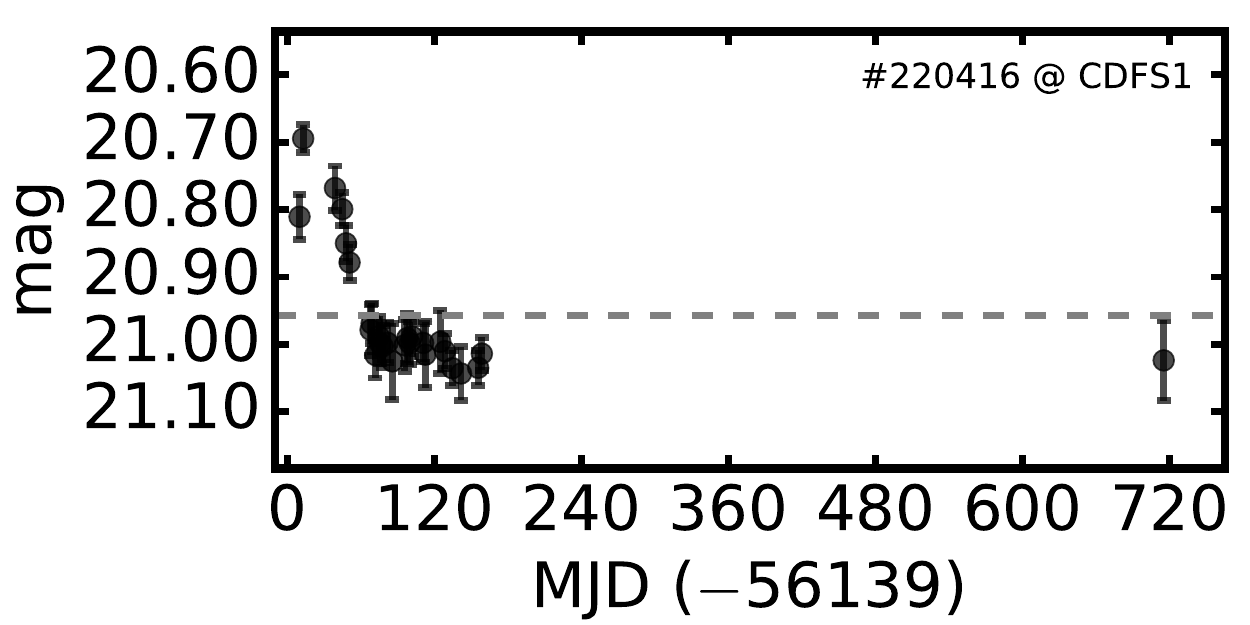}
\includegraphics[width=0.23\textwidth]{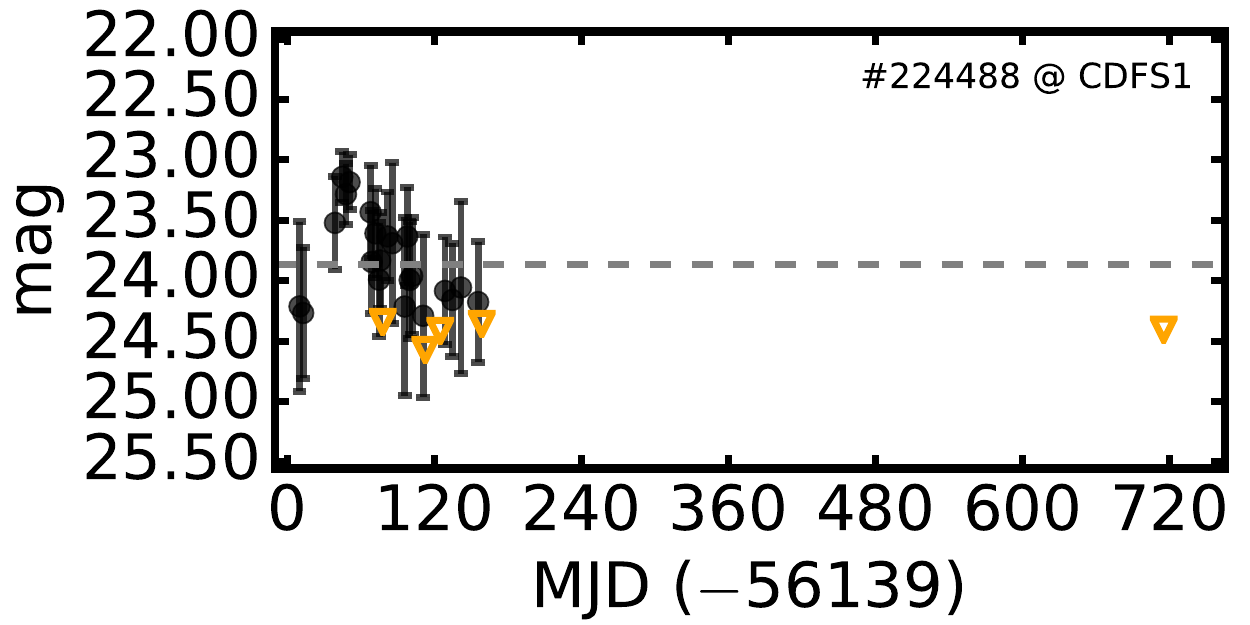}
\includegraphics[width=0.23\textwidth]{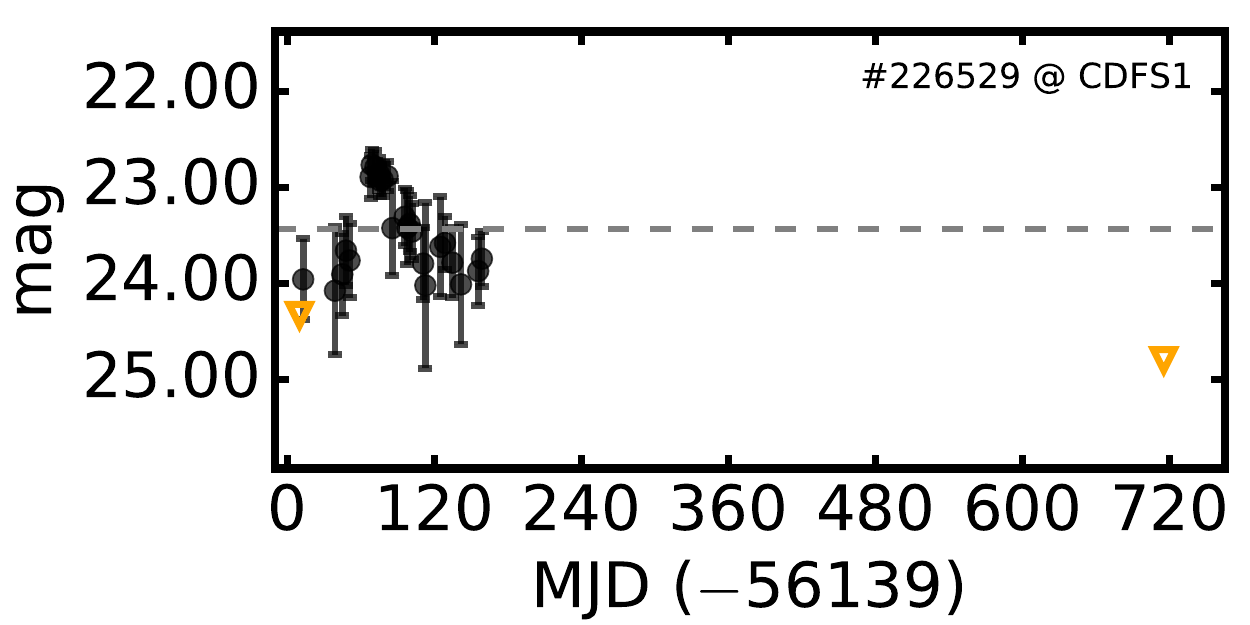}
\includegraphics[width=0.23\textwidth]{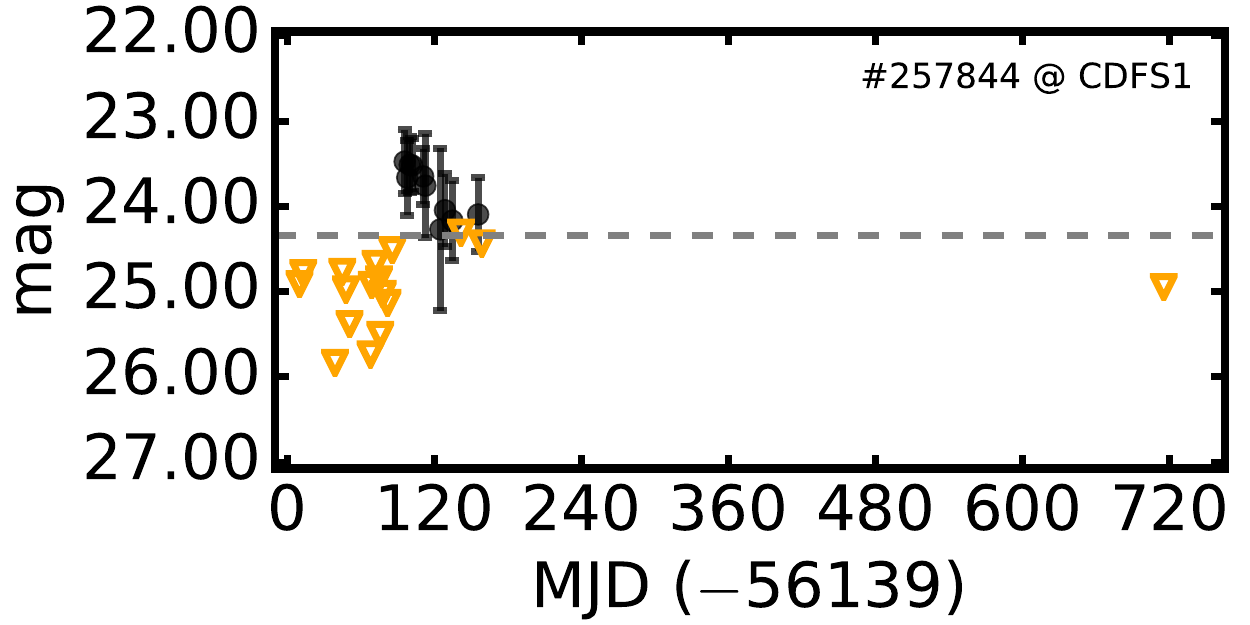}
\includegraphics[width=0.23\textwidth]{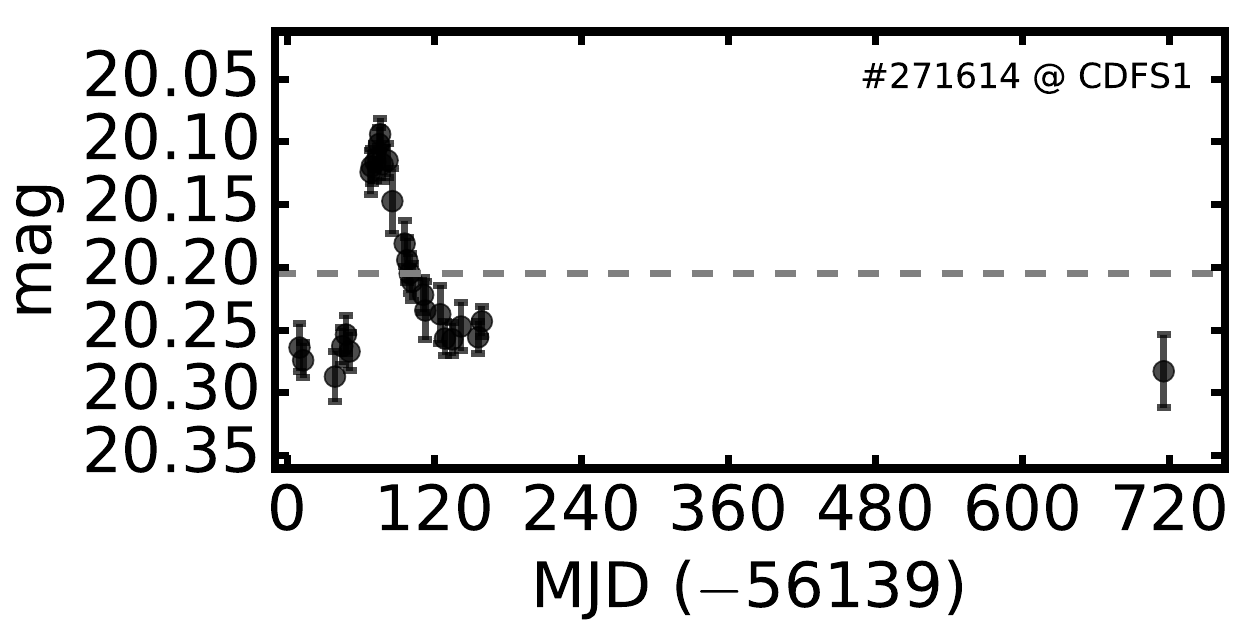}
\includegraphics[width=0.23\textwidth]{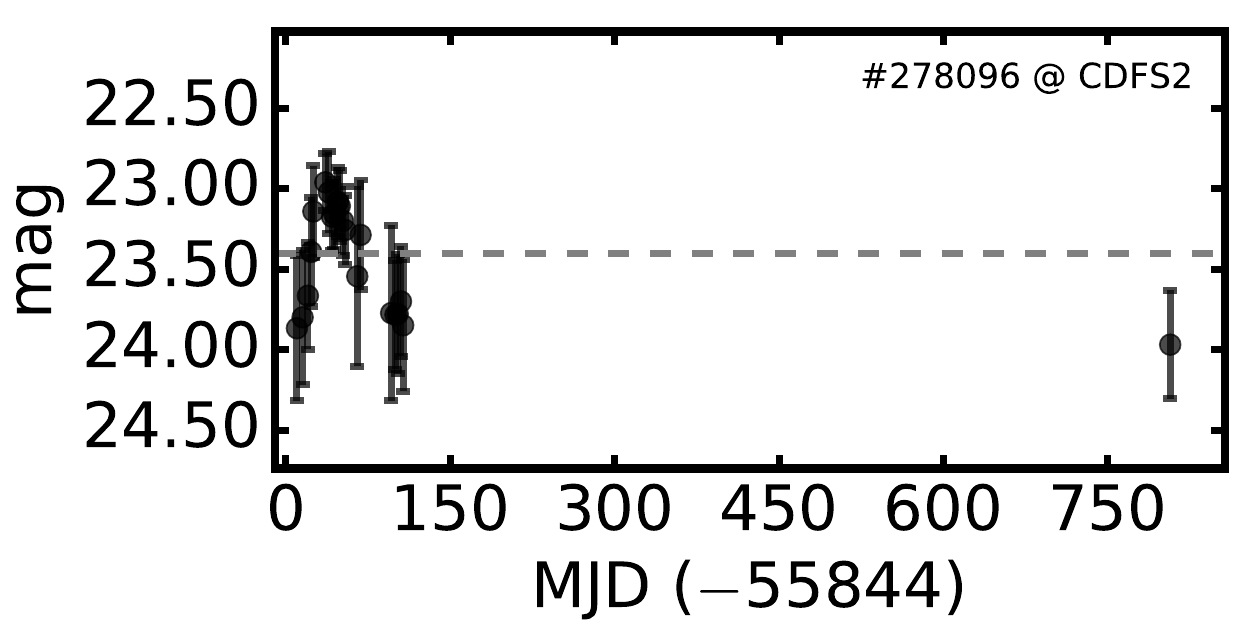}
\includegraphics[width=0.23\textwidth]{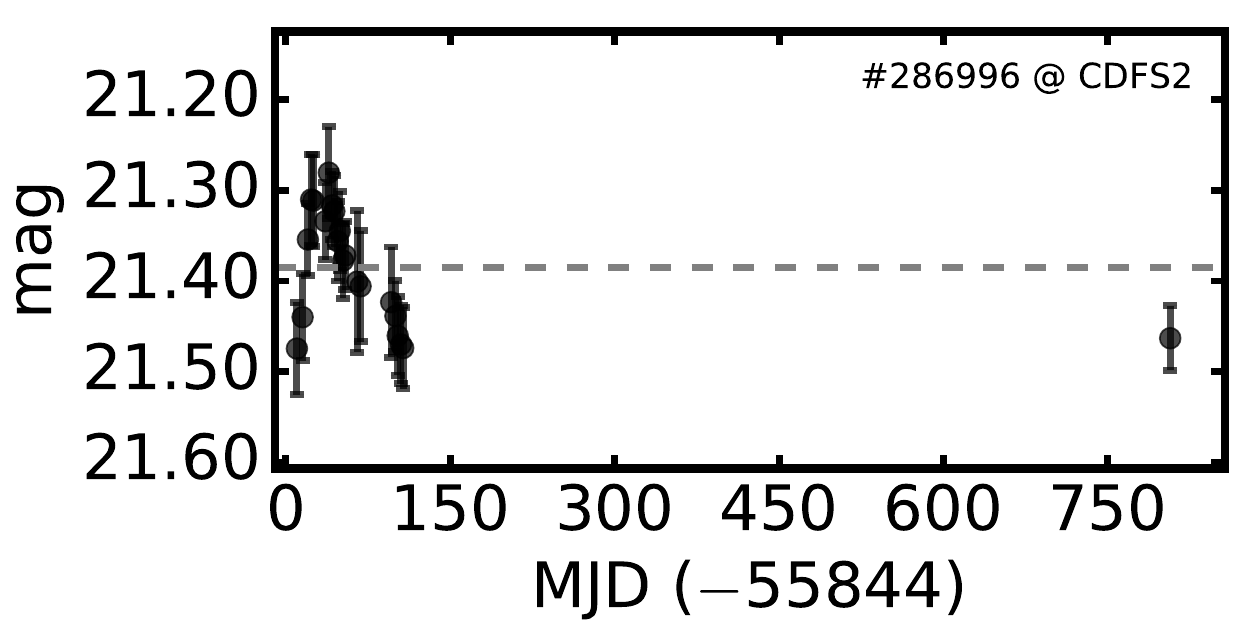}
\includegraphics[width=0.23\textwidth]{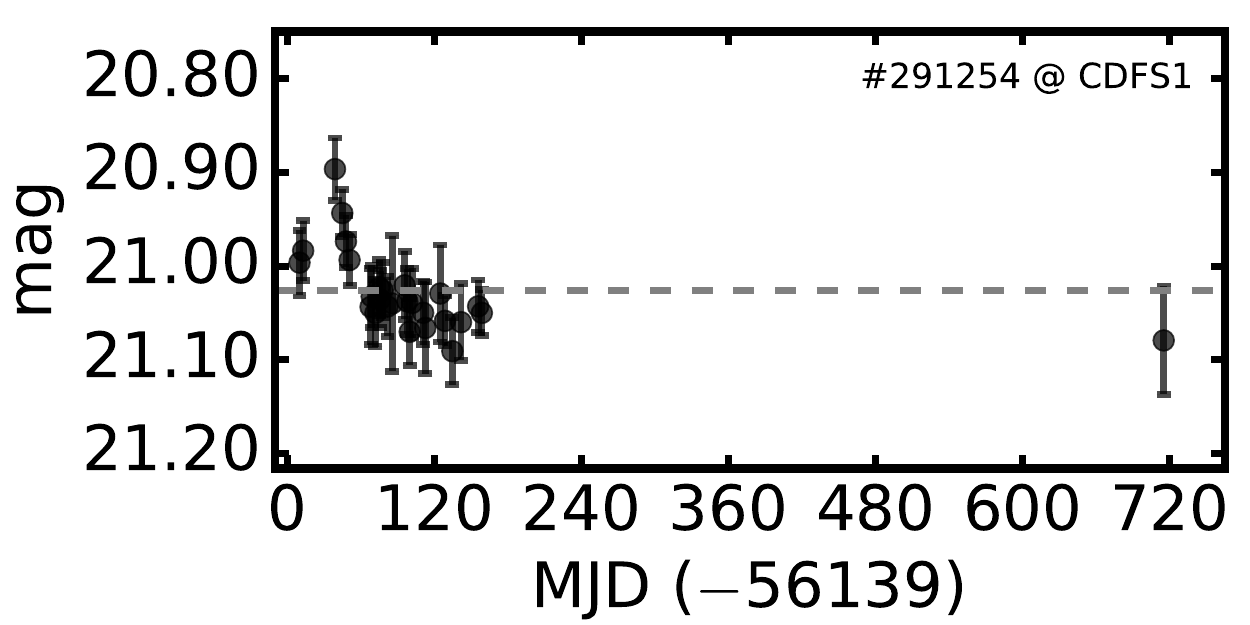}
\includegraphics[width=0.23\textwidth]{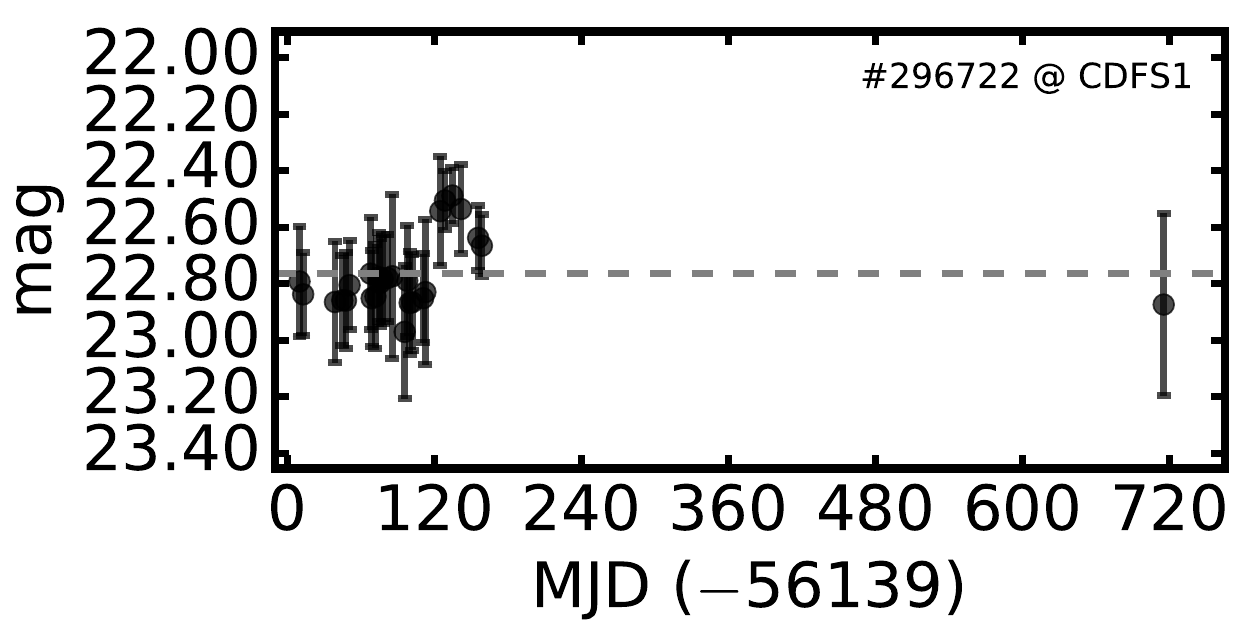}
\includegraphics[width=0.23\textwidth]{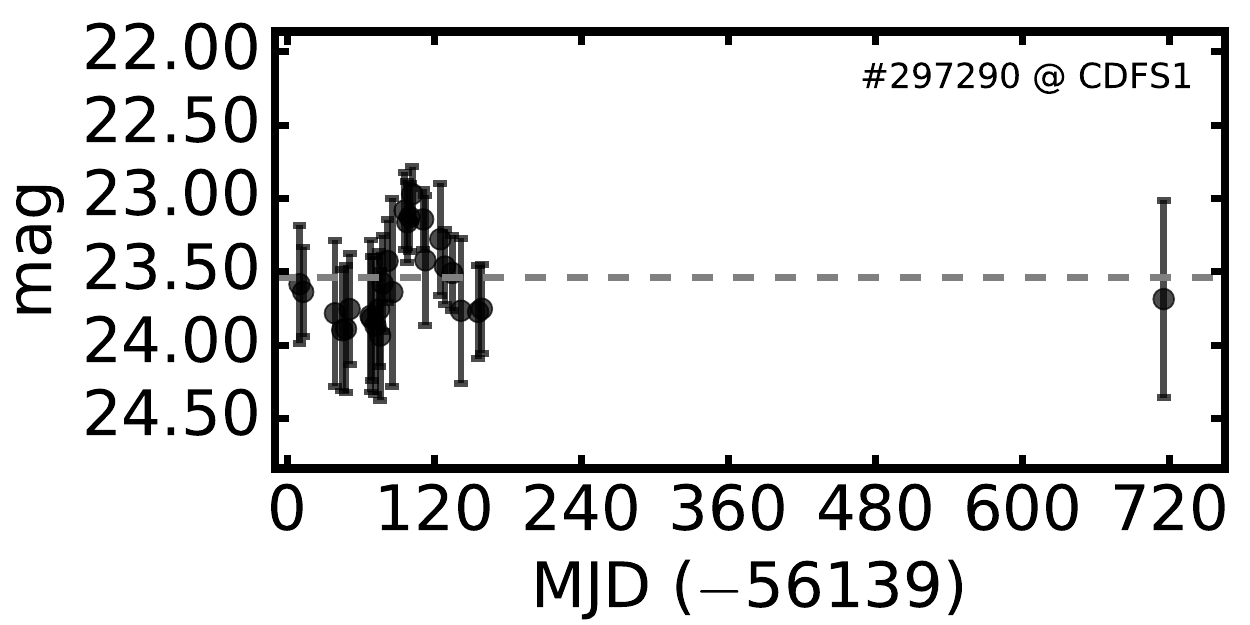}
\includegraphics[width=0.23\textwidth]{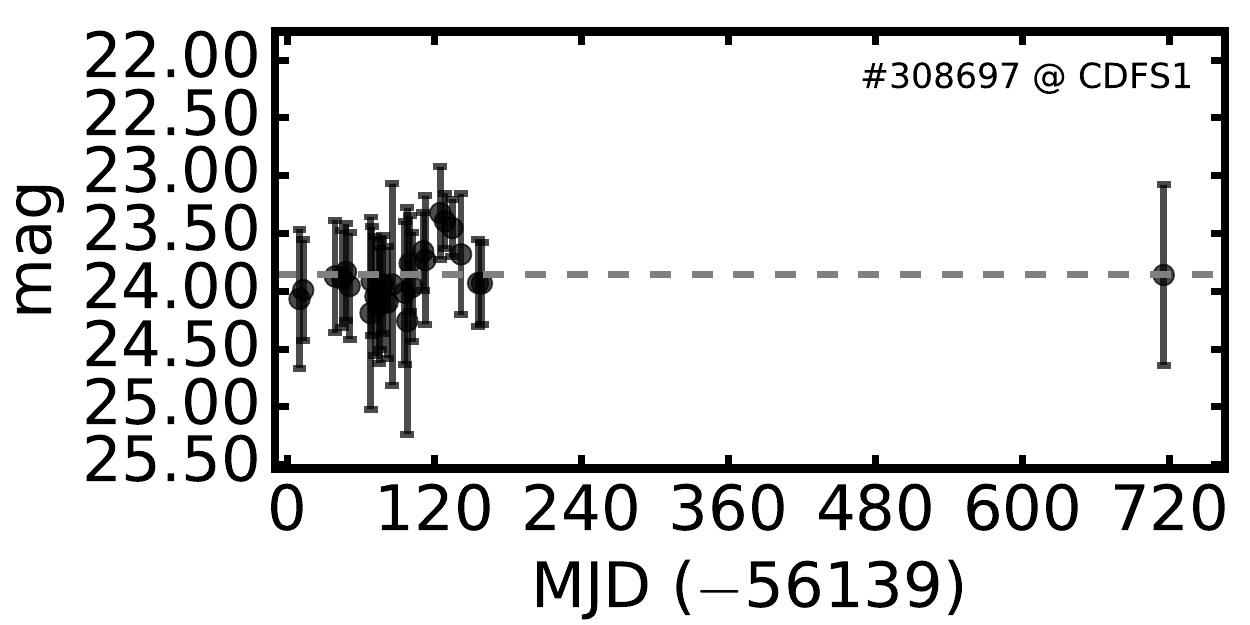}
\includegraphics[width=0.23\textwidth]{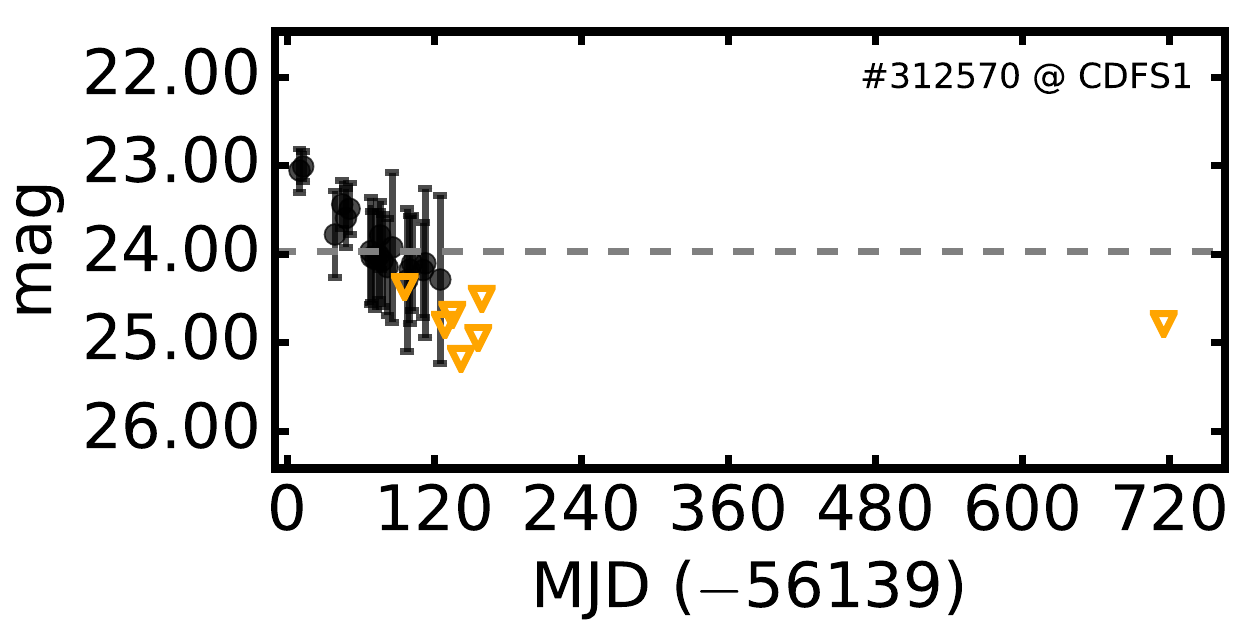}
\includegraphics[width=0.23\textwidth]{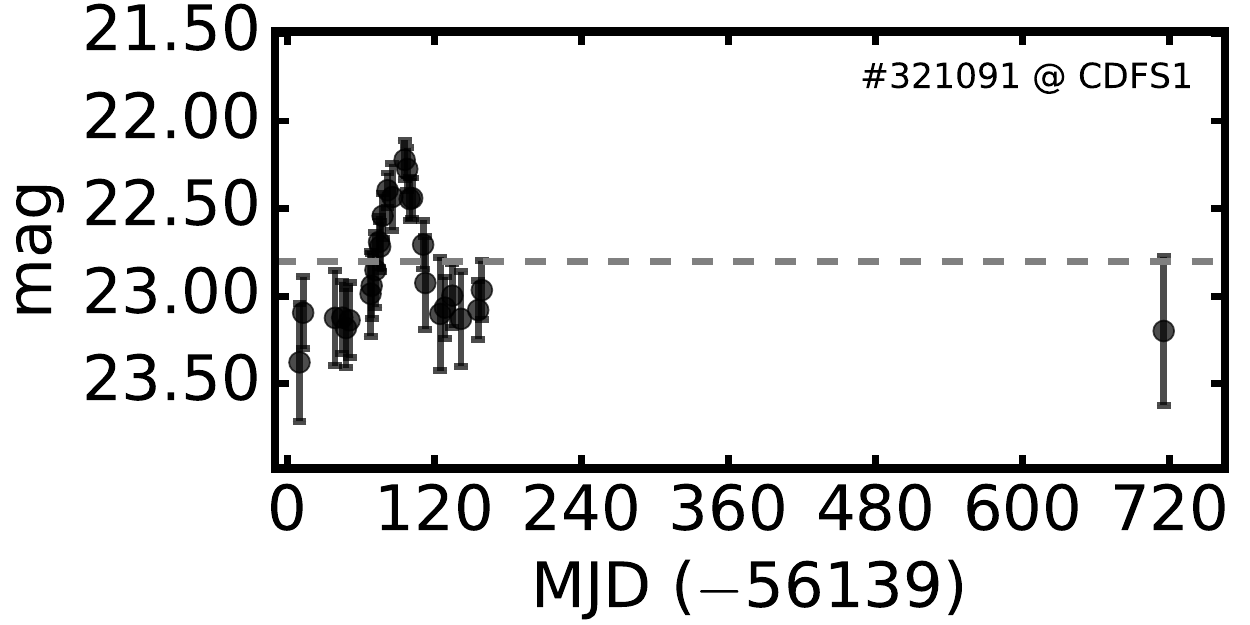}
\includegraphics[width=0.23\textwidth]{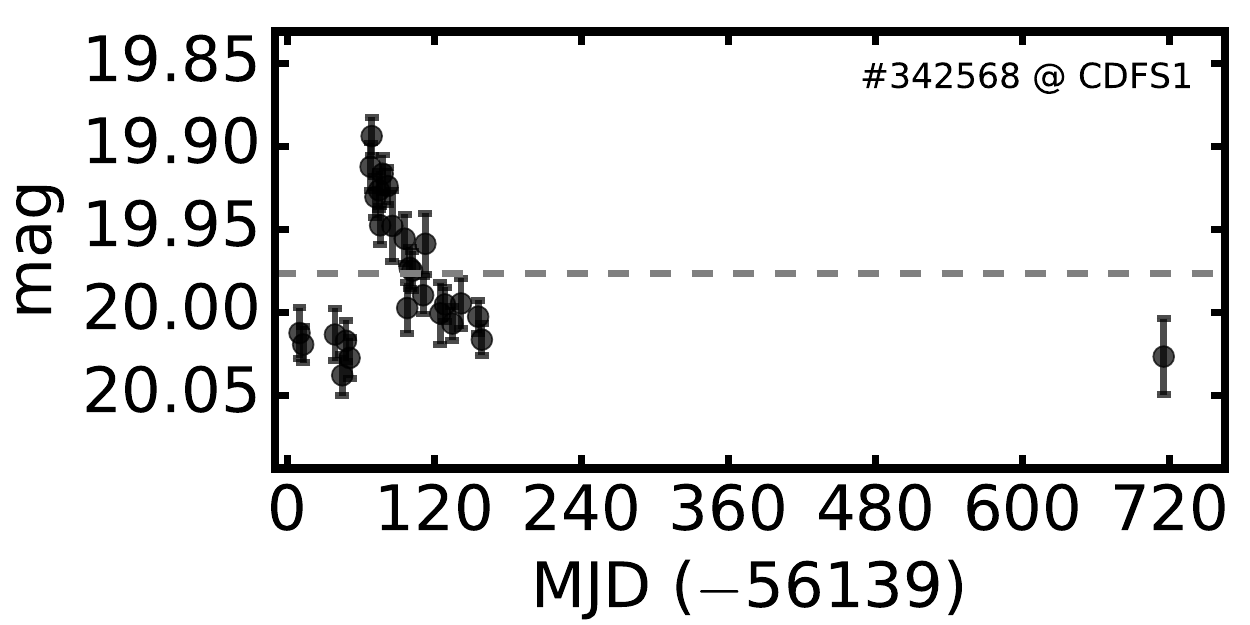}
\includegraphics[width=0.23\textwidth]{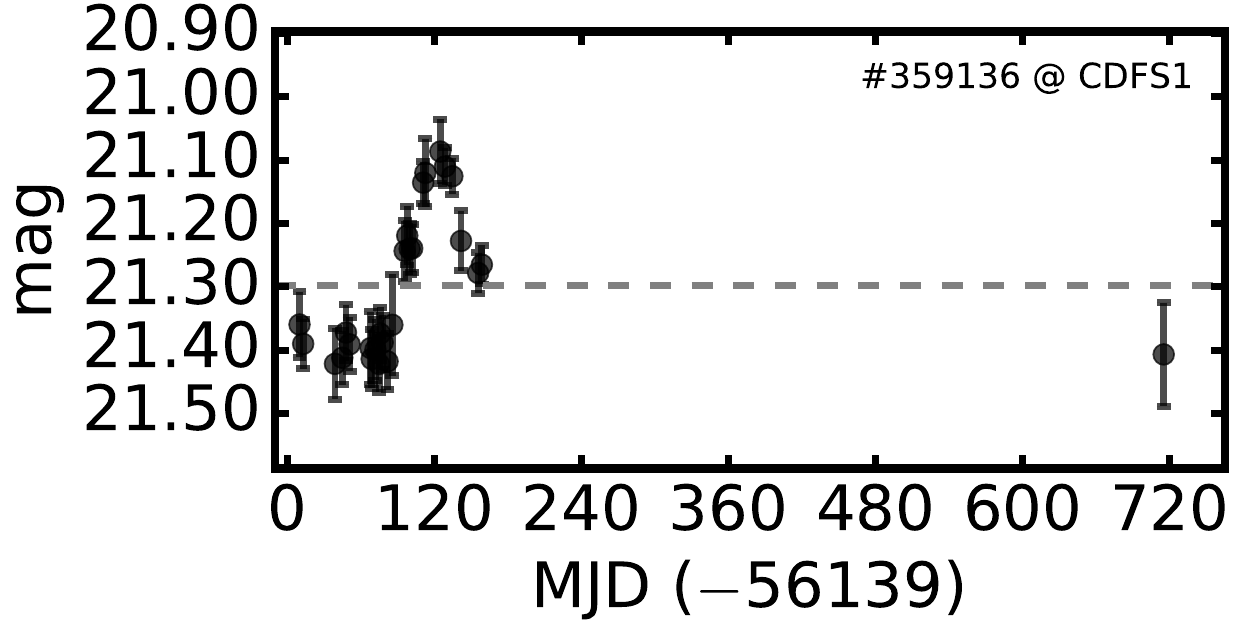}
\includegraphics[width=0.23\textwidth]{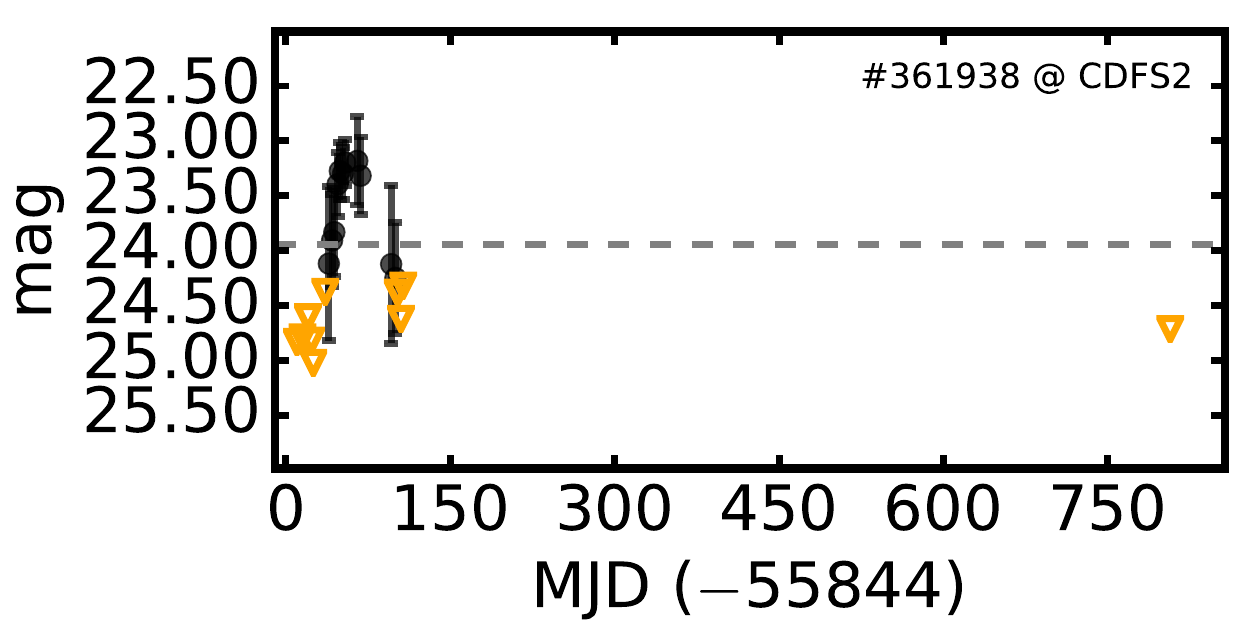}
\includegraphics[width=0.23\textwidth]{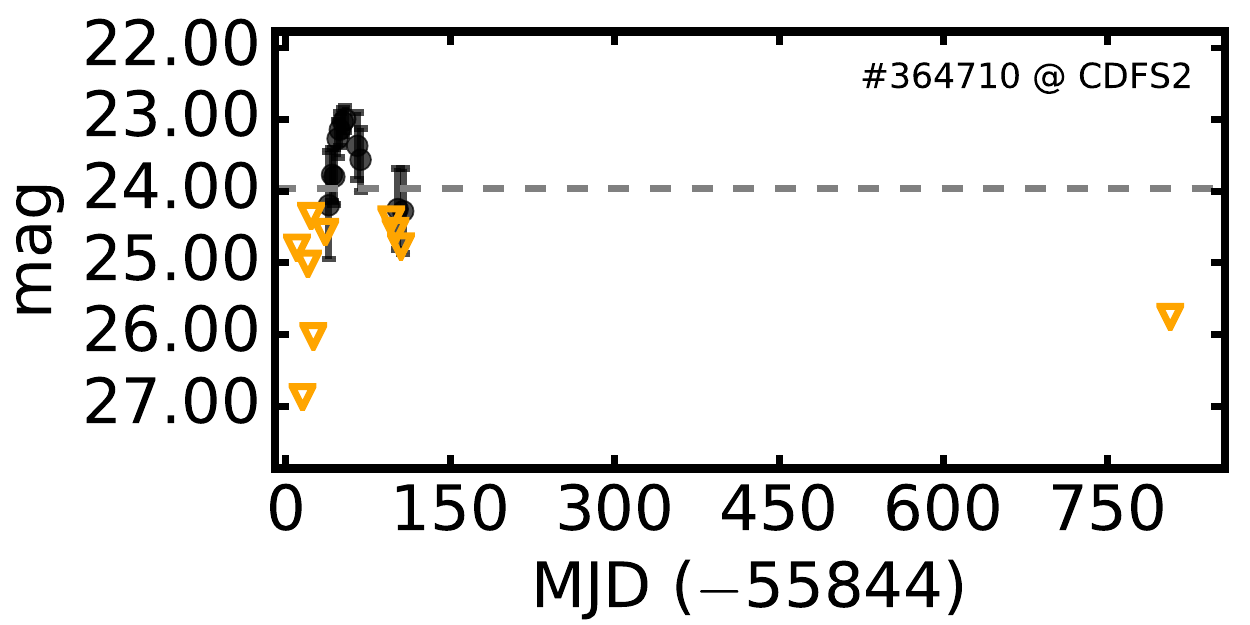}
\includegraphics[width=0.23\textwidth]{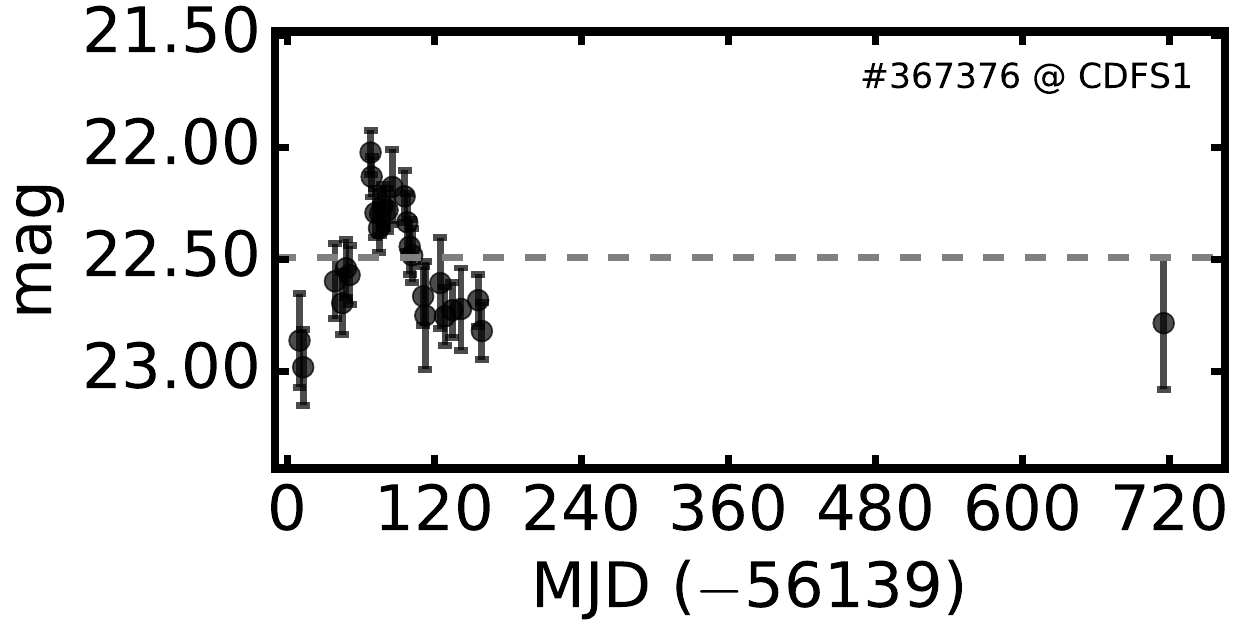}
\caption{Light curves of the transients identified by visual inspection. The orange triangles are 
measures below the 5$\sigma$ limiting magnitude. The labels are the same as in Figure~\ref{fig:agns}.}
\label{fig:trans}
\end{figure*}

\begin{figure*}
\centering
\includegraphics[width=0.96\textwidth]{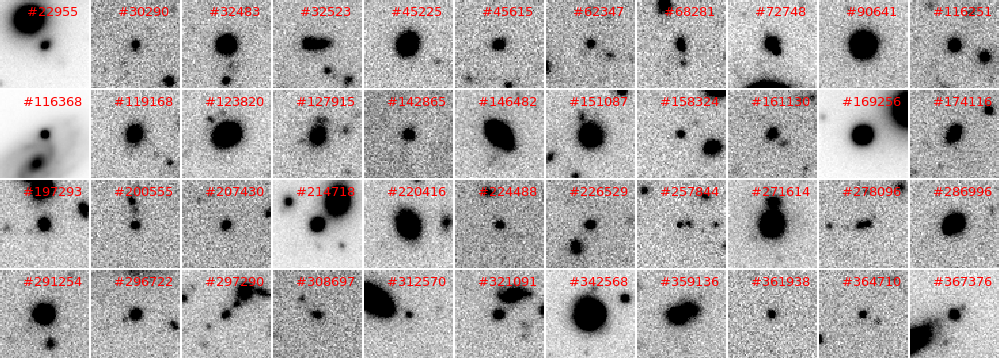}
\includegraphics[width=0.96\textwidth]{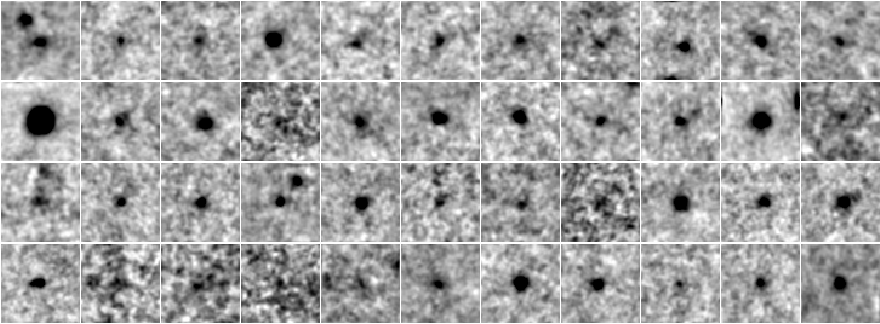}
\caption{\textit{Top four panels}: The $r$-band image stamps of the transients and their host galaxies. 
These stamps are generated by average-stacking all the epochs. The size of each stamp is $65\times65$ pixels, 
corresponding to $13\times13$\,arcsec$^2$. The attached number represents the ID in the \textit{clean} sample. 
\textit{Bottom four panels}: The corresponding difference images of the transients at their peak brightness.}
\label{fig:tstm}
\end{figure*}

%\begin{figure}
%\centering
%\includegraphics[width=0.48\textwidth]{hscImg.png}
%\caption{Image stamps of object \#158324 from HSC observations. They are 
%$g$-band, $r$-band, $i$-band and $z$-band images from left to right with Kron magnitudes 
%of 25.23, 25.34, 24.72, and 24.06\,mag, respectively.}
%\label{fig:hscstm}
%\end{figure}

\section{Results}\label{sec:res}
The CDFS field has also been targeted by a number of photometric surveys, such as  
the \textit{GALEX} ultraviolet survey \citep{2005ApJ...619L...1M}, the Dark Energy Survey 
(DES; \citet{2016MNRAS.460.1270D}), the deep Hyper Suprime-Cam survey (HSC, 
\citet{2019RNAAS...3a...5N}), the Pan-STARRS1 optical survey, the VIDEO near-infrared 
survey \citep{2013MNRAS.428.1281J}, the Spitzer SERVS and SWIRE mid/far-infrared surveys 
\citep{2012PASP..124..714M,2003PASP..115..897L}, the HerMES submilimeter survey 
\citep{2012MNRAS.424.1614O} and the ATLAS radio survey \citep{2006AJ....132.2409N,
2015MNRAS.453.4020F}, producing  a wealth of imaging data with large wavelength coverage. 
Spectroscopic observations in the field includes the 2dF Galaxy Redshift Survey 
\citep{2001MNRAS.328.1039C}, the 6dF Galaxy Survey \citep{2004MNRAS.355..747J,
2009MNRAS.399..683J}, VVDS ``Deep" survey \citep{2005A&A...439..845L,2013A&A...559A..14L}, 
the VANDELS  survey \citep{2018A&A...616A.174P}, the VUDS  survey \citep{2017A&A...600A.110T}, 
and the ongoing DEVILS survey \citep{2018MNRAS.480..768D}. The central region of the field was 
also observed by a series of deep Chandra and XMM-Netwon X-ray surveys \citep{2017ApJS..228....2L,
2016ApJS..224...15X,2011A&A...526L...9C}, and the CANDELS survey \citep{2011ApJS..197...35G,
2011ApJS..197...36K}. A number of time-domain surveys and data, such as the \textit{Catalina} real-time 
transient survey \citep{2009ApJ...696..870D}, the \textit{SkyMapper} transient survey \citep{2017PASA...34...30S}, 
the \textit{Gaia} variable star catalog \citep{2018A&A...618A..30H}, the Pan-STARRS1  variable 
source catalog \citep{2016ApJ...817...73H}, the Dark Energy Survey Supernova Program 
(DES-SN; \citet{2019ApJ...874..106B}), and the SUDARE-VOICE variability-selected AGN 
sample \citep{2015A&A...579A.115F,2020arXiv200102560P} \textit{etc.}, are also publicly available in  this field.

The multi-band surveys and samples offer crucial datasets to study the physical properties of 
the variables and the host galaxies of transients. Because of the small sky coverage of the SUDARE-VOICE 
survey and limited number of objects in the \textit{clean} sample, it is possible for us to visually inspect all the light 
curves. Quantitatively, we firstly calculate the average magnitude and the standard deviation $\sigma$ 
of each light curve. One object is identified as variable if at least three sequential epochs deviate from the 3$\sigma$ 
region of the average magnitude. In total, we select 207 objects with significant variations. It is found that almost 
80\% of them display AGN-like light curves with aperiodic magnitude variations. Besides, we identify 44 transients, 
each of which shows  significant single peak and dramatic magnitude change in the light curve. To demonstrate 
the effectiveness of the \texttt{drap} method, we show some examples in this section.  More detailed analyses of 
the objects in combination with other multi-band data will be presented in our follow-up work.

We match the \textit{clean} sample with the Million Quasars (MILLIQUAS) 
catalogue (v6.1\footnote{http://www.quasars.org/milliquas.htm}; \citet{2015PASA...32...10F}) 
using a radius of 1.0 arcsecond, obtaining 366 AGNs in total. Figure~\ref{fig:agns} shows the light curves 
of three confirmed AGNs\footnote{The three AGNs are labeled as \texttt{Descrip}=Q in the Million Quasars 
(MILLIQUAS) catalogue, meaning that they are type-I broad-line core-dominated quasars.} with different 
brightness and redshifts (top panel) and three AGN candidates (bottom panel) selected by our visual inspection. 
The AGN (ID \#187389 in the \textit{clean} sample) in the middle of top panel has observations in total 76 epochs 
spanning about three years. The AGN candidate (ID \#198895) in the right of bottom panel, with SNR of 46.0 
in the mosaic image,  presents significant magnitude variations which is as large as about 1.2\,mag. 
Meanwhile, we also show the light curves of the 44 transients in Figure~{\ref{fig:trans}}. The orange 
triangles represent the magnitudes below the 5$\sigma$ detection limit of point source. Figure~\ref{fig:tstm} 
shows the corresponding image stamps generated by average-stacking all the available epochs, as well as the 
difference images of these transients at their peak brightness. It can 
be seen that most of the transients show relatively complete light curves from the starting of the event  
to fading. A large fraction of them are expected to be supernovae. To confirm that, we match these transients 
with the supernova sample detected in the CDFS1-2 sub-fields which used the same dataset 
\citep{2015A&A...584A..62C}. For the 22 transients identified in the two sub-fields, only 6 transients 
(ID\#197293, \#286996, \#291254, \#297290, \#308697 and \#367376) are not classified as supernova by \citet{2015A&A...584A..62C}, 
of which the transient \#197293 only detected during the observation of CDFS4 sub-field. Among the 44 
transients, the one \#158324 shows the largest magnitude variation of about 3.0\,mag although we missed 
the observation of its peak. 
For this transient, we do detect its host galaxy in the HSC deep image (the observation was taken between 
January 2015 and March 2017 after the explosion) with $r$-band magnitude of 
25.34\,mag which is very faint but still consistent with our measurement as shown in the light curve. However, 
there are no near-infrared detections in the VIDEO $JK_{s}$ bands (the limiting magnitudes for the two bands 
are 23.98\,mag and 22.79\,mag, respectively). Since there is no spectroscopic observation during the explosion 
and it now has disappeared in the sky, it will be challenging to classify this transient and investigate its properties 
into much detail.

Figure~{\ref{fig:magrms}} shows the standard deviation $\sigma$ and average magnitude of each light curve 
in the entire \textit{clean} sample. The orange 
squares show the AGNs from the Million Quasar Catalog \citep{2015PASA...32...10F}, and the 
magenta stars represent the 44 transients identified by visual inspection. The running median of 
the individual $\sigma_{i}$ and its standard deviation $\sigma_{[\sigma_{i}]}$ are calculated  in a 
given magnitude bin with width of 0.5\,mag. The dashed cyan line represents the 
3.0$\sigma_{[\sigma_{i}]}$ threshold. It can be seen that one transient is below the threshold because 
only the transients with large magnitude variations tend to be selected by our visual inspection. 
For the AGN sample, however, we find that most of them fainter than 22.0\,mag are below the 
threshold, meaning that the intrinsic dispersions of their light curves are comparable to the overall 
measured uncertainties.  Further investigation indicates that most of them (72.4\%) are X-ray 
detected AGNs \citep{2016ApJS..224...40W}. While the X-ray detected AGNs with magnitude 
brighter than 22.0\,mag only account for 21.5\%. This result indicates that a fraction of AGNs could 
be missed out by only variability-selected method, as discussed in previous works of the SUDARE-VOICE 
collaboration \citep{2015A&A...574A.112D,2015A&A...579A.115F,2019A&A...627A..33D,2020arXiv200102560P}.

\section{Summary}\label{sec:sum}
In this work, we use the SUDARE-VOICE $r$-band imaging data to extract the light curves of the detected 
objects in CDFS field.  The total on-sky time for this field spans over four years, distributed over 
four adjacent sub-fields CDFS1-4. The multi-epoch $r$-band observations were taken with a 
cadence of about 3-4 days, avoiding the ten days around the full moon. Besides, this field has 
also been covered by many multi-band surveys, such as  the GALEX survey, the DES, the deep 
HSC survey, the Pan-STARRS1 survey,  the VIDEO survey and other infrared/radio surveys. The 
abundant datasets provide crucial information for studying the properties of the astronomical 
objects in this sky region.

To measure the light curves, careful image reduction are performed. 
We firstly stack the individual exposures for a given epoch to increase the SNR of the objects. 
The accuracy of the astrometric calibration reaches to 0.06 arcsec along both right ascension and 
declination axes. Then we photometrically calibrate the zeropoints between different epochs so that 
the final accuracy of the photometric calibration is better than 0.02\,mag. These calibrated epochs 
are combined, after rejecting several epochs with poor observational conditions, for object detections. 
Finally, 210,530 objects with high SNR and photometric quality are selected for light curve extractions. 
In addition, unbiased photometry between different epochs requires accurate PSF modeling. Taking 
into account the significant spatial variations of the PSF, we split each epoch into 4$\times$2 sub-images 
with uniform size, and construct the  PSF model for each sub-image individually of which the spatial 
variation is described by polynomial interpolation. 

For each object, the image stamps of all available epochs and corresponding PSF models are 
extracted. We introduce a new method, namely \texttt{drap}, to 
measure the light curves of these objects. The mathematics of this method is quite straightforward.  
It can moderately average out the difference of PSFs between different epochs, and suppress 
the background fluctuations.  We estimate the photometric uncertainty of the light curves by taking the 
noise  correlation into consideration, and perform detrending correction to eliminate the systematic 
biases due to the inaccurate image reduction and PSF modeling. We visually inspect the light curves 
to select variable objects. As expected, most of the variable objects are AGNs with aperiodic and 
long-term magnitude variations.  We identify 44 transients with significant magnitude variations. 
For the 22 transients in CDFS1-2 fields, 16 of them are classified as supernova by 
\citet{2015A&A...584A..62C}, meaning that most of the transients we identified are supernovae. 
We will perform further studies on these objects in combination with multi-band data in the follow-up work.

%The main purpose of this paper is to describe the procedures to extract the $r$-band light curves of 
%the detected objects. The classification of them in combination with multi-band data, and specific 
%studies of the AGNs properties will be presented in our follow-up works. 

\section*{acknowledgements}
DZL thanks Zhenya Zheng and Chenggang Shu for their helpful discussions and comments.
This work is supported by the Launching Research Fund for Postdoctoral Fellow from the 
Yunnan University with grant C176220200 and the China Postdoctoral Science Foundation with 
Grant No. 2019M663582.  
ZHF acknowledges the support
of National Nature Science Foundation of China (NSFC) under the grants 11933002, 11333001, 
and 11653001. 
LPF acknowledges the support from NSFC grants  11722326, 11673018 \& 11933002, STCSM 
grant 18590780100, 19590780100, 188014066, the Innovation Program 2019-01-07-00-02-E00032 
and  Shuguang Program 19SG41  supported by SMEC. 
GC acknowledges the SWIFAR visiting fellow program under which he had a fruitful visit to the 
South-Western Institute for Astronomy Research, Yunnan University.
MV and LM acknowledge support from the Italian Ministry of Foreign Affairs and
International Cooperation (MAECI Grant Number ZA18GR02) and the South African
Department of Science and Technology's National Research Foundation (DST-NRF
Grant Number 113121) as part of the ISARP RADIOSKY2020 Joint Research Scheme. 
Support for G.P. is provided by the Ministry of Economy, Development, and Tourism's Millennium 
Science Initiative through grant IC120009, awarded MAS.

This work has made use of data from the European Space Agency (ESA) mission
{\it Gaia} (\url{https://www.cosmos.esa.int/gaia}), processed by the {\it Gaia}
Data Processing and Analysis Consortium (DPAC,
\url{https://www.cosmos.esa.int/web/gaia/dpac/consortium}). Funding for the DPAC
has been provided by national institutions, in particular the institutions
participating in the {\it Gaia} Multilateral Agreement.

\newpage
\appendix
\section{Minimizing the PSF variations by average stacking}\label{appen}
We stack the individual PSFs to derive the master PSF $\mathrm{P_{0}}$ by 
\begin{eqnarray}
\mathrm{P}_{0} = \sum_{i}^{n}\mathrm{P}_{i}/n,
\end{eqnarray}
where $\mathrm{P}_{i}$ is the PSF of the $i$th stamp and $n$ is the total number of stamps. 
Similarly, excluding the PSF $\mathrm{P}_{j}$ of the $j$th stamp, then we can generate 
the stacked PSF $\tilde{\mathrm{P}}_{j}$. It is noted that  both the individual and stacked PSFs 
have been normalized so that the sum of all the pixels is equal to one. Since the stacking is performed 
pixel by pixel, for simplicity but without loss of generality, we can instead analyze the behavior of the 
stacked PSFs in an arbitrary pixel position $(x,\,y)$, where $x$ and $y$ represent the pixel indices of 
the two-dimensional PSF matrix. For a specified pixel position $(p,\,q)$, the intensities of the master 
PSF $\mathrm{P_{0}}$ and $\tilde{\mathrm{P}}_{j}$ can be calculated respectively by 
\begin{eqnarray}
\mathrm{P}_{0}^{pq} = \sum_{i}^{n}\mathrm{P}_{i}^{pq}/n \quad \mathrm{and} \quad 
\tilde{\mathrm{P}}_{j}^{pq} = \sum_{i \neq j}^{n}\mathrm{P}_{i}^{pq}/(n-1),
\end{eqnarray}
where $\mathrm{P}_{i}^{pq}$ is the intensity of the  $i$th PSF at position $(p,\,q)$. 
Through simple mathematical transformation, we can find that $\mathrm{P}_{0}^{pq}$ and 
$\tilde{\mathrm{P}}_{j}^{pq}$ satisfies 
\begin{eqnarray}
n\mathrm{P}_{0}^{pq} = (n-1)\tilde{\mathrm{P}}_{j}^{pq} + \mathrm{P}_{j}^{pq},
\end{eqnarray}
namely,
\begin{eqnarray}
\mathrm{P}_{0}^{pq} - \tilde{\mathrm{P}}_{j}^{pq} = \frac{\mathrm{P}_{j}^{pq} - \mathrm{P}_{0}^{pq}}{n-1}.
\end{eqnarray}
It proves that the two PSFs  $\mathrm{P}_{0}$ and $\tilde{\mathrm{P}}_{j}$ are almost identical 
if $n$ is much larger than the difference of the pixel values between $\mathrm{P}_{j}$ and 
$\mathrm{P}_{0}$. For the VOICE data, the mean difference of the central values between the normalized 
$\mathrm{P}_{j}$ and $\mathrm{P}_{0}$ is 0.03, meaning that $|\mathrm{P}_{0}^{pq} - \tilde{\mathrm{P}}_{j}^{pq}|\lesssim10^{-3}$ 
when $n$\,$\simeq$\,20.

\end{document}